\documentclass[useAMS,usenatbib]{mn2e}

\usepackage{graphicx}
\usepackage{subfigure}
\usepackage{caption}
\usepackage{appendix}
\usepackage{amsmath}
\usepackage{arydshln}
\usepackage{multirow}

\title[Chaotic behaviour in pulsars]{Evidence for chaotic behaviour in pulsar spin-down rates}
\author[A. D. Seymour and D. R. Lorimer ]{A. D. Seymour$^{1}$\thanks{E-mail:
aseymour@mix.wvu.edu} and D. R.
Lorimer$^{1,2}$ \\
$^{1}$Department of Physics, West Virginia University, Morgantown, WV 26505, U.S.A.\\
$^{2}$National Radio Astronomy Observatory, Green Bank, WV 24944, U.S.A.}
\begin{document}

\date{Accepted 2012 September 20. Received 2012 September 18; in original form 2012 June 21}


\maketitle

\label{firstpage}

\begin{abstract}
We present evidence for chaotic dynamics within the spin-down rates of 17 pulsars originally presented by Lyne et al. Using techniques that allow us to re-sample the original measurements without losing structural information, we have searched for evidence of a strange attractor in the time series of frequency derivatives for each of the 17 pulsars. We demonstrate the effectiveness of our methods by applying them to a component of the Lorenz and R\"ossler attractors that were sampled with similar cadence to the pulsar time series. Our measurements of correlation dimension and Lyapunov exponent show that the underlying behaviour appears to be driven by a strange attractor with approximately three governing non-linear differential equations. This is particularly apparent in the case of PSR B1828$-$11 where a correlation dimension of $2.06\pm0.03$ and a Lyapunov exponent of $(4.0\pm0.3)\times10^{-4}$ inverse days were measured. These results provide an additional diagnostic for testing future models of this behaviour. 
\end{abstract}


\section{Introduction}
	Pulsars are spinning neutron stars whose emission is thought to be driven by their magnetic fields \citep[see, e.g.,][]{Manchester:1977fk}. Their 
	dynamics exhibit a wide ranging degree of stability, with the fastest spinning and oldest `millisecond pulsars' generally being the most predictable. 
	 \cite{1996A&amp;A...308..290P} showed that these millisecond pulsars can be as stable as atomic clocks on large time-scales. 
	
	Because pulsars are so stable, it is surprising when we see them misbehave. Departure from normal behaviour, characterized by steady emission and 
	rotation, can occur in a variety of ways. Some pulsars have nulling events, where the emission seems to 
	turn off for a while and then suddenly turn back on \citep{BACKER:1970fk}. Even more extreme behaviour can be seen in the intermittent pulsar 
	B1931+24, which behaves like a normal pulsar for five to ten days, then is undetectable for about 25 days \citep{Kramer28042006}.
	Recently even longer-term intermittency (spanning hundreds of days) has been reported in two other pulsars \citep[][Lorimer et. al. 2012 in prep.]{0004-637X-746-1-63}. Finally, 
	another related class of pulsars known as Rotating Radio Transients (RRATs) seem to sporadically turn their emission on and off on a wide range 
	of time-scales. 
	
	It is even more shocking when we see changes in the dynamics of a pulsar. Though pulsars are expected to gradually spin-down over time, due 
	to an energy loss from magnetic braking \citep{GOLD:1969uq}, other changes in the dynamics are highly unexpected. These fluctuations often give 
	us insight to the interior and the environment of a pulsar. One such phenomenon, known as a `glitch', is a sudden discrete increase 
	in rotation that has been attributed to superfluid vortices within the interior of the pulsar \citep{ANDERSON:1975fk}. 
	
	Since the dynamical fluctuations in pulsars were unexpected, a bulk of the irregularities were considered to be `timing noise' \citep{1009-9271-6-S2-31}. 
	When these irregularities were observed over a 20 yr time span, large time-scale fluctuations in the spin-down rate became clear. \cite{DataHome} 
	describe the irregularities as `quasi-periodic' and were able to relate them to the pulse shape. There have been several different
	 processes proposed for these fluctuations from precession \citep{2012MNRAS.420.2325J} to non-radial modes \citep{2041-8205-728-1-L19}. 
	 Yet, the mechanisms that govern these fluctuations and their connection to the pulse shape is still a mystery. Quasi-periodicities are often a sign of a 
	non-linear chaotic system. Previous chaotic studies on pulsars \citep{1990ApJ...353..588H,1998AAS...192.6802D,0004-637X-519-1-291} focused 
	on emission abnormalities and timing noise for particular pulsars. 
		
	In this paper we search for chaotic behaviour within the spin-down rate of 17 pulsars presented by \cite{DataHome}. In Section 2 we give a 
	brief introduction to chaotic systems and their behaviours. In Section 3 we present techniques to form an evenly sampled time series with the 
	same structural information as an unevenly measured series. In Section 4 we demonstrate methods to search for different chaotic characteristics 
	within a time series, and test our algorithm with known chaotic systems. In Section 5 we discuss our results and their implications. The techniques 
	presented here are very general and can be used on almost any time series. Therefore, we have written this paper explicitly in the hope that these 
	techniques will be more approachable, and that they will be used more frequently within the pulsar community.  
  
\section{Chaotic Behaviour} \label{sec:CB}
	In everyday conversation, the word chaos is often interchangeable with randomness, but in dynamical studies these are two distinct ideas. 
	Chaos is continuous and deterministic with underlying governing equations, while randomness is more complex and uncorrelated; 
	values at an earlier time have no effect on the values at a later time.
	
	One of the characteristics of chaos has been colourfully described by Edward \cite{Lorenz:1993vn} as the `butterfly effect'. Lorenz asks, 
	\emph{`Does the flap of a butterfly's wings in Brazil set off a tornado in Texas?'} This is used to illustrate an instability, where a system 
	is highly sensitive to initial conditions. The consequence is that if there is a small displacement in the initial conditions, the 
	difference between the two scenarios will grow exponentially to cause significant changes at a later time. This instability does not arise 
	in linear dynamics and is a chaotic phenomenon in non-linear systems \citep{1992scma.conf..411S}. We will utilize this chaotic trait in 
	Section \ref{sec:BE}.
	
\begin{figure} 
	 \centering
		\mbox{
		\subfigure[]
	  		{\includegraphics[width=.45\linewidth, trim =30 15 30 15, clip=true]{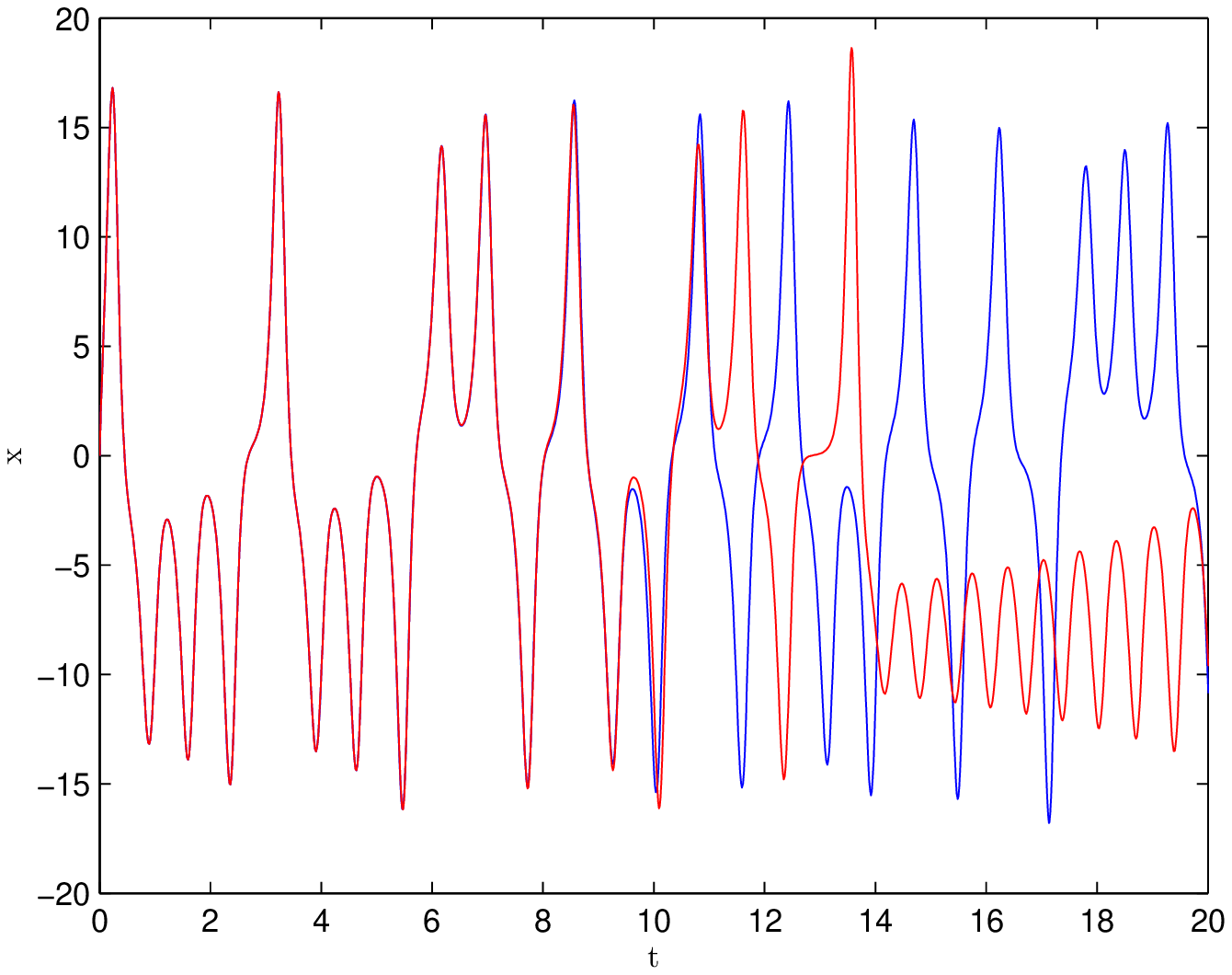}
	     		\label{fig:SIC}}
		\quad
		\subfigure[]
	  		{\includegraphics[width=.45\linewidth,trim = 30 15 30 15, clip=true]{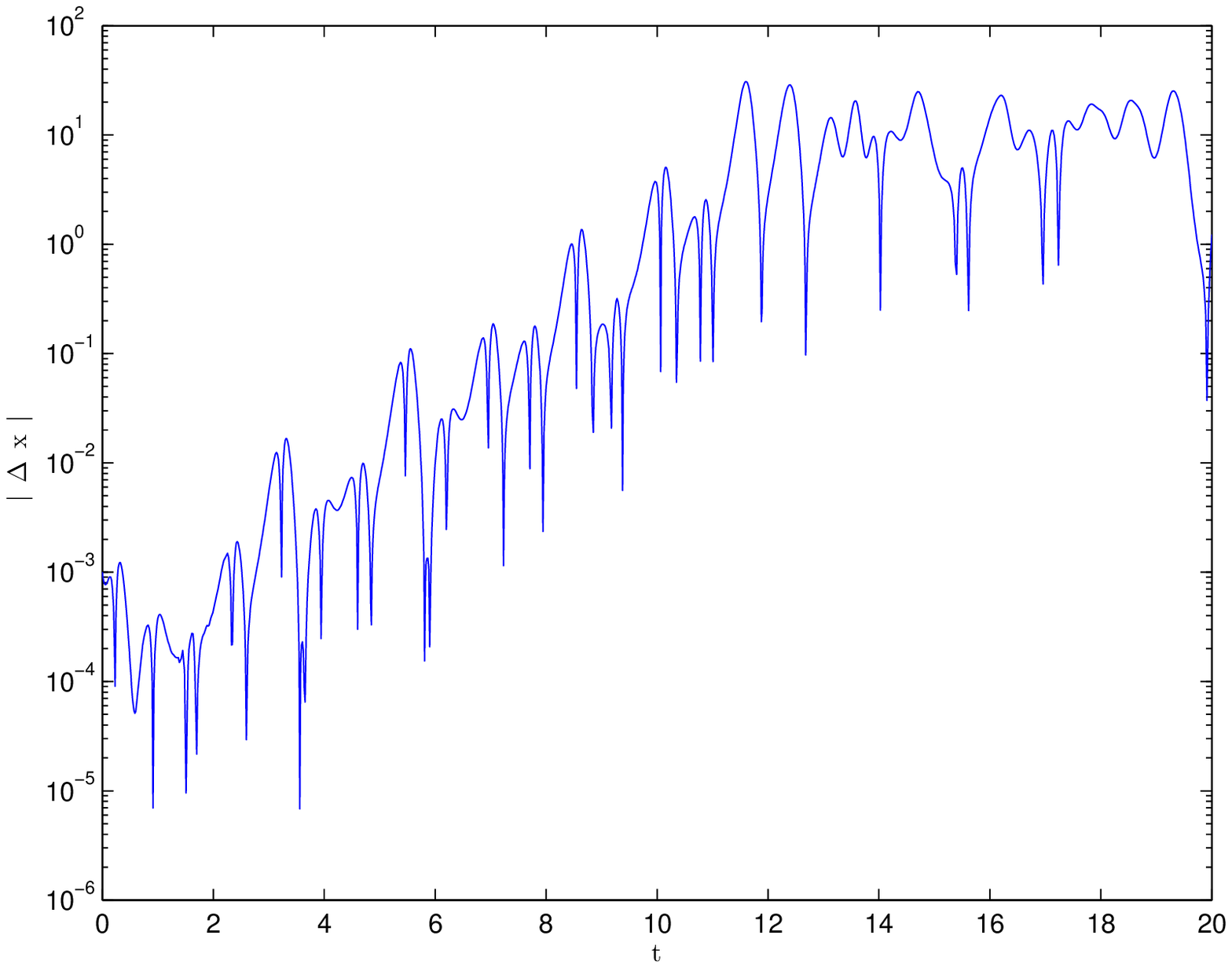}
	     		\label{fig:DeltaX}}
		}
		\caption{$(a)$ The $x$ component of the Lorenz equations whose initial conditions differed by $10^{-3}$ in $x$. $(b)$ The difference between 
		the two scenarios in \emph{(a)} over time. An overall increasing exponential trend can be seen.} 
		\label{fig:SICex}
	\end{figure}

	An example of this behaviour is shown in Fig. \ref{fig:SICex} for the Lorenz system of equations:
	\begin{equation}
		\begin {array}{l}
			\displaystyle \dot{x}= \sigma ( y - x ) \\
			\displaystyle \dot{y}= x ( \rho - z ) - y \\
			\displaystyle \dot{z}= xy - \beta z. \\
		\end{array}
		\label{eq:Lorenz}
	\end{equation}
	Here $\sigma$, $\rho$, and $\beta$ are positive parameters. \cite{Lorenz:1963fk} derived this system from a simplified model of 
	convection rolls in the atmosphere. In the original derivation $\sigma$ is the Prandtl number, $\rho$ 
	the Rayleigh number and $\beta$ has no proper name but relates to the height of the fluid layer. As for the governing variables, 
	$x$ is proportional to the intensity of the convective motion, $y$ is proportional to the temperature difference of the acceding and descending currents, 
	and $z$ is proportional to the distortion of the vertical temperature profile from linearity \citep{Lorenz:1963fk}. Since then, the Lorenz 
	equations have appeared in a wide range of physical systems. 
	
\begin{figure*} 
	 \centering
		\mbox{\subfigure[]
	  		{\includegraphics[width=.4\linewidth,trim = 50 0 75  25,clip=true]{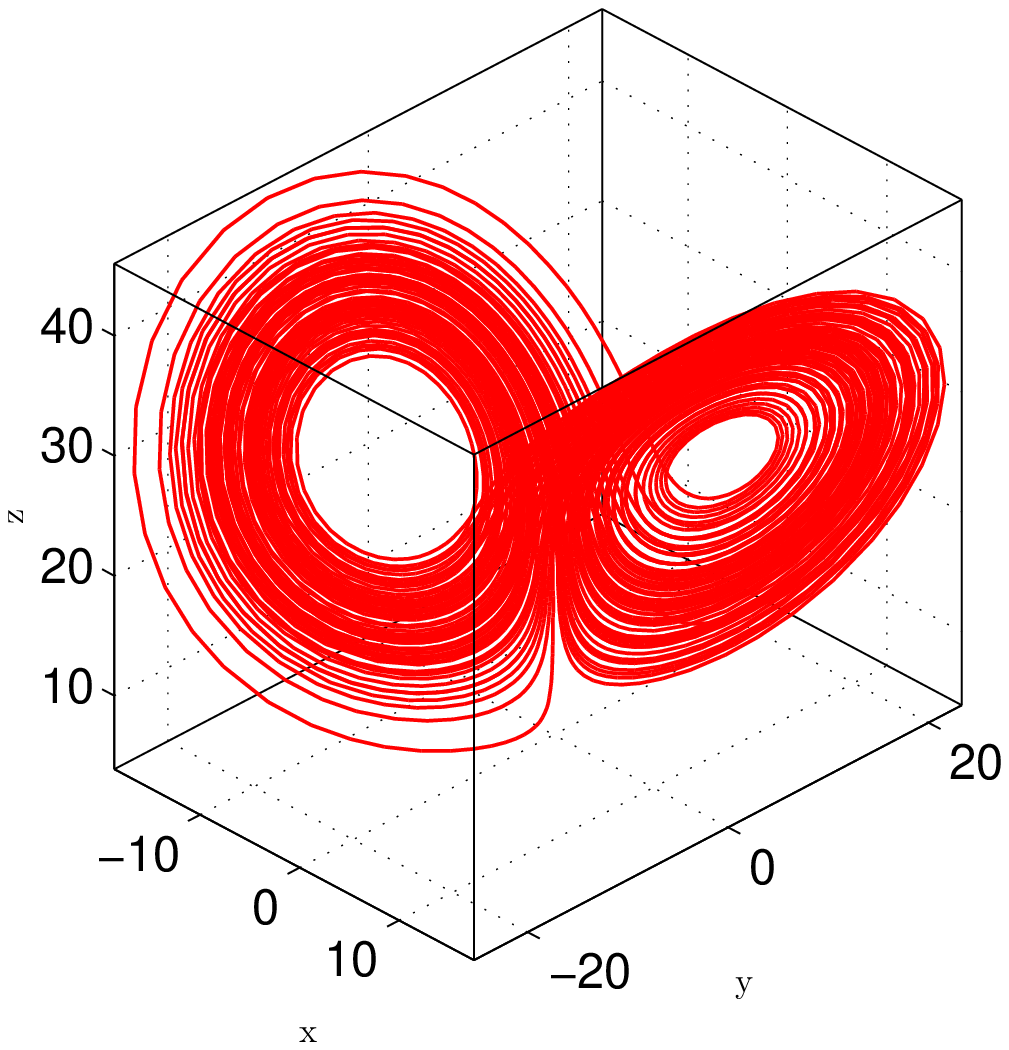}
	     		\label{fig:LorenzAtt}}
		\quad
		\subfigure[ ]
	  		{\includegraphics[width=.4\linewidth,trim = 50 0 75  25,clip=true]{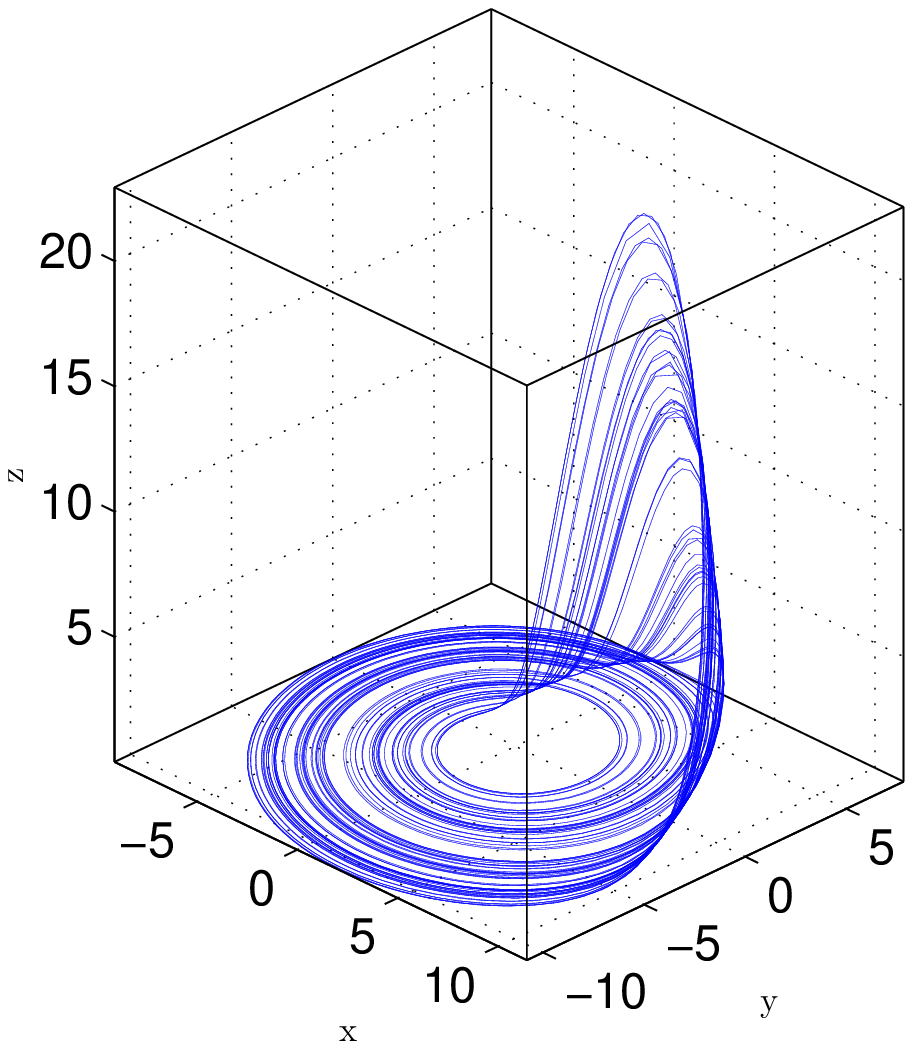} 
	     		\label{fig:RosslerAtt}}
		}
	\caption{Three dimensional visualization of two chaotic attractors. $(a)$ The Lorenz attractor from $t= 0$ to $100$ with $\sigma=10$, $\rho=28$, 
	$\beta=8/3$ and $x_{o}=0$, $y_{o}=10$, $z_{o}=10.2$. $(b)$ The R\"ossler attractor from $t= 0$ to $500$ with $a=0.2$, $b=0.2$, $c=5.7$ 
	and $x_{o}= -1.887$, $y_{o}= -3.5$, $z_{o}= 0.09789$} 
	\label{fig:Att}
	\end{figure*}

	It is important to note that there are no analytical solutions to most non-linear equations. Often to produce a solution, as used in Fig. \ref{fig:Att}, the 
	system is marched forward with small enough time steps that enable linear relationships to be used to simulate a function. We can see from the 
	governing equations that any function that is produced will be highly dependent on the other variables. When these time functions are plotted 
	with respect to each other, as seen in Fig. \ref{fig:Att}(a), they trace out a rather odd surface. 
	
	The Lorenz equations are not the only system in which this occurs. \cite{Rossler} was in search of a simpler set of equations with similar chaotic 
	behaviour to the Lorenz attractor. He came up with with the following three equations with only one non-linear term $zx$:
	\begin{equation}
		\begin{array}{l}
			\displaystyle \dot{x}=- y - x \\
			\displaystyle \dot{y}= x + ay \\
			\displaystyle \dot{z}= b - z ( x - c ). \\
		\end{array}
		\label{eq:Rossler} 
	\end{equation}
	Here $a$, $b$, and $c$ are parameters. Again, plotting the variables against each other produces a different bizarre surface, seen in Fig. 
	\ref{fig:RosslerAtt}. Though the R\"ossler equations started solely as a mathematical construct, analogous behaviour has been seen in 
	chemical reactions \citep{doi:10.1021/ar00144a002}.
	 
	The complex shapes that the dynamics form are known as `strange attractors'. They are called `attractors' because, regardless of the initial
	 conditions, the functions will converge to a path along these surfaces. They are `strange' because they are fractal in nature. The dimension 
	 of the attractor will be a non-integer that is less than the number of equations. We will discuss this more in Section \ref{sec:MD}.
	
\begin{figure}
	 	\begin{tabular}{c c}
			\includegraphics[width=.5\linewidth,trim=20 10 30 20, clip=true]{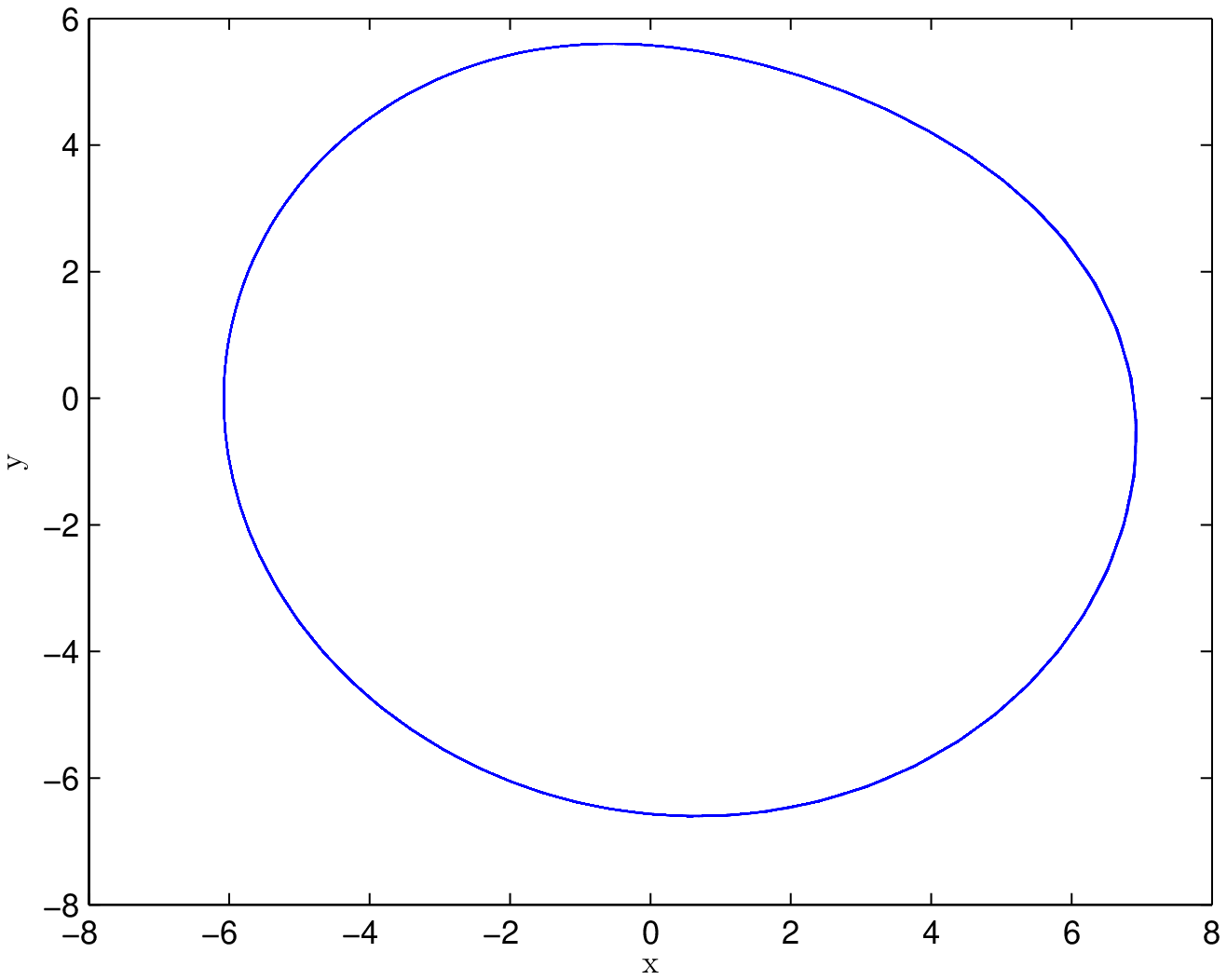} &
	     			\includegraphics[width=.5\linewidth,trim=20 10 30 20, clip=true]{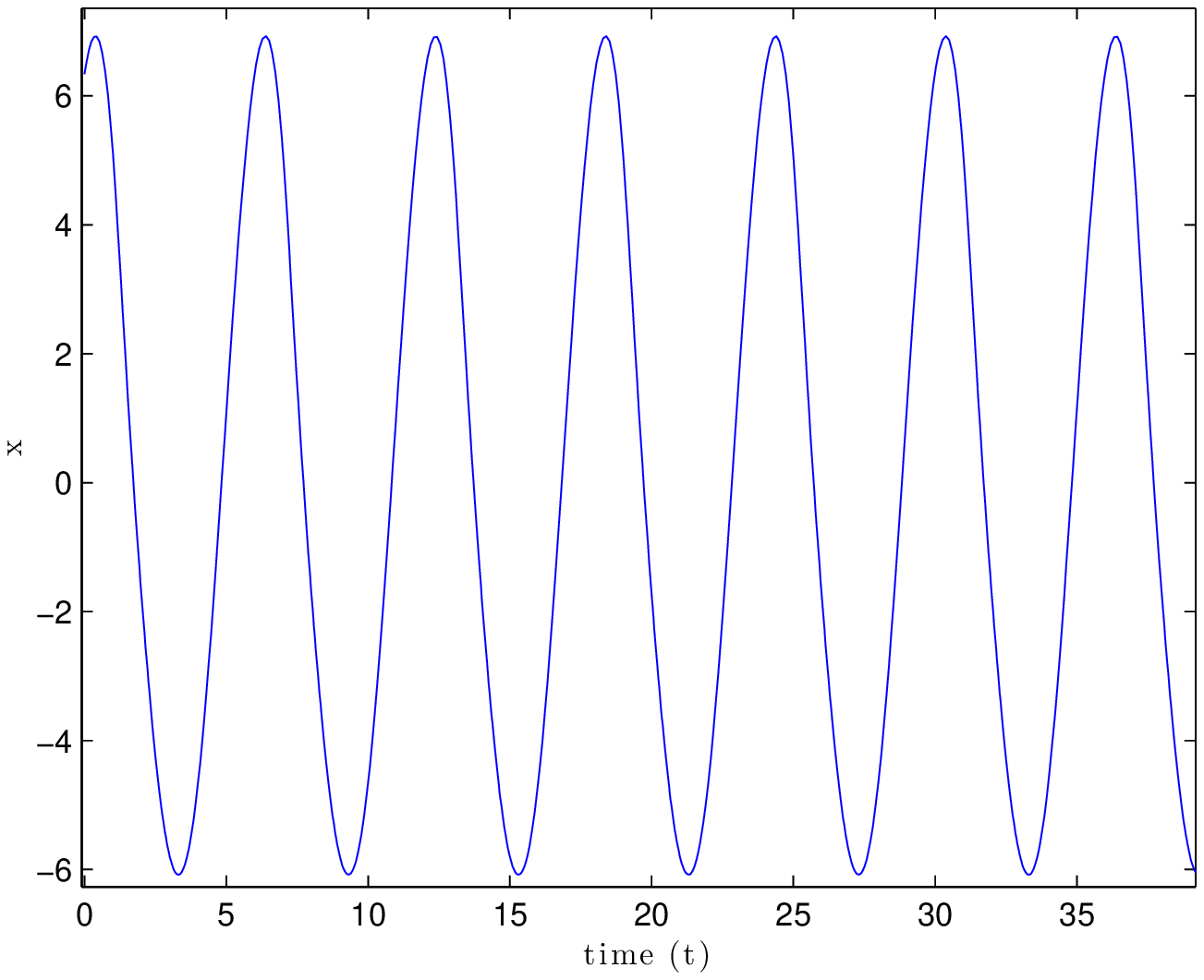}\\
			\multicolumn{2}{c}{Periodic $c=4$}\\ 	
			\includegraphics[width=.5\linewidth,trim=20 10 30 20, clip=true]{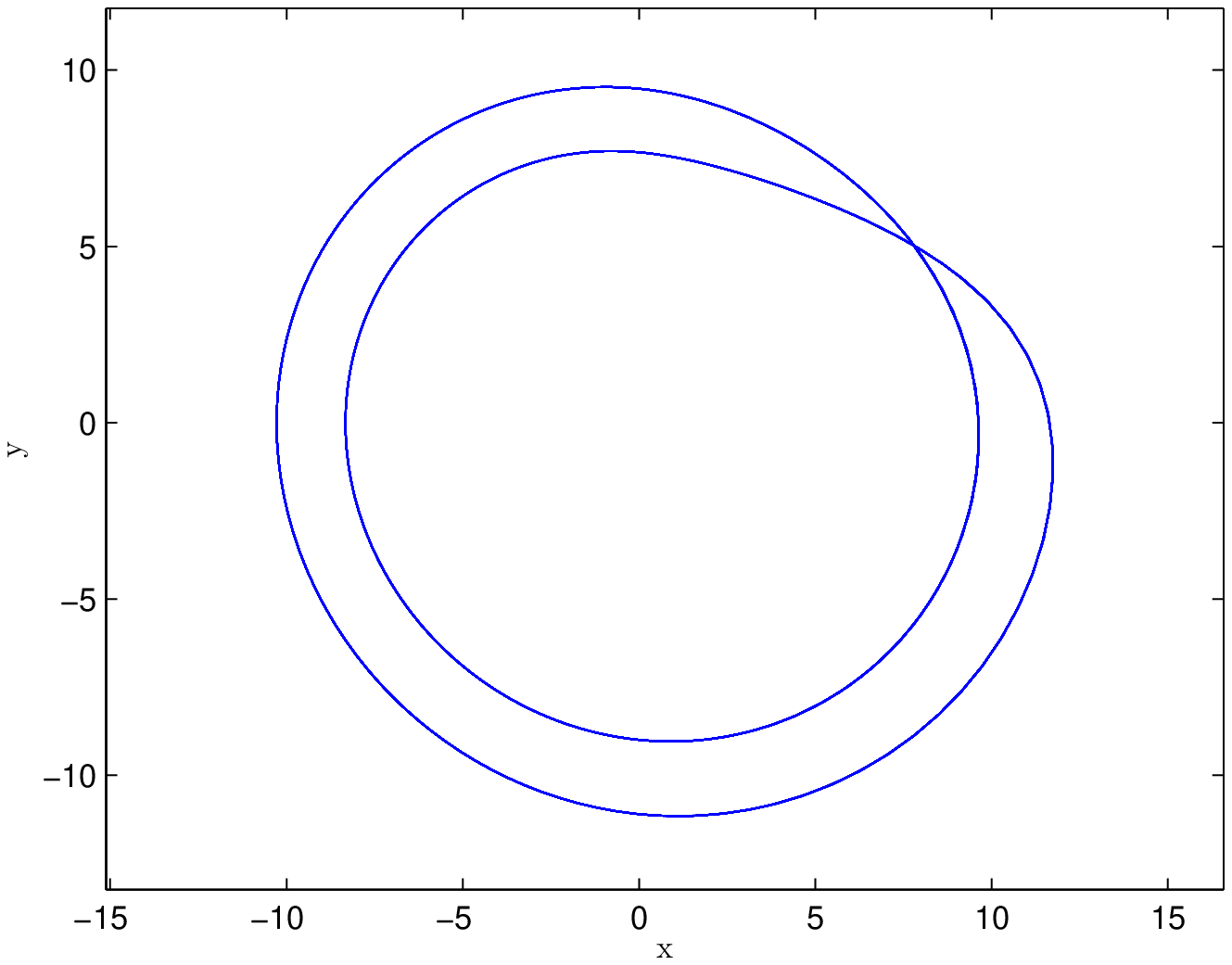} & 
	     			\includegraphics[width=.5\linewidth,trim=20 10 30 20, clip=true]{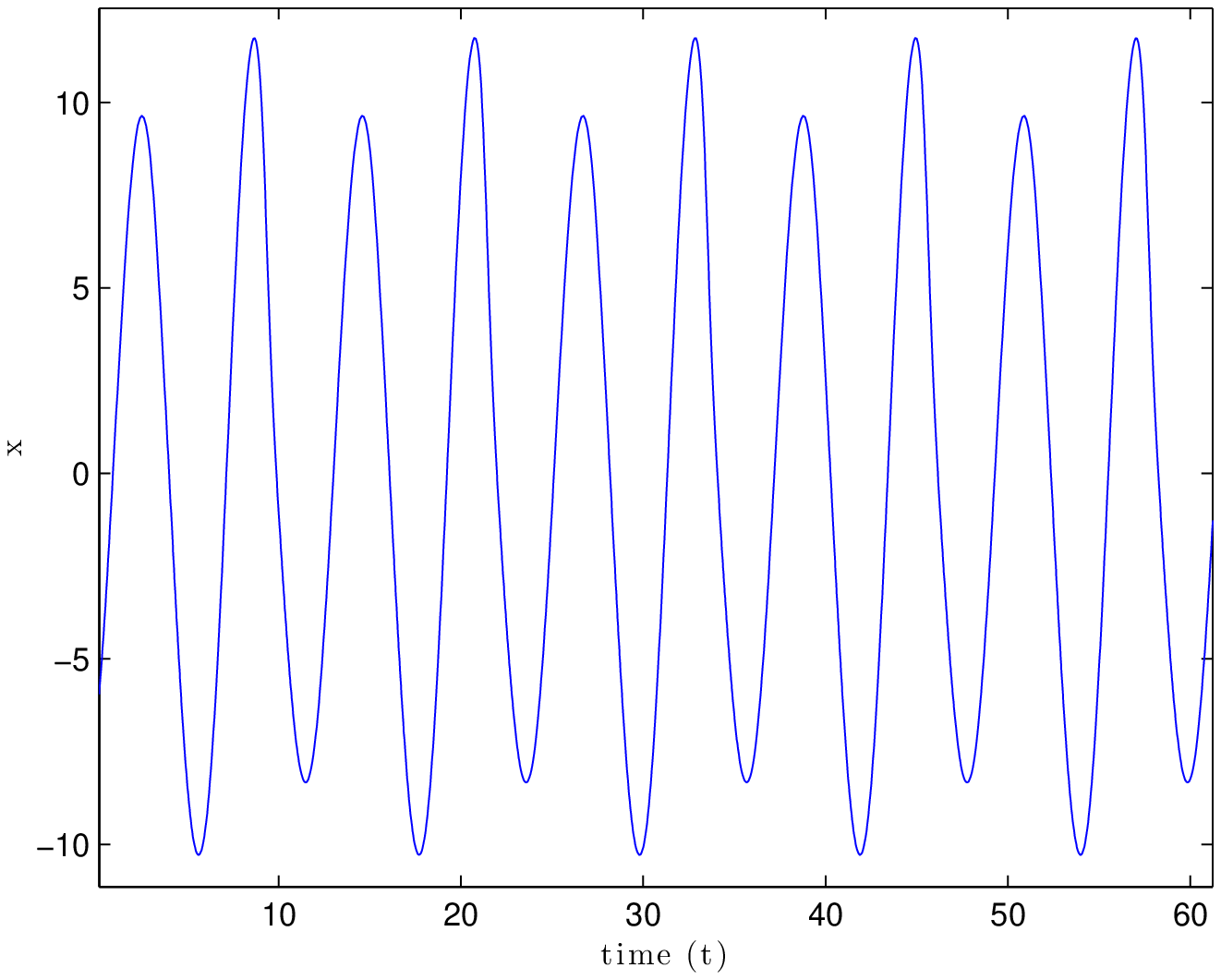}\\
			\multicolumn{2}{c}{Period two $c=7$}\\
			$\bullet$ & $\bullet$ \\
			$\bullet$ & $\bullet$ \\
			$\bullet$ & $\bullet$ \\
			\includegraphics[width=.5\linewidth,trim=20 10 30 20, clip=true]{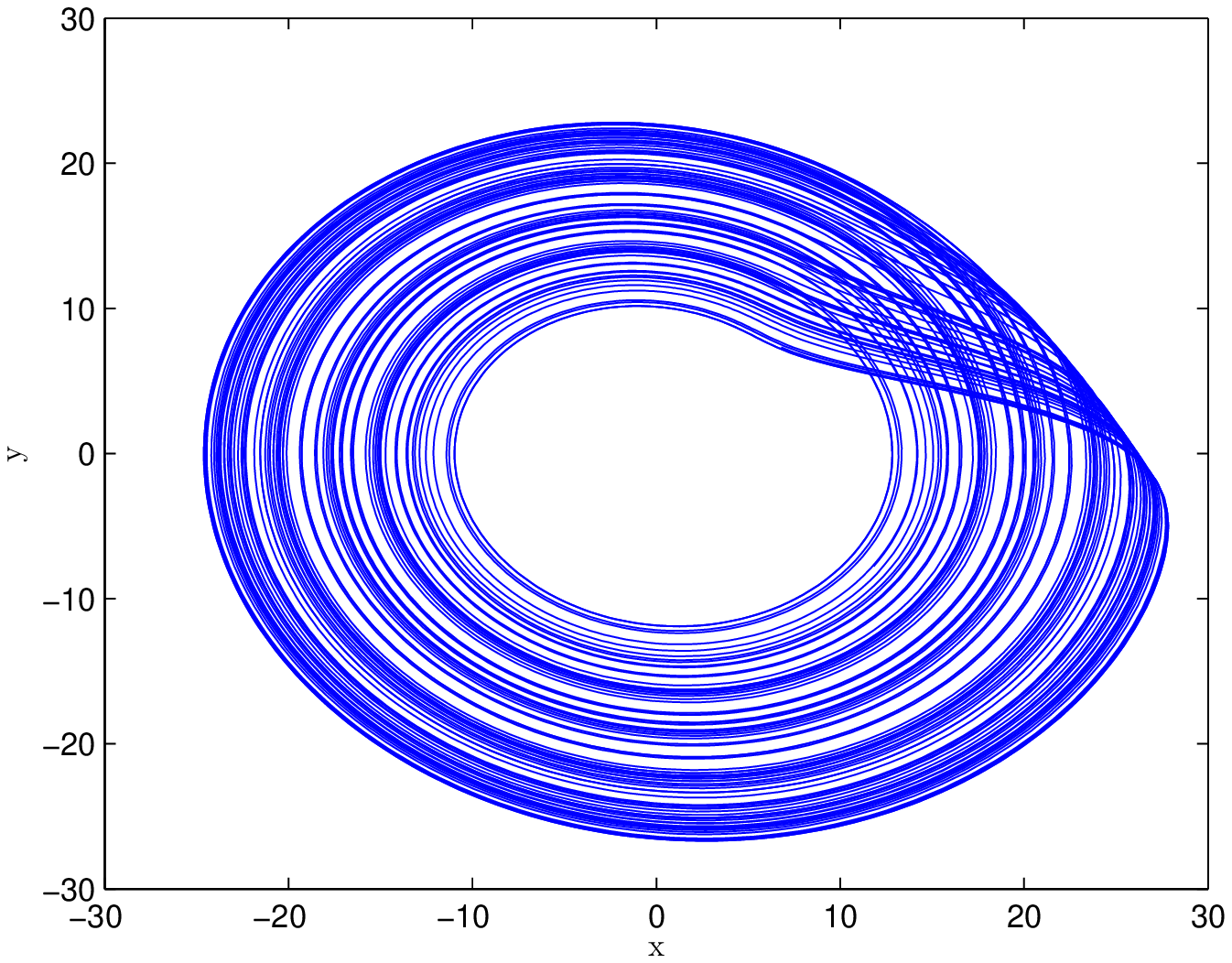} & 
	     			\includegraphics[width=.5\linewidth,trim=20 10 30 20, clip=true]{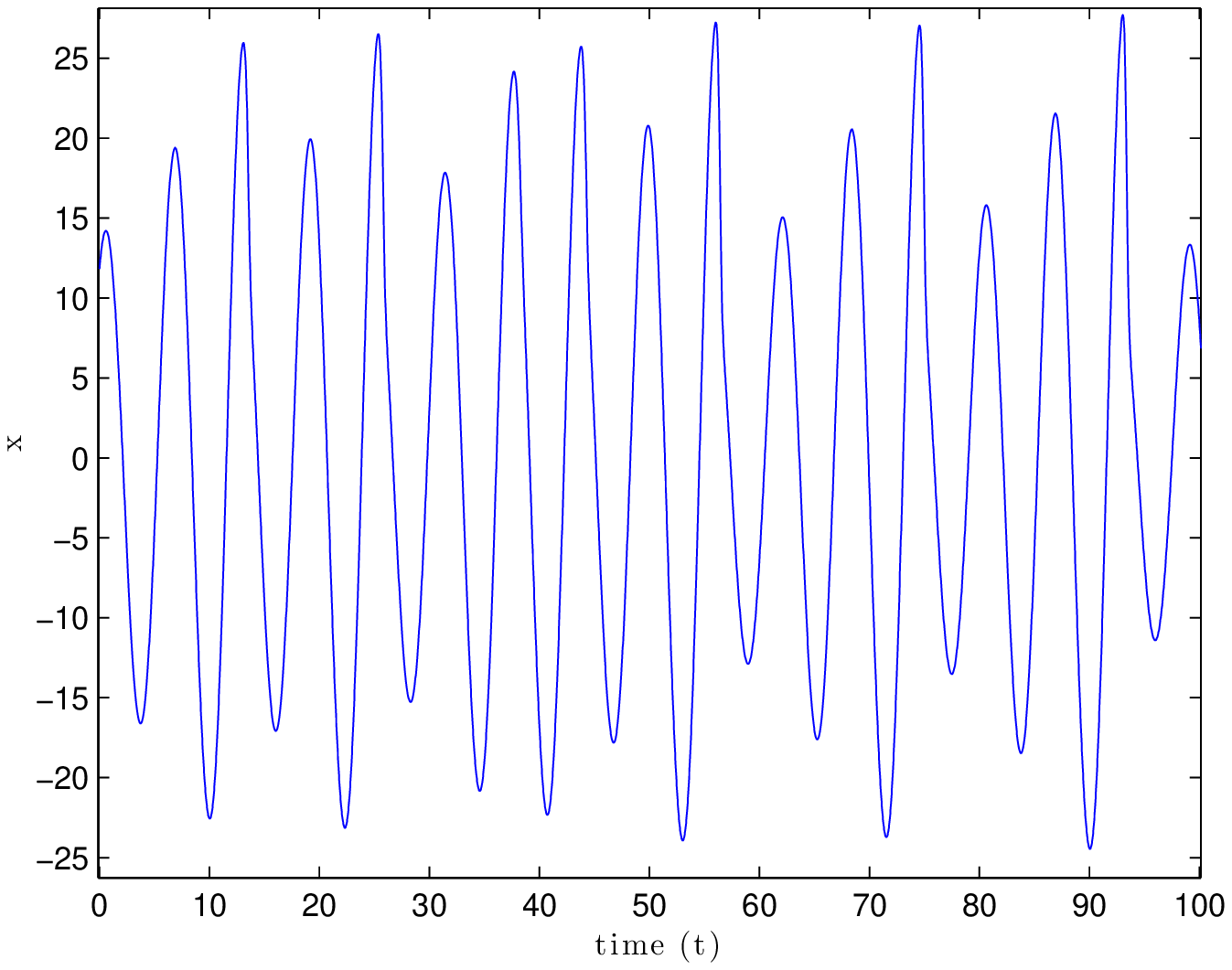}\\
			\multicolumn{2}{c}{Chaotic $c=18$}\\ 
		\end{tabular}		
	
	\caption{The route to chaos for the R\"ossler equations with $a=b=0.1$ }
	\label{fig:Route}
	\end{figure}

	The convergent path along the attractor is controlled by the parameters in the governing equations. When a control parameter is slowly increased, 
	the system exhibits a series of behaviours \citep{1992scma.conf..411S}. This series is known as the `route to chaos'. The behaviours usually unfold
	 as: \emph{constant $\Rightarrow$ periodic $\Rightarrow$ period two $\Rightarrow$ $\dots$ chaos} \citep{CambridgeJournals:4334828}, as 
	 demonstrated in Fig. \ref{fig:Route}, where period two is a repeating path that travels twice around the attractor. As the parameter is increased, this 
	behaviour continues to where the path cycles three, four, or more times to return to the same location, until it suddenly becomes chaotic, where 
	the path will never repeat. 

\section{Linear Analysis}
	We wish to search for chaotic and non-linear behaviour in the spin-down rate presented in fig. 2 of \cite{DataHome}. There they isolated a 
	subset of 17 pulsars with prominent variations in their frequency derivatives. These time series are ideal for non-linear studies because they 
	directly relate to the dynamics of a pulsar. Before non-linear analysis can be done, we need to compensate for some their limitations.   

\subsection{Mind the gap} \label{sec:MtG}
	
	 When dealing with large time-scales, such as those encountered in astronomy, it is not always feasible to record data at regular intervals. This 
	 produces a times series that sporadically samples a continuous phenomenon. When the time spacing between two points in the series 
	 is relatively small, little information is lost about the continuum inside that region. If the spacing is large, more information is lost, 
	 which can cause a change in the structure of the data.
	 
\begin{figure*}
	 \centering
		\subfigure[]
	  		{\includegraphics[width=.4\linewidth]{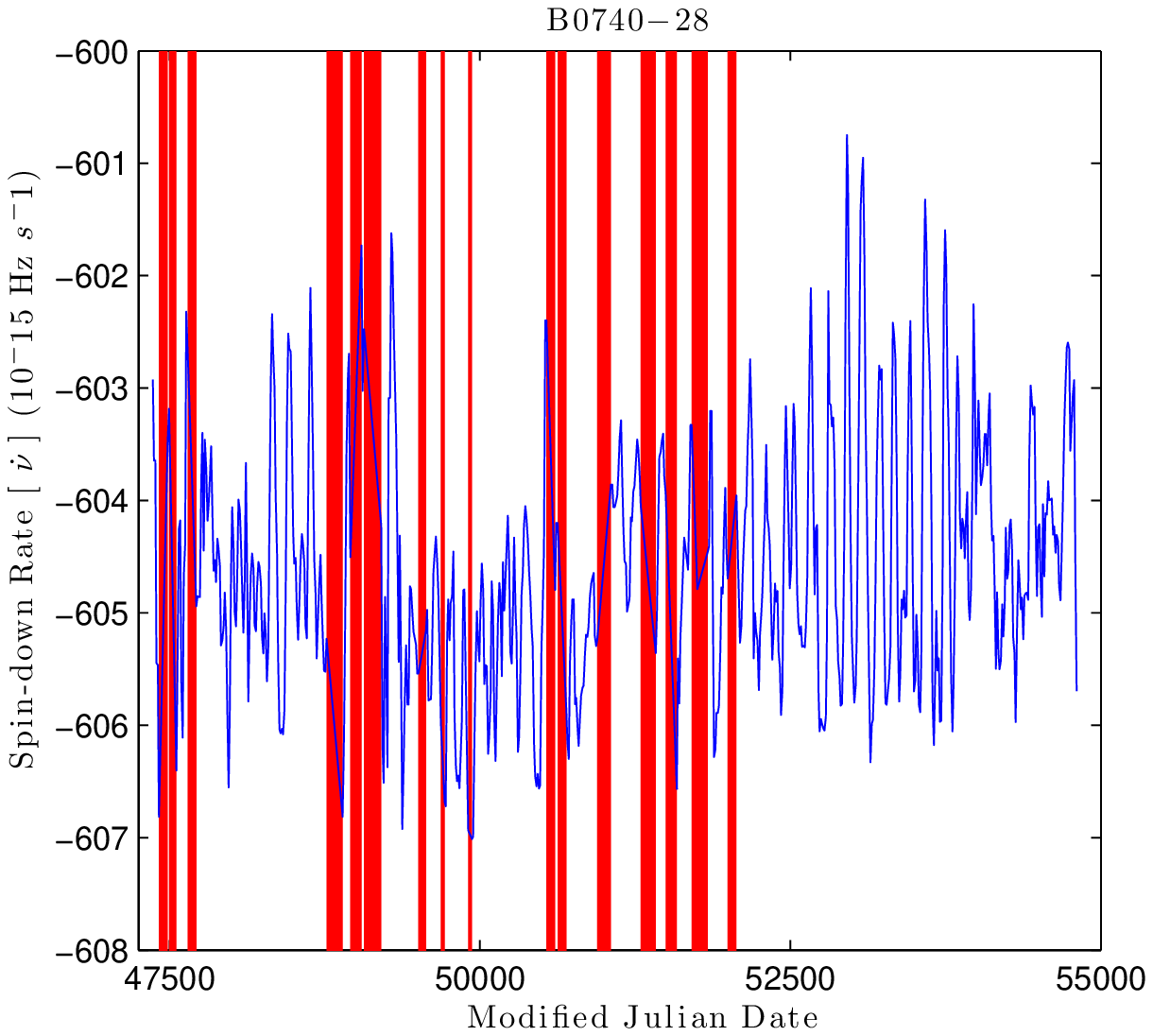}
	     		\label{fig:SegSer}}
		\subfigure[ ]
	  		{\includegraphics[width=.4\linewidth]{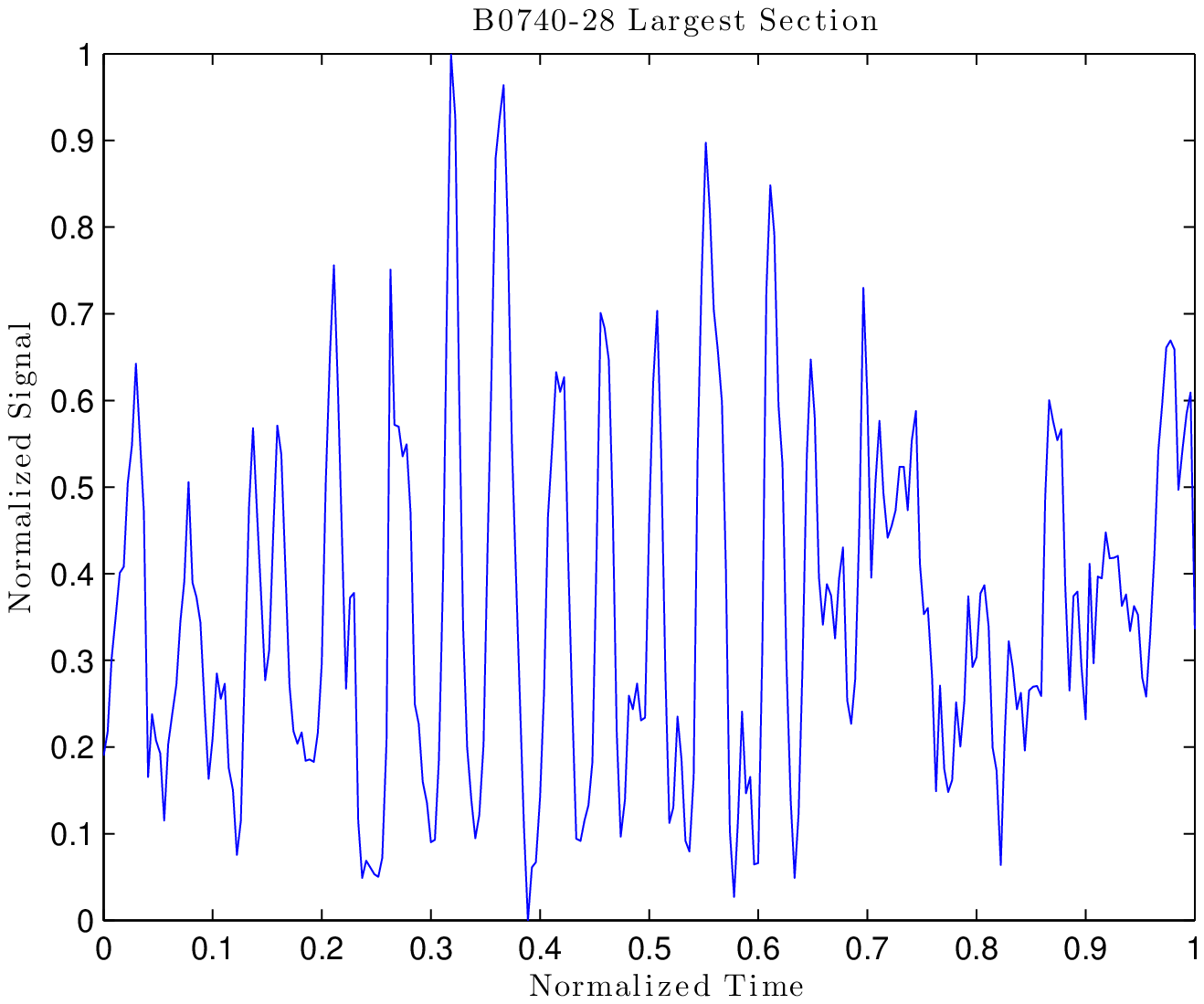} 
	  		\label{fig:Normalized}}
	\label{fig:Start}
	\caption{$(a)$ The time series recorded for PSR B0740$-$28. The vertical red bars highlight the gaps that are greater than three times 
	the mode spacing. $(b)$ The largest section for PSR B0740$-$28 after being normalized on both axes.} 
	\end{figure*}

	 To avoid such changes, we would like to analyze the largest section in the time series that best samples the phenomenon. We start by finding 
	 the statistical mode of the spacing, which gives us the step size that is the closest to being evenly sampled. We then compare this with the 
	 spacings in the series. If a gap is greater than three times the mode spacing, we assume that this region has significant information loss. The 
	 time series is now broken into several sections that are separated by these large gaps, an example of which can be seen in Fig. \ref{fig:SegSer}. We 
	 extract the longest section of data which is then normalized on both axes for more efficient computing, as seen in Fig. \ref{fig:Normalized}.
	
	\begin{figure}
	 \centering
		\subfigure[]
	  			{\includegraphics[width=.45\linewidth]{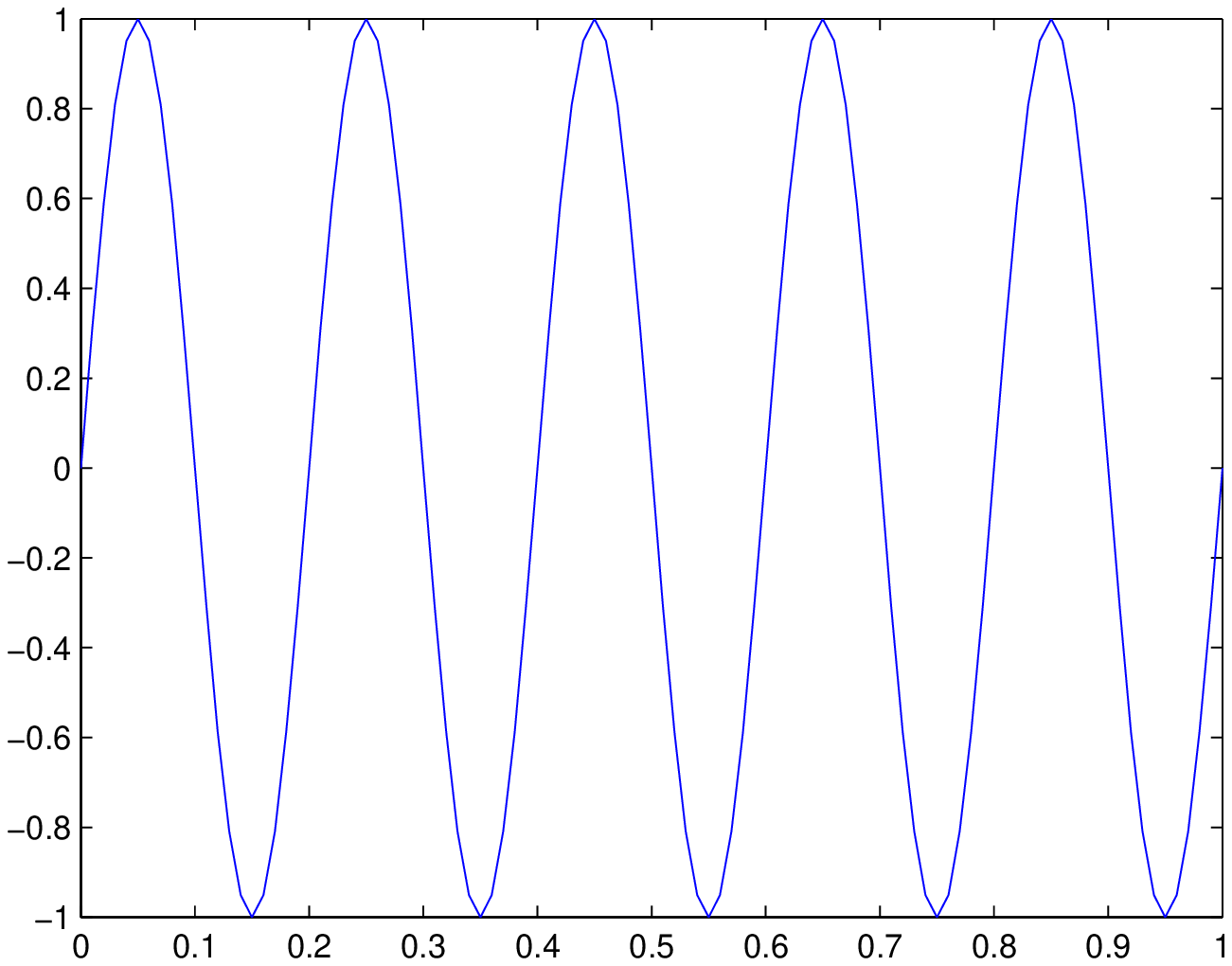}
	     			\label{fig:Sine}}
		 \subfigure[]		 	
			 	{\includegraphics[width=.45\linewidth]{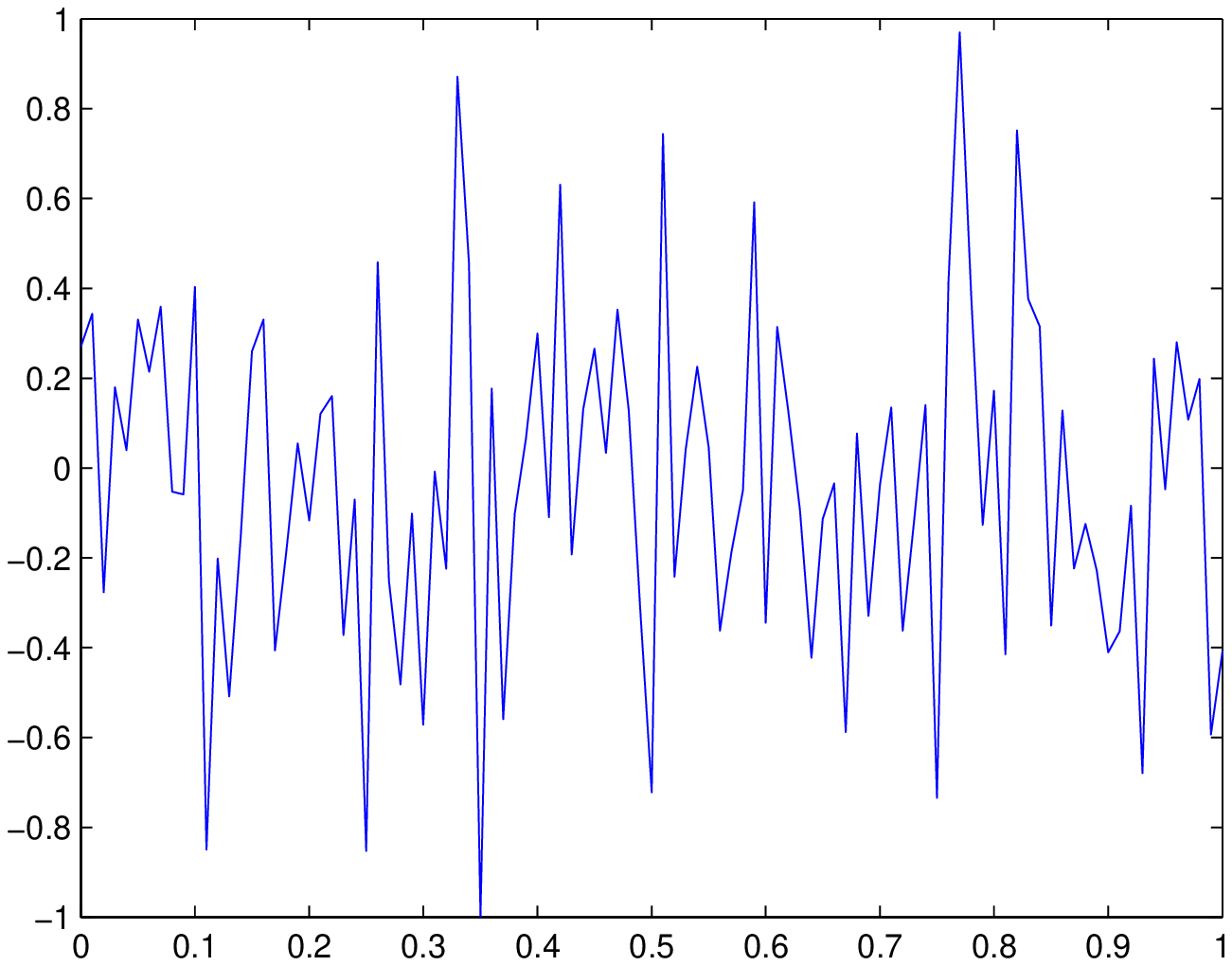}
				\label{fig:Random}}

	\caption{ $(a)$ Sine function: an example of a continuous function. $(b)$ Random time series with the same number of data points as the sine wave. }
	\label{fig:TP}
	\end{figure}

\subsection{Turning point analysis} 
	 
	 We want to ensure that the new time series is depicting a phenomenon and is not solely a consequence of noise. If the time series samples a continuous
	 function well, then we should expect a smooth curve with a large number of data points between maxima and minima, or what are known as turning
	 points. If the time series samples a random distribution, then we should expect several rapid fluctuations which will cause a large number of turning
	  points. We can see an example of this in Fig. \ref{fig:TP}, where the well sampled continuous sine wave has very few turning points, while the random 
	  series, with the same number of data points, has considerably more. 
	 
\begin{figure}
		\centering
		\begin{tabular}{cc}
			$u_{1}, u_{2}, u_{3}$ & \includegraphics[width=0.05\linewidth ]{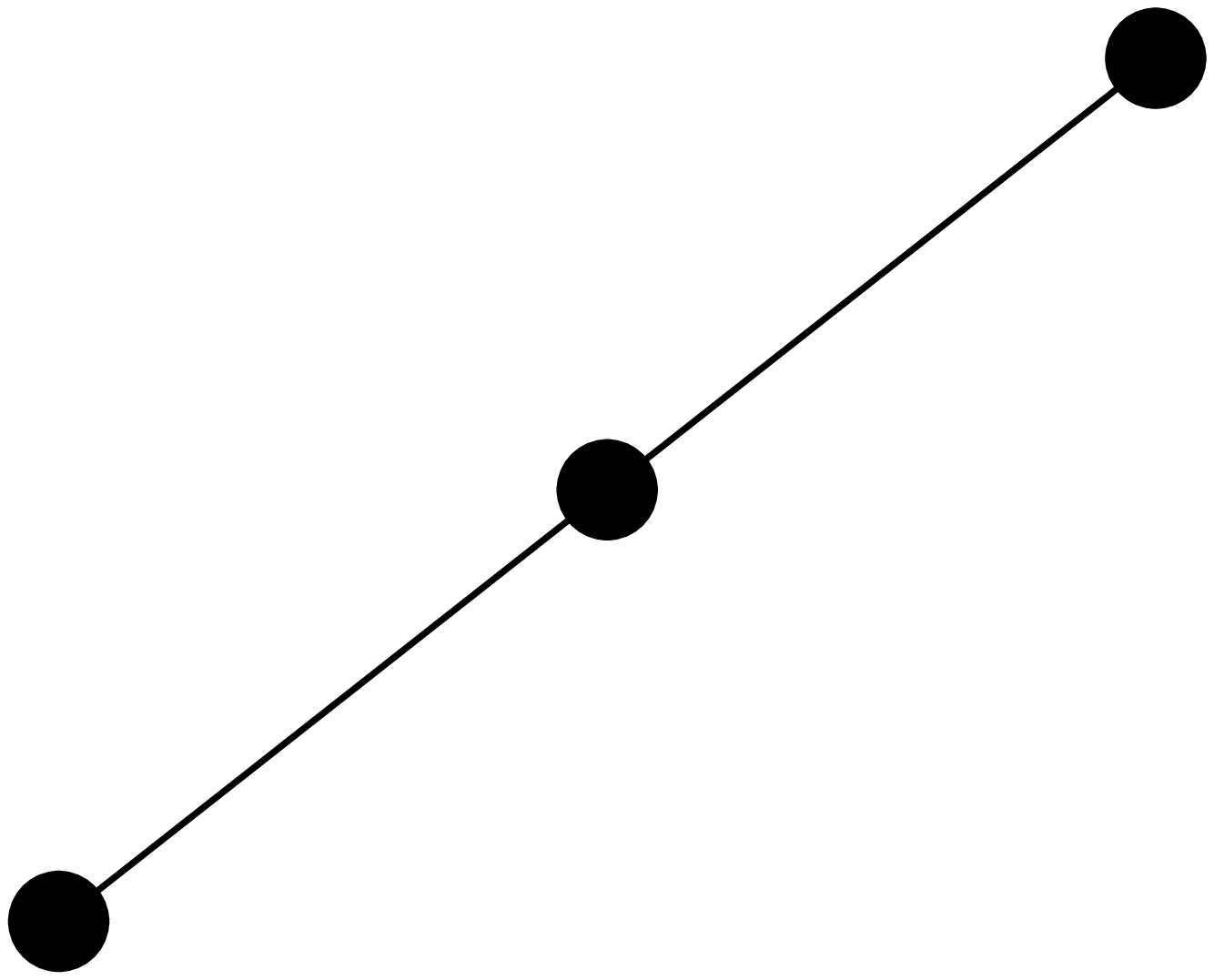}\\
			\\
			$u_{1}, u_{3}, u_{2}$ & \includegraphics[width=0.05\linewidth]{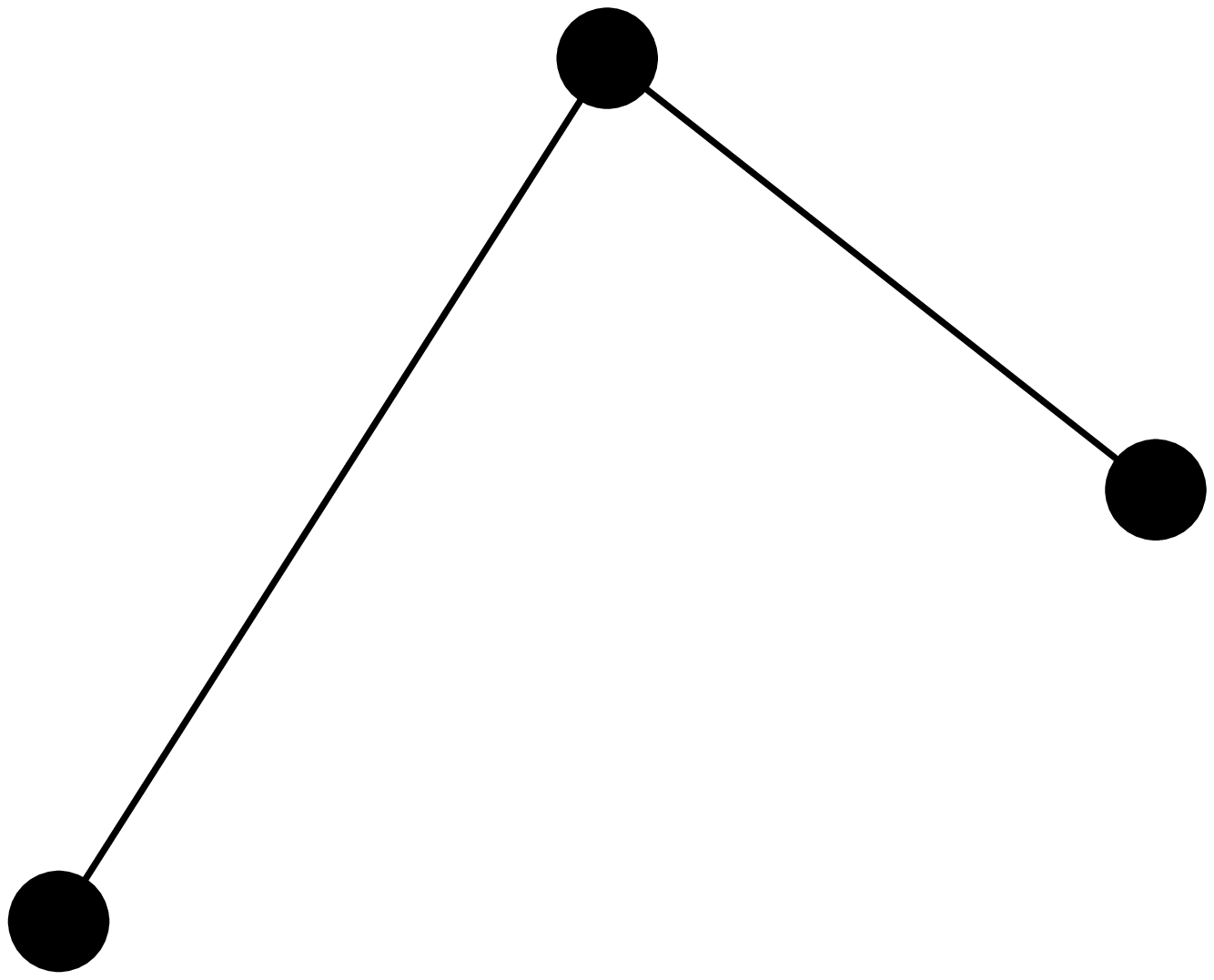}\\
			\\
			$u_{2}, u_{1}, u_{3}$ & \includegraphics[width=0.05\linewidth]{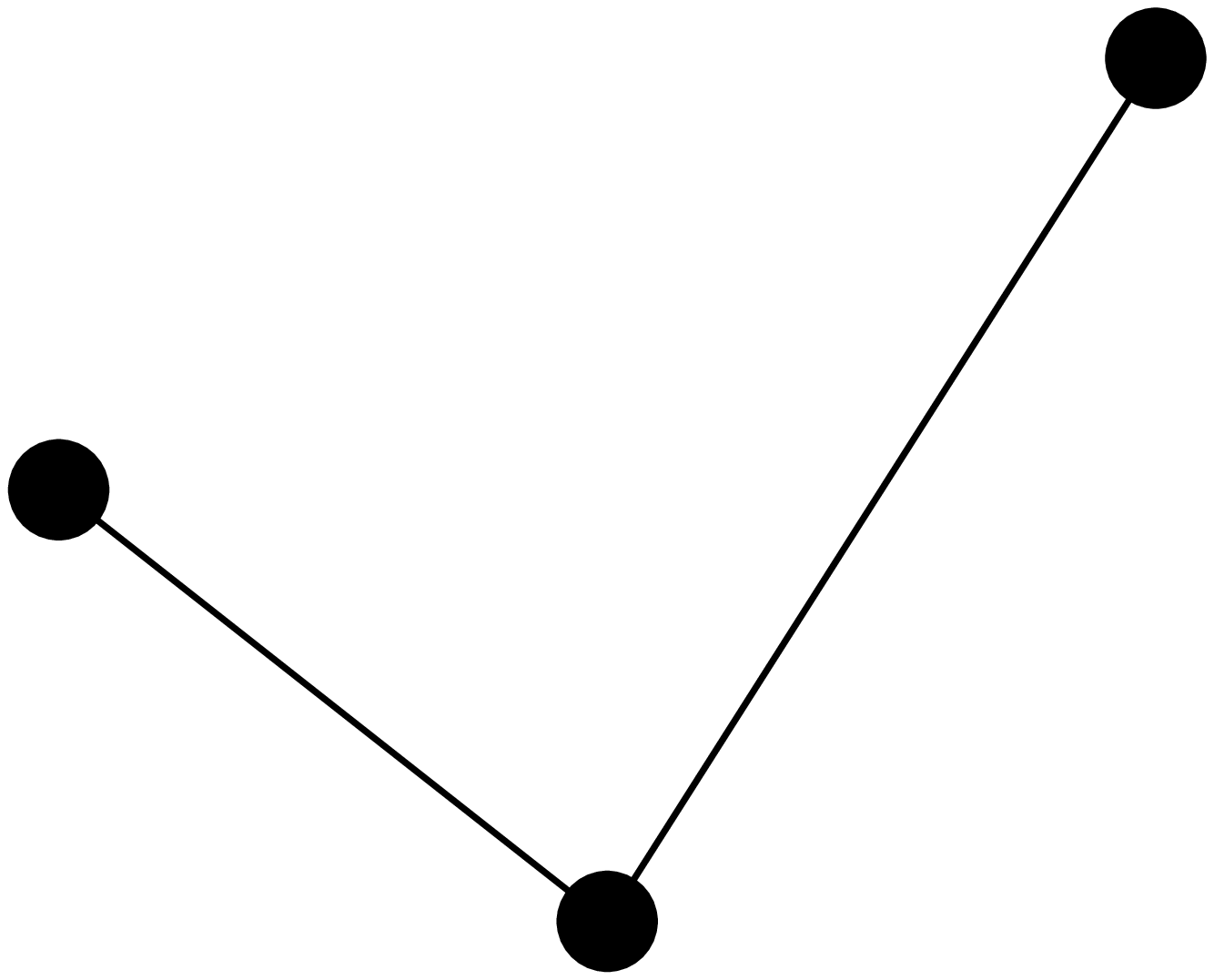}\\
			\\
			$u_{2}, u_{3}, u_{1}$ & \includegraphics[width=0.05\linewidth]{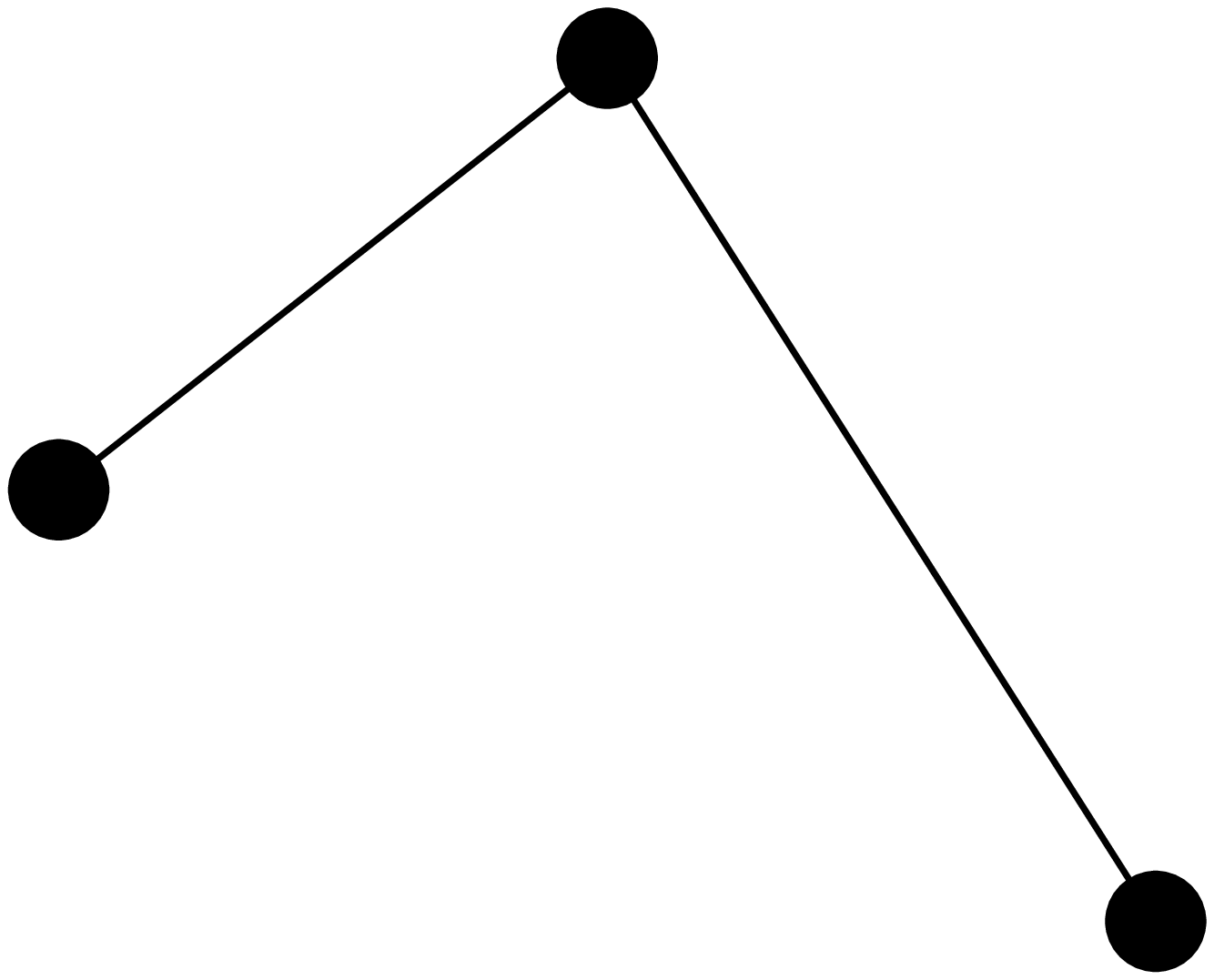}\\
			\\
			$u_{3}, u_{1}, u_{2}$ & \includegraphics[width=0.05\linewidth]{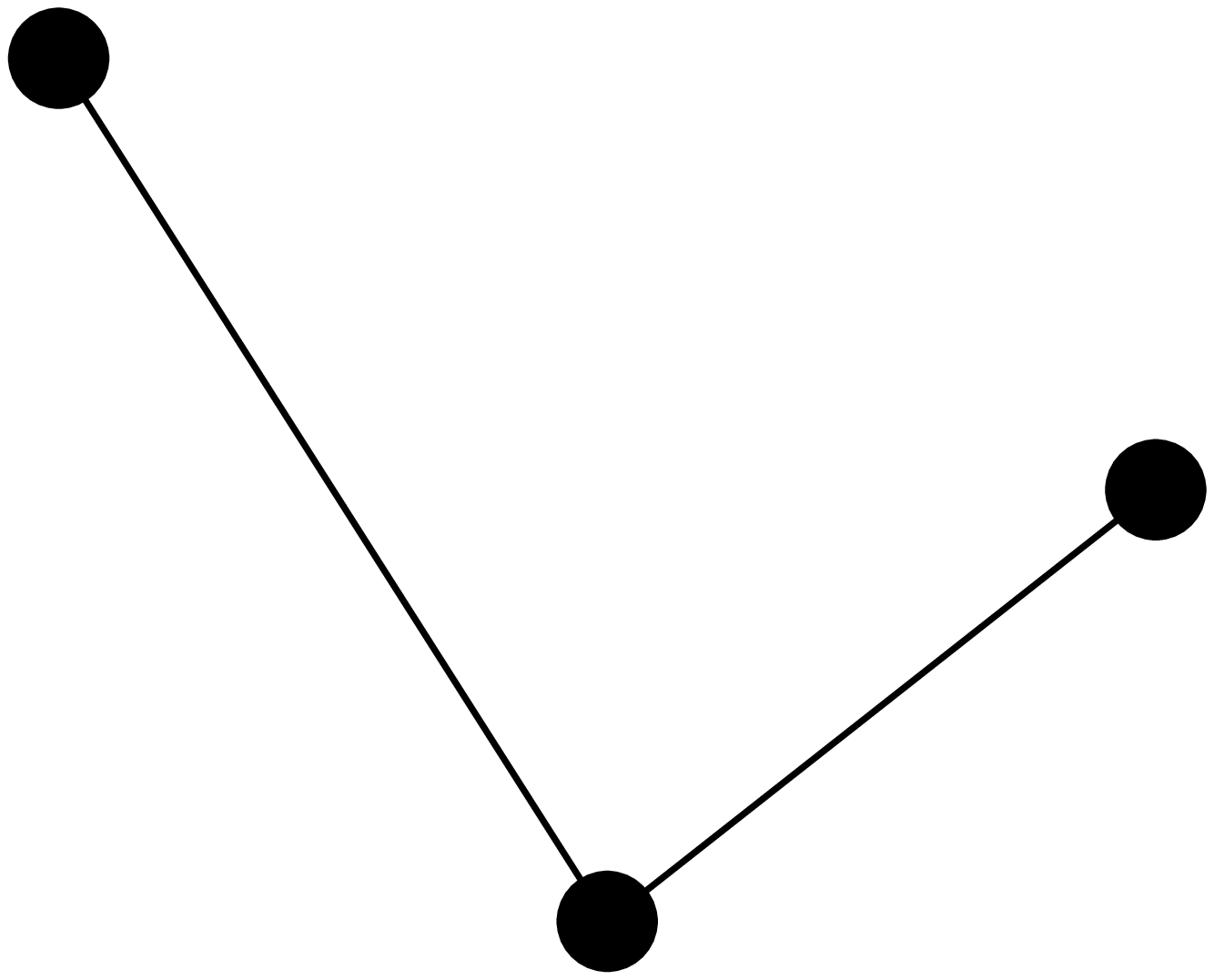}\\
			\\
			$u_{3}, u_{2}, u_{1}$ & \includegraphics[width=0.05\linewidth]{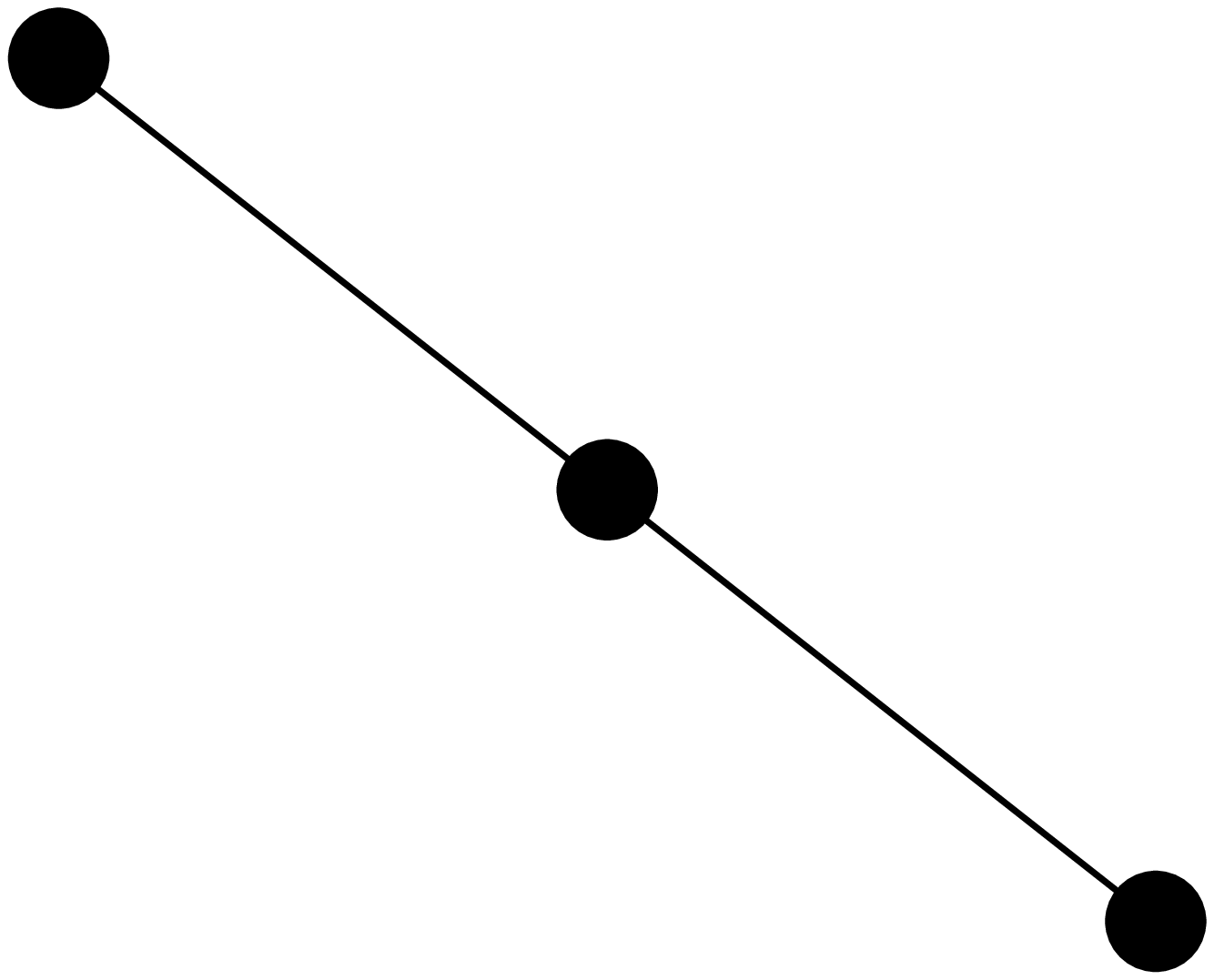}\\
		\end{tabular}
	\caption{Illustrations showing the combinations of data points for which $u_{1}< u_{2}< u_{3}$. See text for details.}
	\label{fig:AllCom}
	\end{figure}

	In order to compare a series to noise, we need to have some expectation of the number of turning points that should be in a random series. It 
	takes three consecutive data points to create a turning point. When randomly sampling, it is virtually impossible to have two values that are 
	exactly the same. Therefore, the values of the three data points will have the relationship, $u_{1}< u_{2}< u_{3}$ \citep{BookRedStats}. These 
	values can be rearranged in six different combinations, as seen in Fig. \ref{fig:AllCom}, were we find four that will produce a turning point 
	\citep{BookRedStats}. 
	 This means that any enclosed data point will have a 2/3 probability of being a turning point. The first and last data points in a series of size 
	 $n$ are the only unenclosed points. Thus, there are $(n-2)$ possible locations where a turning point can occur. This leads to an 
	 expectation value  
	 \begin{equation}
		\mu_{T}=\frac{2}{3}(n-2)
		\label{eq:mu}
	\end{equation}
	and, as described in Appendix \ref{sec:App}, a variance 
		\begin{equation}
			\sigma_{T}^{2}=\frac{16n-29}{90}.
		\end{equation}
	
	If the number of turning points is greater than the expected value, the series is fluctuating more rapidly than expected for a random 
	time series \citep{BookYellowStats}. On the other hand, a value less than the expected value indicates an increase in the number of data 
	points between each turning point.
	
	To be confident that the signal is not a result of random fluctuations, we will only keep the series that have a total number of turning points that are 
	five standard deviations less than the expectation value of a series of the same size. If a time series has a total number of turning points greater 
	than this value, it is indistinguishable from noise and will not be used. This does not necessarily imply that there is not a phenomenon present, rather 
	that the data may not sufficiently sample the phenomenon.  
	\begin{table}
	\begin{tabular}{ l r r r r c }
		\hline
		 &&&&& $\#$ of $ \sigma_{T}$'s away \\ 
		Pulsars & $T$ & $n$ & $\mu_{T}$ & $\sigma_{T}$ &T is from $\mu_{T}$\\
		\hline
		B1828$-$11&         56&      259&    171.33&   6.76&    $-$17.06\\
		B0740$-$28&	     105&      270&    178.67&	6.90&    $-$10.67\\
		B1826$-$17&	      34&  	123&	80.67&	4.64&	$-$10.05\\
		B1642$-$03&	      53&     154& 	101.33&	5.20&	$-$9.29\\
		B1540$-$06&	      19&      74& 	48.00&	3.58&	$-$8.10\\
		J2148+63&	      25&      81&     52.67&	3.75&	$-$7.37\\
		B0919+06&	      31&      90&	58.67&	 3.96& 	$-$6.99\\
		B1714$-$34& 	       7&  	  39& 	24.67&	2.57&	$-$6.87\\
		B1818$-$04&	      23& 	  73&	47.33&	3.56&	$-$6.84\\
		J2043+2740&	      12&      44&	28.00&	2.74& 	$-$5.84\\
		B1903+07&         30& 	  79&	51.33&	3.70&	$-$5.76\\
		B0950+08&	      29&       76&	49.33&	3.63&	$-$5.60\\
		B1907+00&	      32&       80&	52.00&	3.73&	$-$5.36\\ \hdashline
		B2035+36&	      14&       43&	27.33&	2.71&	$-$4.93\\	
		B1929+20&	      11&       33&	20.67&	2.35&	$-$4.11\\
		B1822$-$09&	      56&	 108&	70.67&	4.34&	$-$3.38\\
		B1839+09&    	  13&	  31&	19.33&	2.28&	$-$2.78\\
	\end{tabular}
	\caption{The 17 pulsars listed in order of their distance away from $\mu_{T}$. Here $T$ is the total number of turning points in the extracted series. $n$ is the total number of data points in that series. $\mu_T$ is the expected number of turning point for a random series of size $n$, and $\sigma_T$ is the standard deviation of the expected value. The dashed line marks the five standard deviation threshold. }
	
	\label{tab:Ratio}
	\end{table}	
	
	In Table \ref{tab:Ratio} we have listed the values for the time series that was extracted from the corresponding pulsar. Unfortunately, 
	the last four pulsars failed to pass the detection threshold, and are excluded from any further analysis.  

\subsection{Cubic spline}
	Most signal processing techniques assume that a time series is evenly sampled, and when the series is spaced randomly these algorithms 
	severely increase in complexity. Therefore we would like to form an evenly sampled series that has the same structure as our own. Since we 
	have selected a section that is nearly evenly spaced and we are reasonably confident that a signal is resolved, we can use a cubic spline 
	interpolation to approximate the structure in between data points. After the data have been splined, we then resample with a step size equal 
	to the statistical mode of the original spacing. This is done with the intent of limiting a significant increase in the time resolution. 

\subsection{Fourier transform}
	Having more data points will be beneficial in upcoming analyses, but caution must be used to avoid introducing structures or signals that are 
	not present in the series. If we take the Fourier transform of the data, this will give us an amplitude and a phase at each frequency below the 
	Nyquist frequency. It is important to note that we receive both amplitude and phase only when the data are evenly sampled. These can in 
	return be used to construct an approximate time function for the series. This function can then be sampled at any rate without adding any 
	significant non-pre-existing frequencies as described below. 

\subsubsection{ Noise reduction} \label{sec:Nr}
	Even though the turning point analysis convinced us that our series is not governed by noise, this does not mean that there is no noise in 
	the data. Before generating the new time series, we take the opportunity at this point to perform noise reduction. 
	
	The simplest method is a low pass filter, but a decision needs to be made as to where to place the cut-off frequency. One can view our time 
	series as the addition of two series: a signal series plus a noise series of equal size. From our turning point analysis, we know the expected 
	number of total turning points for the noise series, and on average we should expect the minima and maxima to be equally spaced. Therefore, 
	on average the noise should contain a frequency
	\begin{equation}
		f_{\rm noise}=\frac{\mu_{T}}{2\Delta t_{\rm total}},
	\end{equation} 
	 were $\Delta t_{\rm total}$ is the total time recorded. We then extend the frequency to a value corresponding to two more standard deviations 
	 below the expected number. Our data have been normalized in time, so that they only cover a single time unit. This leads us to set an angular 
	 cut-off frequency 
	\begin{equation}
		\omega_{max}=\pi(\mu_{T}-2\sigma_{T}).
	\end{equation} 
	
\begin{figure*}
			 \centering
			\subfigure[]
		  		{\includegraphics[width=.4\linewidth]{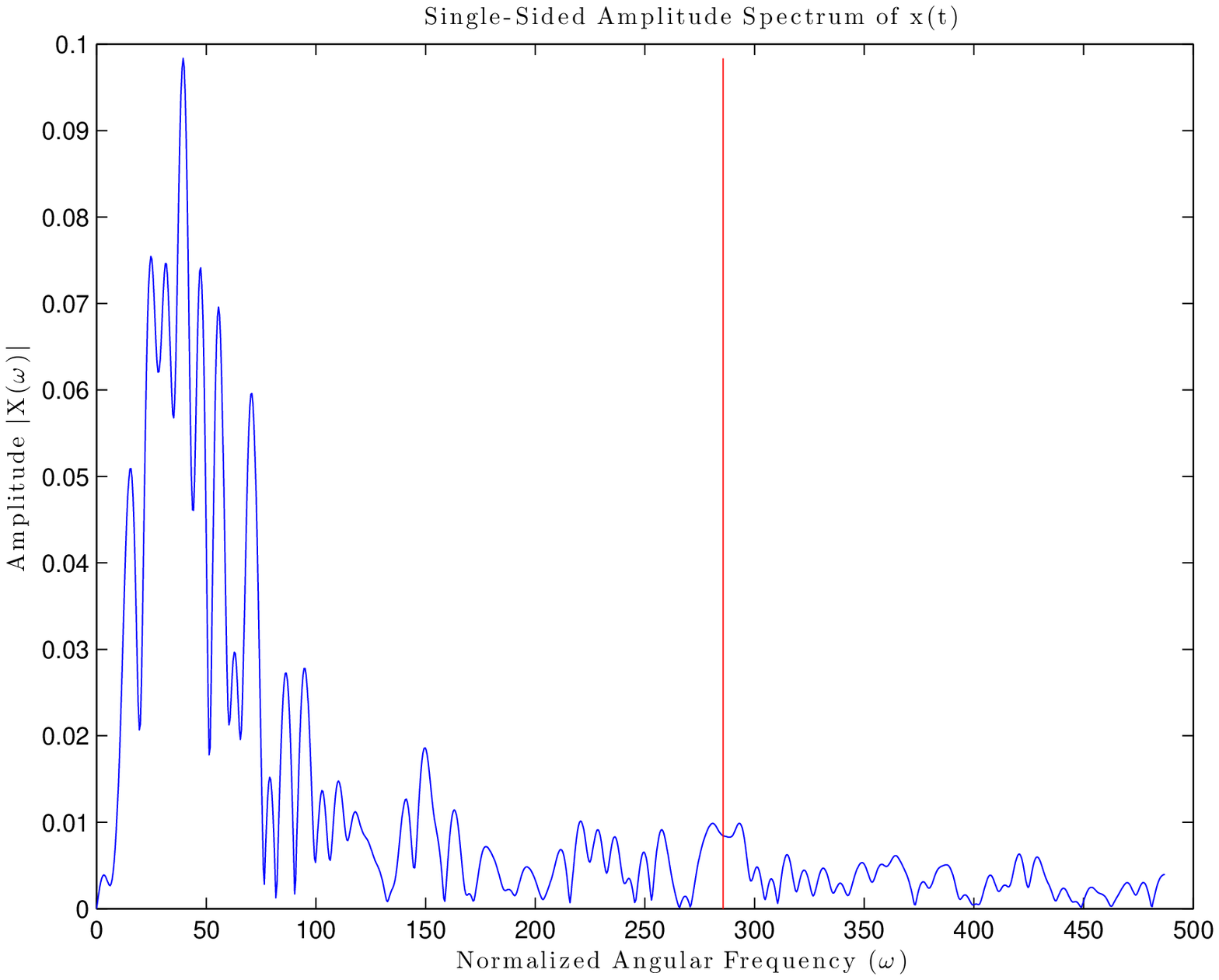}
		     		}
			\subfigure[ ]
		  		{\includegraphics[width=.4\linewidth]{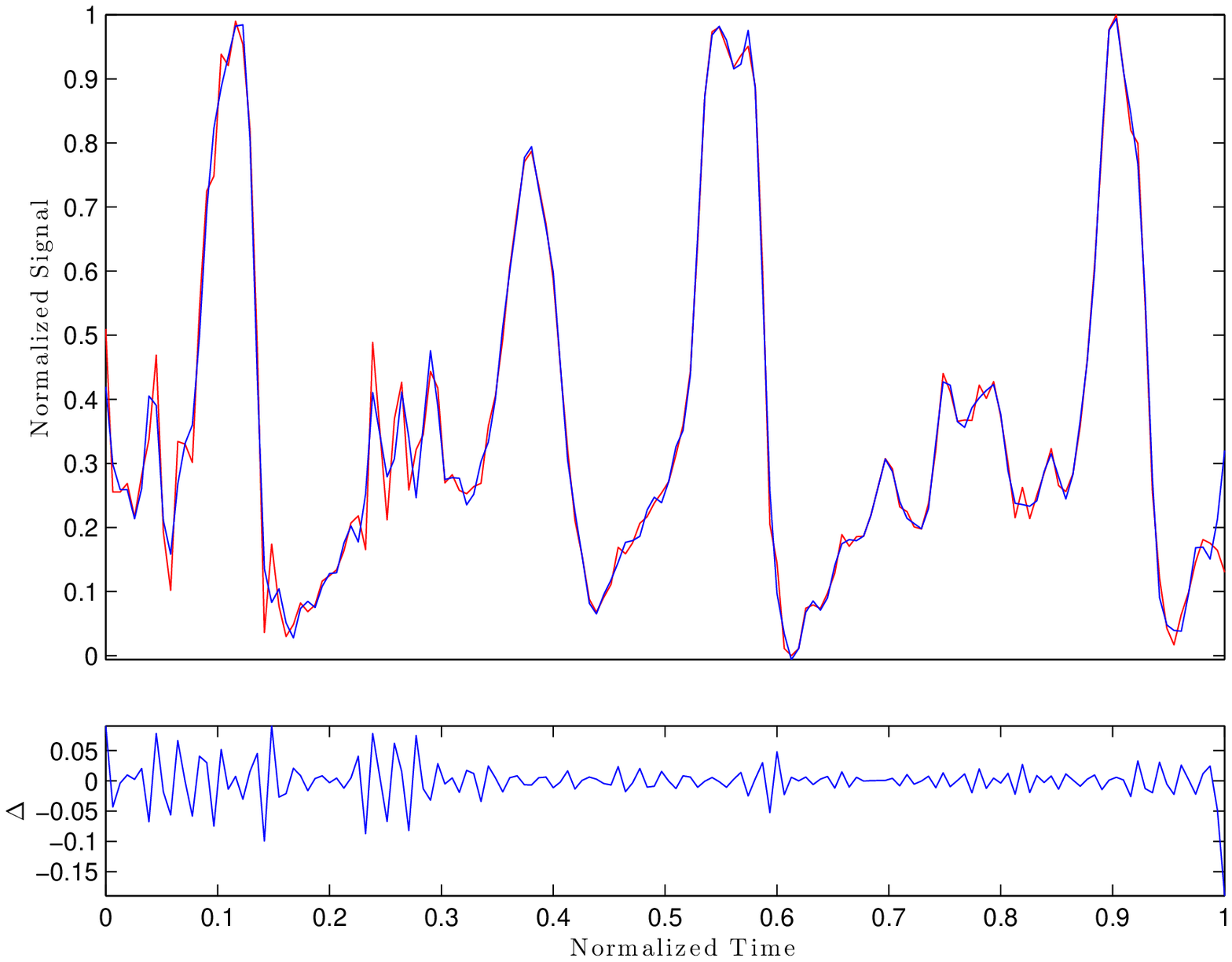} 
		  		}
		
		\caption{$(a)$ The Fourier transform amplitudes of PSR B1642-03. The vertical red line marks the cut-off frequency. $(b)$ $Top$: The pulsar data (red) overlaid  with Fourier time function with the same cadence (blue). $Bottom$: The difference of the two curves. Note that at the end of the series the magnitude of the difference increases sharply.}	
	 \label{fig:FTfit}
	\end{figure*}

	We are now able to approximate a time function by doing a reverse Fourier transform up to the cut-off frequency. Since the Fourier series is 
	a sum of periodic functions, the new time series will, by definition, repeat back upon itself. This can cause noticeable errors towards the beginning 
	and end of the time series, as seen in Fig. \ref{fig:FTfit}, and to avoid this we will not sample within the first and last five percent of the time recorded. We 
	then generate the time series by evenly evaluating the time function until we have 5,550 data points. This leaves us with approximately 5,000 data points 
	within the desired range.  

\section {Non-linear Analysis}
	For the non-linear analysis we use a combination of the \textsc{tisean} software package presented by \cite{TISEAN} and our programs written 
	in the \textsc{matlab} programing language. We intend to have the \textsc{matlab} programs available to the reader on MathWorks file exchange website to 
	help aid in the understanding of the algorithms.
	 
\subsection{Attractor reconstruction} \label{sec:AttRecon}
	With only one observable, it seems that we are unable to reproduce an attractor, but surprisingly the dynamic series contains all the 
	information needed for its reconstruction. 

\subsubsection{Method of delays} 
	We can use embedding theorems developed by \cite{Takens:1981fk} and by ~\cite{Sauer:uq} to reconstruct the attractor. These 
	theorems state that a series of scalar measurements of a dynamical system can provide a one-to-one image of a vector set, the strange 
	attractor, through time delay embedding \citep{TISEAN}. Each element in the vector is the scalar measurement $S(t)$ at a different 
	time as follows 
	\begin {equation}
		{\bf{x}}(t)=[ S(t), S(t+\tau), S(t+2\tau),\dots,S(t+(m-1)\tau)].
	\end{equation}
	The number $m$ of elements in the vector is said to be the embedding dimension \citep{TISEAN}, while $\tau$ is a time delay. 
	
	 \begin{figure}
	 \centering
		\includegraphics[width=\linewidth]{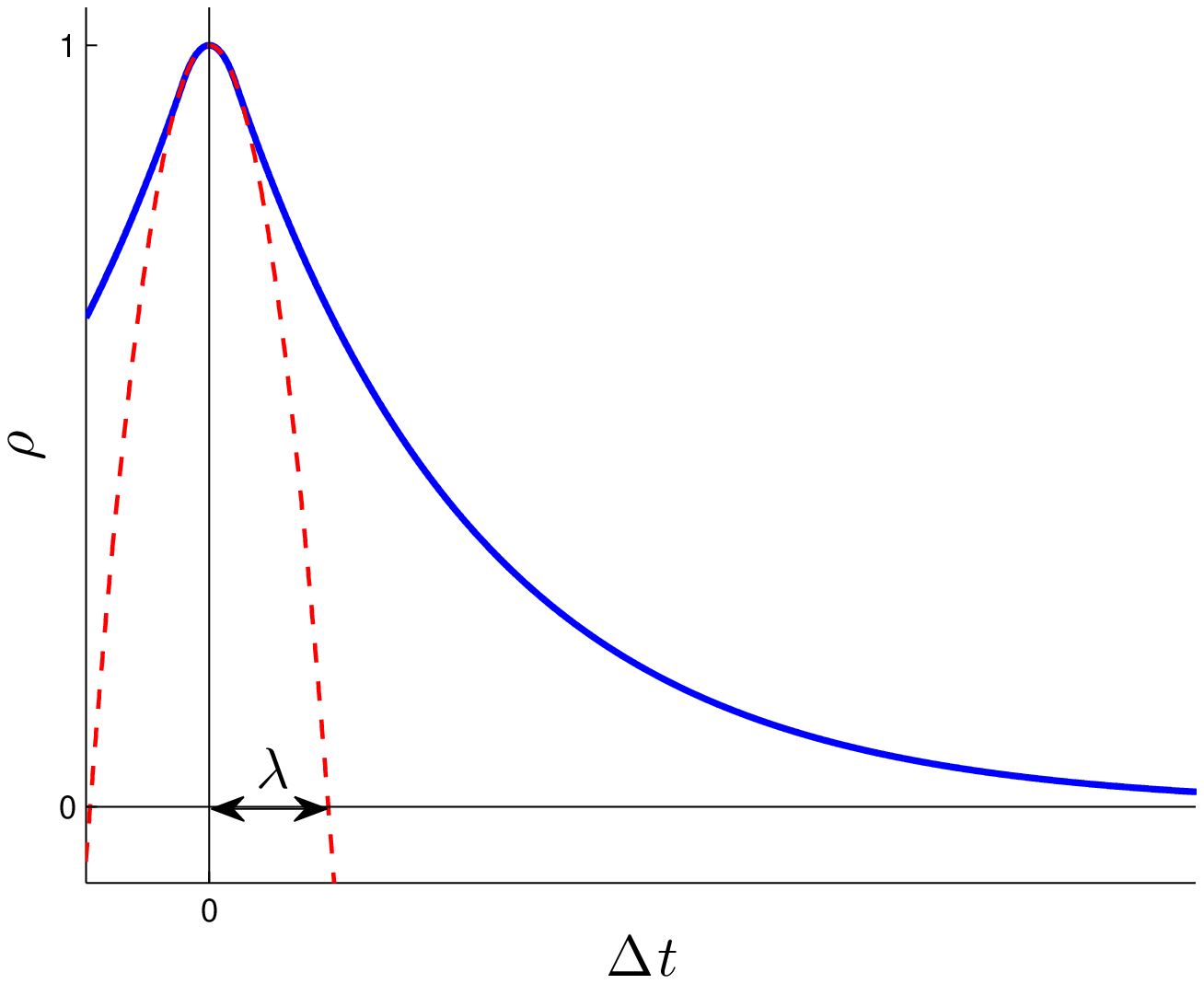}  			
	\caption{An ideal example of autocorrelation coefficients. The blue curve is the autocorrelation coefficients. The red dashed curve is a parabolic fit to the autocorrelation on small time-scales. $\lambda$ is the time-scale estimate where linear relationships are believed to be dominant. }
	\label{fig:AutoCorr}
	\end{figure}
	
	 \begin{figure*}
	 \centering
		\mbox{
			\subfigure[]
	  			{\includegraphics[width=.4\linewidth]{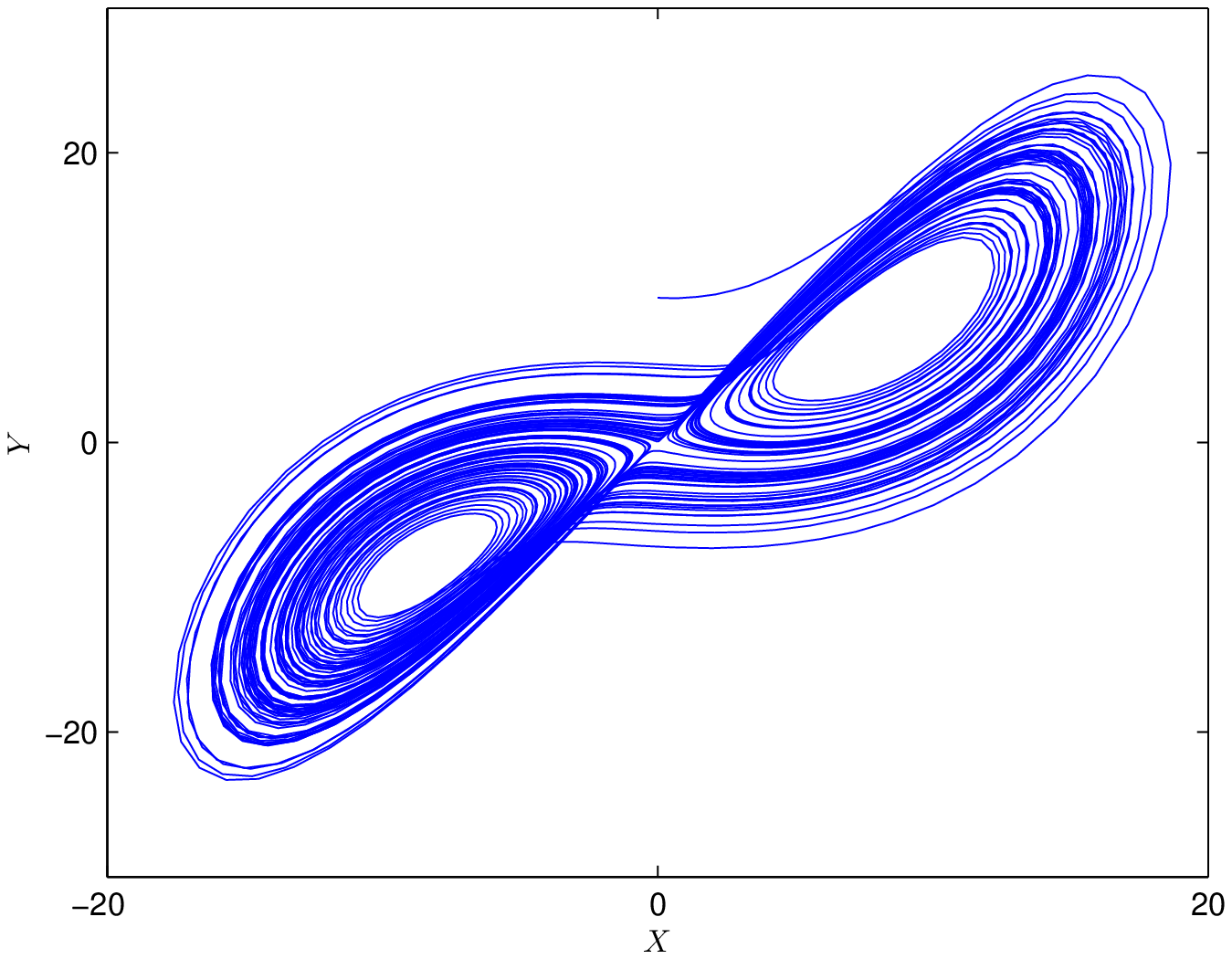}
	     			\label{fig:LorXY}}
			\quad
			 \subfigure[]		 	
			 	{\includegraphics[width=.4\linewidth]{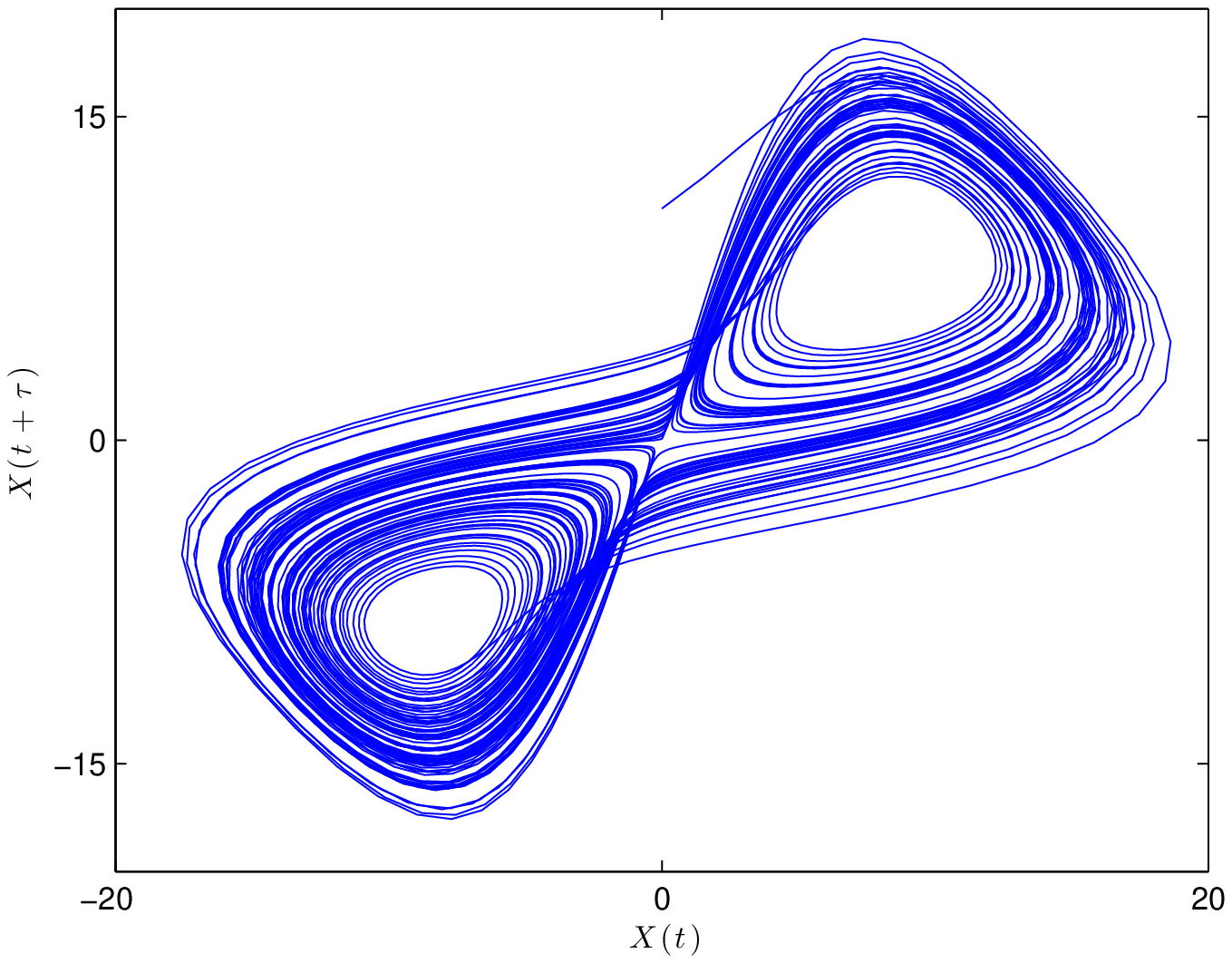}
				\label{fig:LorXTau}}
			}
	\caption{(a) The Lorenz attractor XY plane under same condition as Fig. \ref{fig:LorenzAtt}. (b) The reconstructed attractor from only the X values 
	with $\tau=0.1345$, which was estimated using the autocorrelation coefficients. }
	\label{fig:Delay}
	\end{figure*}

\subsubsection{Time delay} \label{sec:TD}
	It soon becomes evident that picking the proper time delay is crucial. If the time delay is too small, then each element will be very close in value,
	 forming a tight cluster. If the delay is too large, then the elements are unrelated and the attractor information is lost. If we were simulating 
	 a solution to the non-linear equations, we would need to find a time-scale where linear effects are dominant and to make sure that our step size 
	 was within this range. We would like to do the same thing but for the time delay, because the orthogonal axes will differ by a dynamically linear 
	 trigonometric function. 
	
	There are a wide range of algorithms that are used to find this appropriate time delay, but the simplest ones fail to account for non-linear effects 
	 \citep{TISEAN}. The algorithms that do account for these effects are not very intuitive. We decided to use adaptations of turbulent flow techniques 
	 which can be seen in \cite{BookTurbFlow}. 
	
	One way of estimating this linear range is to use the autocorrelation coefficients. We are interested in the structure of the fluctuations and would 
	like to have a zero average signal $s(t)$. This is easily done by subtracting the time average of the series from each scalar measurement. We 
	then generate the autocorrelation coefficients for this new signal, defined as 
	\begin{equation}
		\rho(\Delta t)=\frac{<s(t)s(t+\Delta t)>}{<s(t)^{2}>}.
	\end{equation}
	This function will generate a curve that will start at one and then taper to zero, seen in Fig. \ref{fig:AutoCorr}. 
	
	We can estimate the linear region by fitting a parabola to the small $\Delta t$ region of $\rho(\Delta t)$. The positive root of this parabola is our 
	estimate, $\lambda$, for a linear time-scale. This means that time-scales up to $\lambda$ should be dominated by linear effects, but to make sure we 
	are well into this region, we set the time delay to half of this value. We can see how well this estimate works in Fig. \ref{fig:Delay}, where the 
	topology of the attractor has been conserved. 
	
\begin{figure}
	 \centering
	  \begin{tabular}{c c}
	  	\includegraphics[width=.45\linewidth,trim=1100 0 0 0 ,clip=true]{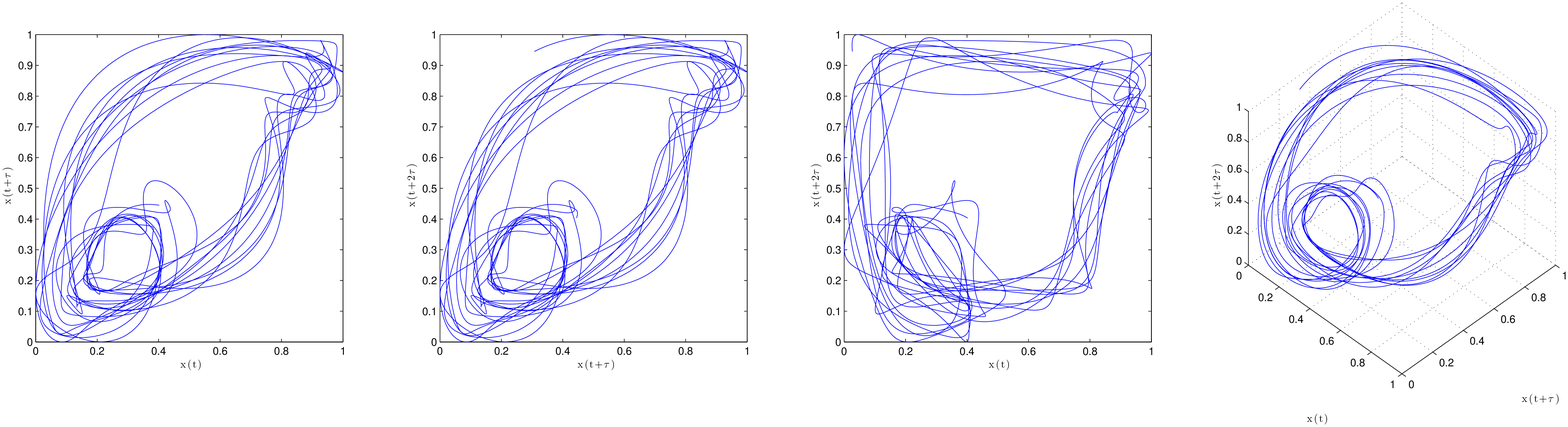} & 
			\includegraphics[width=.45\linewidth,trim=1100 0 0 0 ,clip=true]{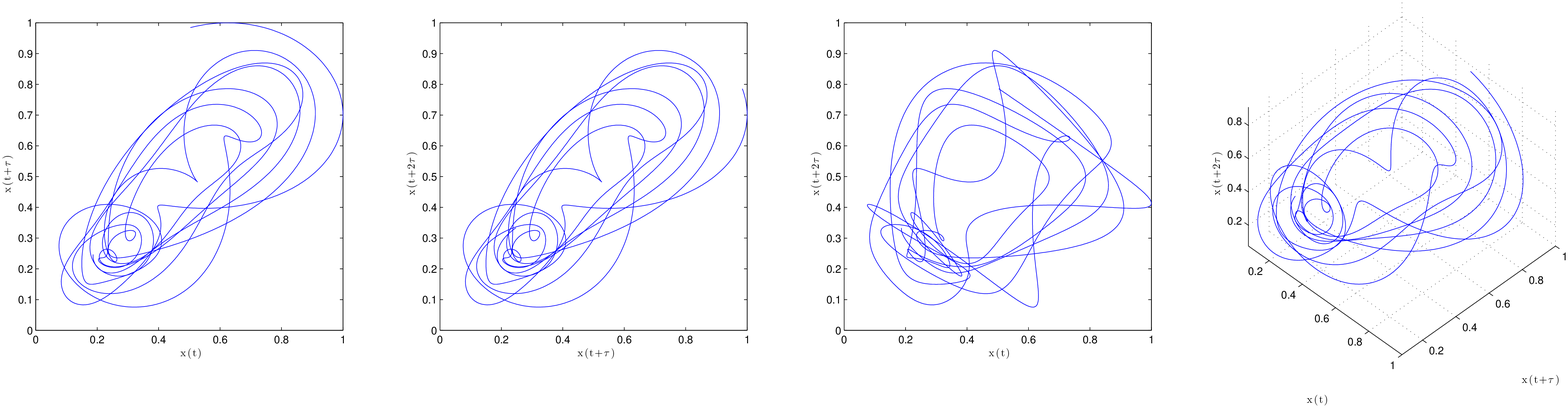} \\
		B1828$-11$ & B1826$-$17\\ 
		\includegraphics[width=.45\linewidth,trim=1100 0 0 0 ,clip=true]{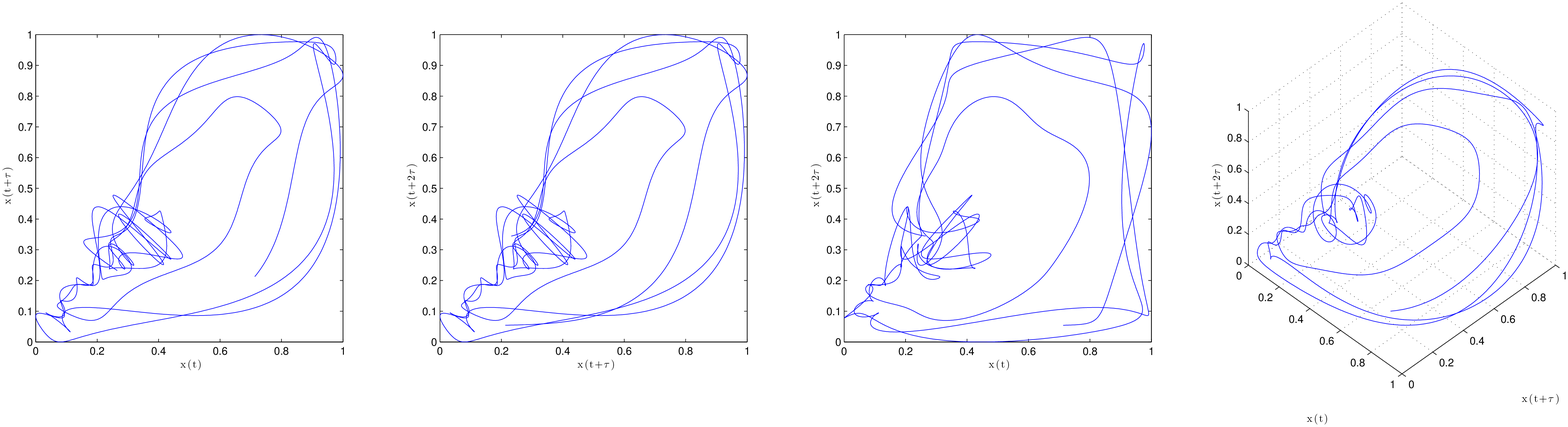} &
			\includegraphics[width=.45\linewidth,trim=1100 0 0 0 ,clip=true]{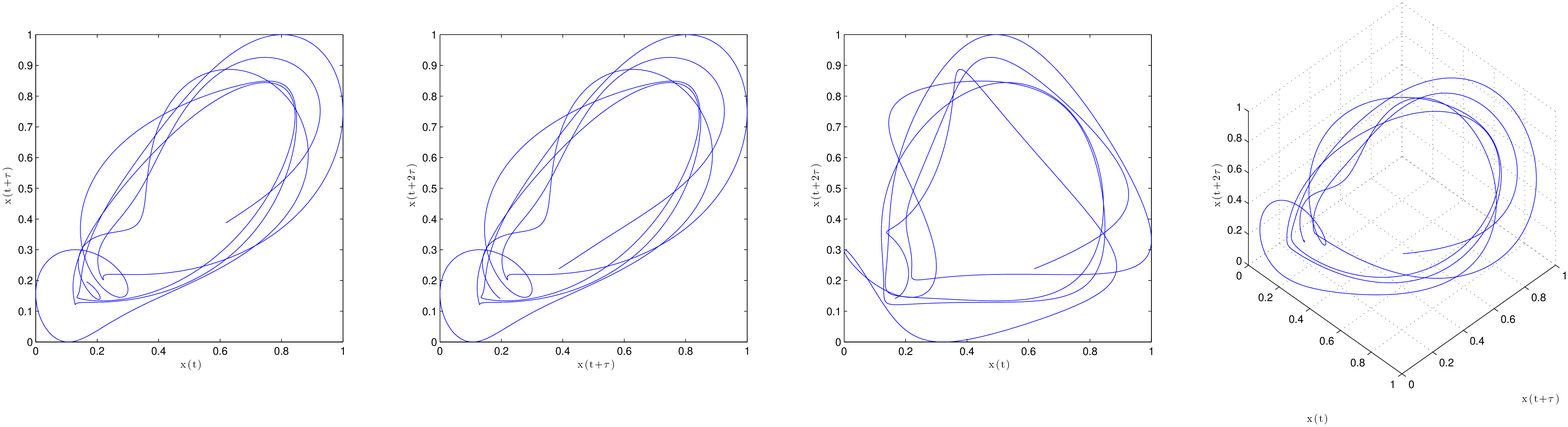}\\
		B1642$-$03 & B1540$-$06\\
		 
	 \end{tabular} 
	 \caption{The topology of the pulsar time series after being embedded in three dimensional space.}
	  \label{fig:PulAtt}
	\end{figure}

	When these techniques are applied to the pulsar time series, a similar topology appears among the best-sampled pulsars, which is seen in Fig. 
	\ref{fig:PulAtt}. This seems to suggest that these are different paths along the same attractor, demonstrating a \emph{route to chaos}. One could see 
	PSR B1540$-$06 as \emph{periodic} behaviour, B1828$-$11\footnote{A linear trend was removed in order to make the time series stationary, 
	because we are solely interested in the fluctuations. When B1828$-$11 is referenced in the rest of the paper, it is implied that this linear trend has been
	removed.} as \emph{period two} behaviour, and B1826$-$17 and B1642$-$03 as 
	\emph{chaotic}. To be sure that this is truly the case, we need to measure the dimension of each topology. 
	
\subsection{Measuring dimensions } \label{sec:MD}
	We often see dimensions as the minimum number of coordinates needed to describe every point in a given geometry \citep{BookStrogatz}. 
	For example, a smooth curve is one dimensional because we can describe every point by one number, the distance along the curve to 
	a reference point on the curve \citep{BookStrogatz}. This definition fails us when we try to apply it to fractals, and we can see why by 
	examining the \emph{von Koch curve}. 
	
\begin{figure}
		\centering
		\includegraphics[width=\linewidth ]{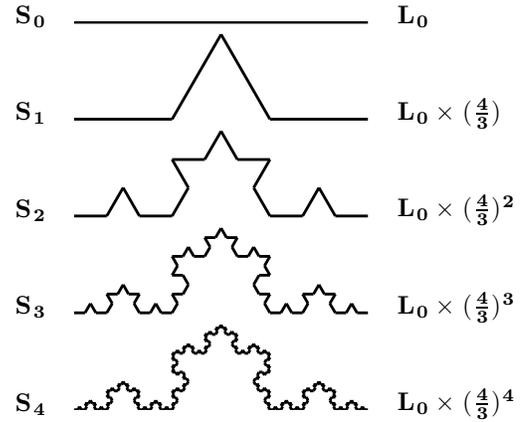} 
		\caption{Iterations leading to the \emph{von Koch curve}. $S_{n}$ denotes the segment on iteration $n$ and the far right column is the 
		total length of the corresponding segment. The \emph{von Koch curve} is when $n\to \infty$.}
	\label{fig:Koch}
	\end{figure}

	Shown in Fig. \ref{fig:Koch}, the \emph{von Koch curve} is an infinite limit to an iterative process. We start with a line segment $S_{0}$ and break
	the segment into three equal parts. The middle section is swung 60 degrees and another section of equal length is added to close 
	the gap to form segment $S_{1}$. This process is repeated to each line segment to produce the next iteration. 
	
	With each iteration, the curve length is increased by a factor of $\frac{4}{3}$. Therefore, the total length at an iteration $n$ will be 
	$L_{n}=L_{0}\times(\frac{4}{3})^n$. When this is iterated an infinite amount of times, it produces a curve with an infinite length. Not only 
	that, there would be an infinite distance between any two points on the curve. 
	
	We would not be able to describe every point by an arc length, but we would be able to describe them with two values, say the Cartesian 
	coordinates. Our original definition would define this as two-dimensional. However, this does not make sense intuitively, since this is not an area. 
	Therefore 	this is something in between, one plus some fraction of a dimension. This is known as a fractal dimension.  

\subsubsection{Correlation dimension} \label{sec:CorrDim}
	There are several different algorithms that are used to measure this fractal dimension, but nearly all depend on a power-law relationship. The most 
	widely used is the correlation dimension, first introduced by \cite{Grassberger1983189}. 
	 
\begin{figure}
		\centering
		\begin{tabular}{c|c|c}
			 Geometry &$N_{x}$ & $\frac{d\ln{N_{x}}}{d\ln{R}} $\\
			\hline
			\multirow{3}{*}{\includegraphics[width=0.15\linewidth ]{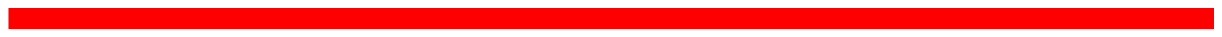}} &&\\& $2R$ & $1$\\ &&\\
			\multirow{3}{*}{\includegraphics[width=0.18\linewidth ]{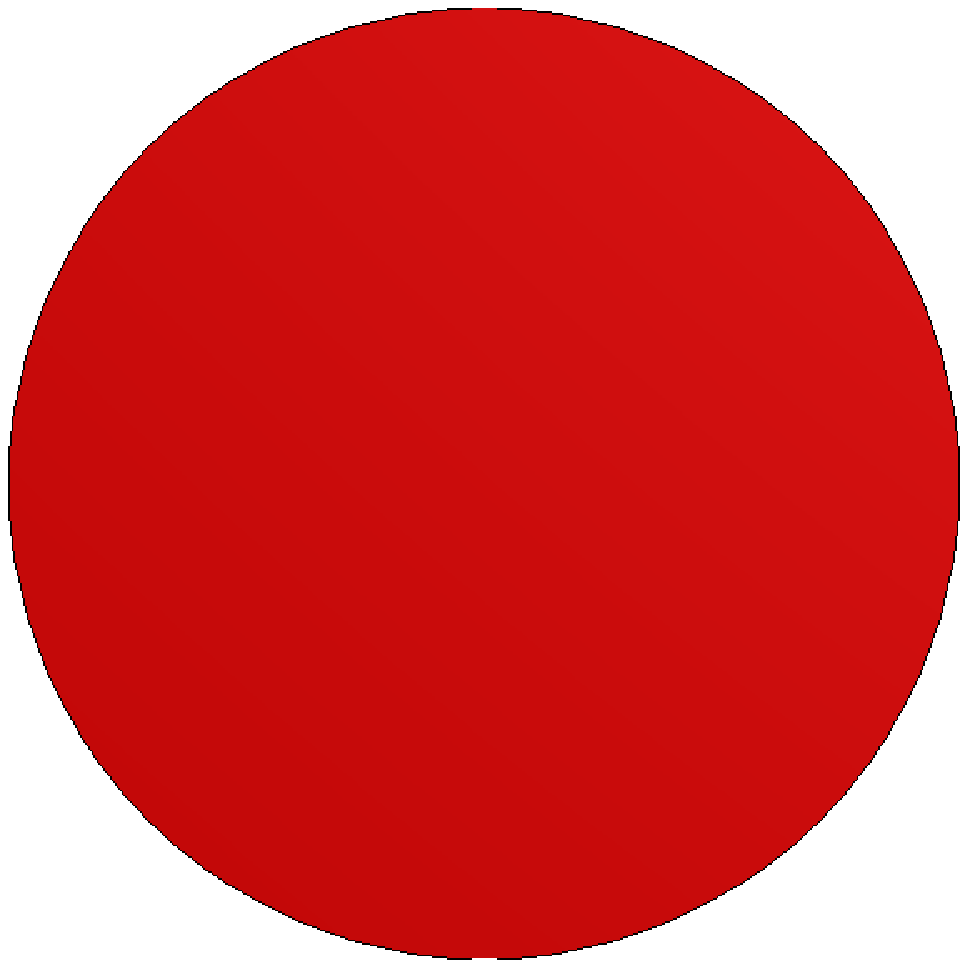}} &&\\ &$\pi R^2$ & $2$\\ &&\\
			\multirow{3}{*}{\includegraphics[width=0.2\linewidth ]{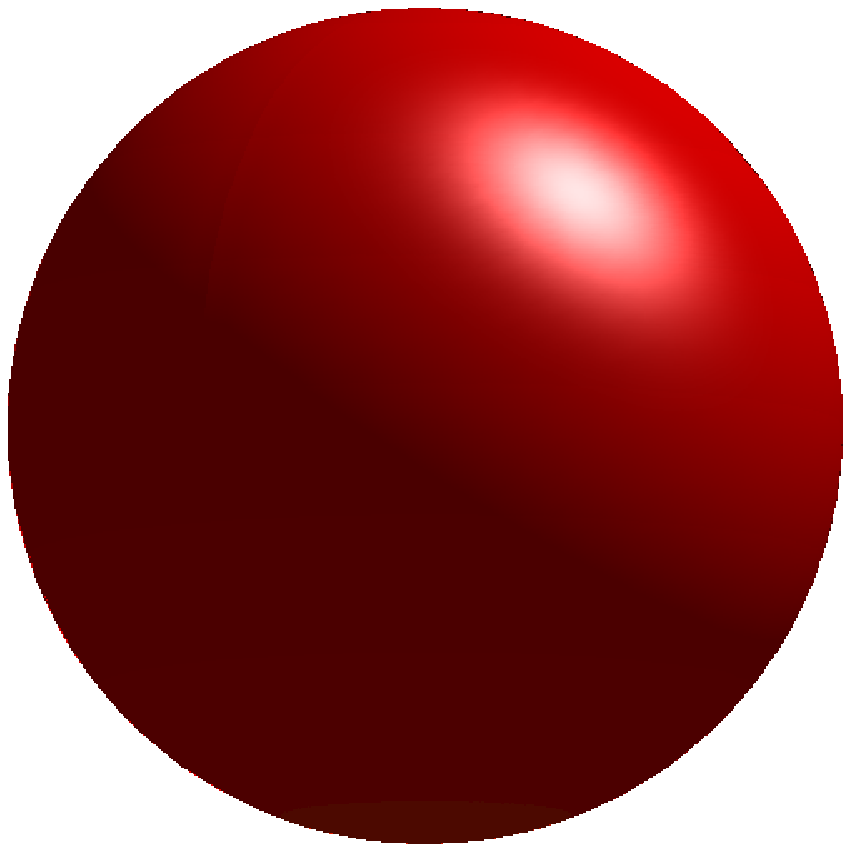}} &&\\ & $\frac{4}{3} \pi R^3$ & $3$\\ &&\\
		\end{tabular}
	\caption{$N_{x}$ is the number of data point within the radius R. $N_{x}$ will be proportional to the equations for each given situation. We 
	can see that $\frac{d\ln{N_{x}}}{d\ln{R}}$ will produce the correct dimension for each of these situations.}
	\label{fig:DimR}
	\end{figure}

	It is calculated by creating a test sphere of radius $R$ centred on a data point located at $\bf{x}$. The number of data points 
	inside the sphere is then counted, $N_{x}(R)$. The radius is slowly increased. As it increases, the number of data points in the sphere grows as a 
	power-law \citep{BookStrogatz}
	\begin{align}
		N_{x}(R) \propto R^d \nonumber \\
		d=\frac{\emph{d}(\ln{N_{x})}}{\emph{d}(\ln{R})}.
	\end{align}
	Here $d$ is referred to as the pointwise dimension at $\bf{x}$. Fig. \ref{fig:DimR} shows this relationship with familiar geometries. 
	
	Due to fluctuations in the sampling density, the pointwise dimension can vary depending on where $\bf{x}$ is located. In order to produce a 
	more self consistent measurement, an average of $N_{x}(R)$ is taken over all data points. This average is known as the correlation sum, 
	$C(R)$, and is often written as 
	\begin{equation}
		C(R)=\frac{2}{N(N-1)}\sum_{i=1}^{N}\sum_{j=i+1}^{N}H(R-\| \bf{x}_i-\bf{x}_j\|) 
		\label{eq:CorrSum}
	\end{equation}
	where $H$ is the Heaviside function, $H(x)=0$ if $x\leq0$ and $H(x)=1$ if $x>0$. $N$ is the total number of locations. $\|$ $\|$ is the 
	magnitude of the vector such that $\| \bf{v} \|= (\bf{v} \cdot \bf{v})^{1/2}$. 
	
	This correlation sum will have the same type of power relation as $N_{x}(R)$. The exponent for $C(R)$ is then properly named the 
	correlation dimension. This is defined as 
	 \begin{equation}
		D=\lim_{N \to \infty}\lim_{R \to 0}\frac{d(\ln{C})}{d(\ln{R})}.
	\end{equation}
	We seldom have the luxury of infinite sample size with infinitely small resolution, and therefore different behaviours occur in the correlation sum.
	
	To estimate $D$, one would plot $\ln{C}$ versus $\ln{R}$. If the power relationship held true for all $R$ we would see a straight line with a slope of
	 $D$. However, due to the finite size of the attractor, all points could lie within a large enough $R$. We can see in Equation \ref{eq:CorrSum} that this 
	 would cause $C(R)$ to converge towards a value of one. At low enough $R$ we will reach a resolution limit, where only the centre data point is inside 
	 the test sphere. This causes the $C(R)$ to converge to zero. Therefore the power-law will only hold true over an intermediate scaling region. 
	 
	 We could safely avoid this resolution limit if 10 points were in each of our test spheres. This would correspond to $C(R)=\frac{10}{N}$, which we use 
	 as our lower limit of the scaling region. To set an upper limit, we say that if 10\% of all points are in the each test sphere, we would start to 
	 approach the scale of the attractor. Therefore, we set our upper bound at $C(R)=0.10$ to close off a rough scaling region.  
	 
	With this definition of dimension, one can see how there is a transition between integers. We can increase the number of data points of a line within 
	$R$ by simply folding that line inside a test radius, perhaps like $S_{1}$ in Fig. \ref{fig:Koch}. In order to have a power-law relationship across all sizes 
	of $R$, we need to have a similar fold on all scales, known as \emph{self similarity}. If we were to steepen the angles of each fold, this would increase 
	the number contained and also increase the dimension. We can do this until the folds are directly on top of one another to `colour in' a two 
	dimensional surface. We could then repeat the process by folding the two dimensional surfaces to transition to three dimensions. Folding like this 
	is seen as the underlying reason why strange attractors have their fractal dimensions and is explored in depth by \cite{0202.55202} and 
	\cite{Grassberger1983189}.  

\subsubsection{Theiler window}
	If the attractor was randomly sampled we would be ready to measure the dimension, but the solutions are one dimensional paths on a 
	multi-dimensional surface. Therefore, we have two competing dimension values which will corrupt our measurement. What we wish to 
	do is to isolate only the attractor dimension, the \emph{geometric correlations}, and remove the path relationships, the \emph{temporal correlations}. 
	
	As difficult as this sounds, \cite{Theiler:1986fk} introduced a rather simple remedy. We exclude the points within a time window, 
	$w\times\Delta t_s$, around the centre of the test radius. \footnote{Here $\Delta t_s$ is the step size of the time series, and $w$ is an integer value.} 
	This is known as the \emph{Theiler window} and changes the correlation sum in the following way 
	
	\begin{equation}
		C(R)=\frac{2}{(N-w)(N-w-1)}\sum_{i=1}^{N}\sum_{j=i+w+1}^{N}H(R-\| \bf{x}_i-\bf{x}_j\|).
		\label{eq:ThWin}
	\end{equation}
	
	We can see that picking the appropriate $w$ is nontrivial. If we pick a window that is too small we fail to remove the temporal correlations, and 
	if the window is too large it would significantly reduce the accuracy of the geometric measurement. 

\subsubsection{Space time separations}
	In order to estimate a safe value for the Theiler window, \cite{Provenzale199231} introduced the \emph{space time separation plot}. It shows 
	the relationship between the spatial and temporal separations, by forming a contour map of the percentage of locations within a 
	distance, for a given time separation.
	
	\begin{figure}
		\centering
		\includegraphics[width=\linewidth ]{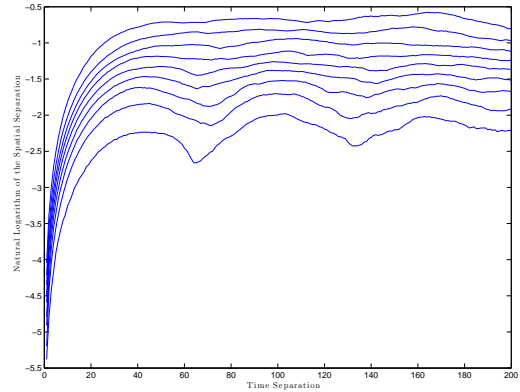} 
		\caption{Space time separation plot, generated using the \textsc{tisean} package, for PSR B1642$-$03 with $m=3$. The contours indicate 
		the percentage of locations within a distance at a given time separation. The different curves correspond to 10\% increases starting with the 
		lowest curve at 10\% to the top curve at 90\%.}
	\label{fig:STP}
	\end{figure}
	
	An example of this is seen in Fig. \ref{fig:STP}, here we used the \emph{stp} program in the \textsc{tisean} package. We can see that the locations with 
	small time separations are close to one another. As the time separations increase, so do their spatial separations, until they reach some asymptotic 
	behaviour. The substantial fluctuations at larger times are contributed to the cycle period of the attractor \citep{BookTISEAN}. Temporal correlations are 
	present until the contour curves saturate \citep{BookTISEAN}. In Fig. \ref{fig:STP} this transition seems to occur around a time separation of 40 steps. 
	This is a similar quantity that appears in all of our data sets, but to be safe we chose a more conservative value of $w=100$ steps for the rest of our 
	analysis. 

\subsubsection{Embedding to higher dimensions}
	We are now ready to measure the dimensions of the topologies that were created in Section \ref{sec:AttRecon}. If these topologies are true 
	geometric shapes, their dimensions will not change when placed in a higher dimensional space. This means that as long as our embedding 
	dimension, $m$, is greater than the attractor dimension, our correlation dimension measurement will be constant. On the other hand, if we were 
	to embed a random distribution, it would be able to occupy the entire space and would have a correlation dimension equal to the embedding 
	dimension. 
	
	Because of this behaviour, we can embed our time series to higher and higher dimensions to see if it will plateau to a constant. We cannot do this 
	forever because the number of location vectors that could be formed drops off as 
	\begin{equation}
		N=n_s-\tau^{*}(m-1),
	\end{equation}
	 where $n_s$ is the total number of scalar values in the original time series and $\tau^{*}$ is the number of time steps, $\Delta t_s$, corresponding 
	 with the time delay, $\tau$. This drop off causes a slight reduction in our statistical accuracy as we continue to higher dimensions. Therefore, we 
	 limit our embedding dimensions to between 2 and 10.  
	
	Though we have picked a rough scaling region in Section $\ref{sec:CorrDim}$, $C\approx \frac{10}{n_{s}}$ to $0.10$, this does not mean that the 
	true scaling region extends through this entire range. Therefore, we sweep the region with a window size from a third of the total logarithmic range 
	to the entire range, searching for the minimum deviation in the correlation dimension measurements over the embedding dimensions. This 
	ensures that we are picking the `flattest' possible section over a significant scale. 
	
\begin{figure} 
		\centering
			\mbox{
				\subfigure[]
			  		{\includegraphics[width=.51\linewidth,trim=37 13 45 30,clip=true]{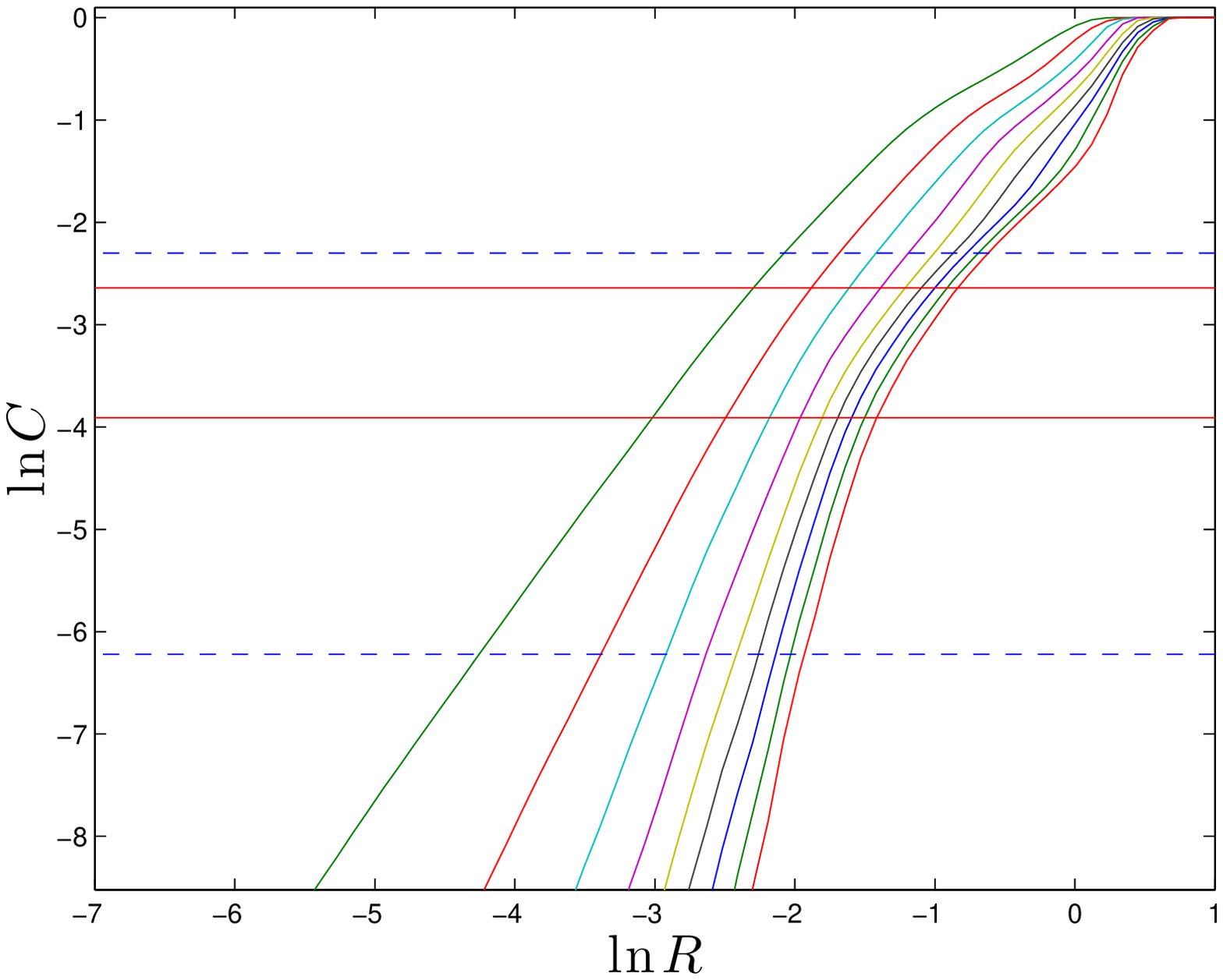}
			     		\label{fig:lnCvrlnR}}
				
				\subfigure[ ]
			  		{\includegraphics[width=.49\linewidth]{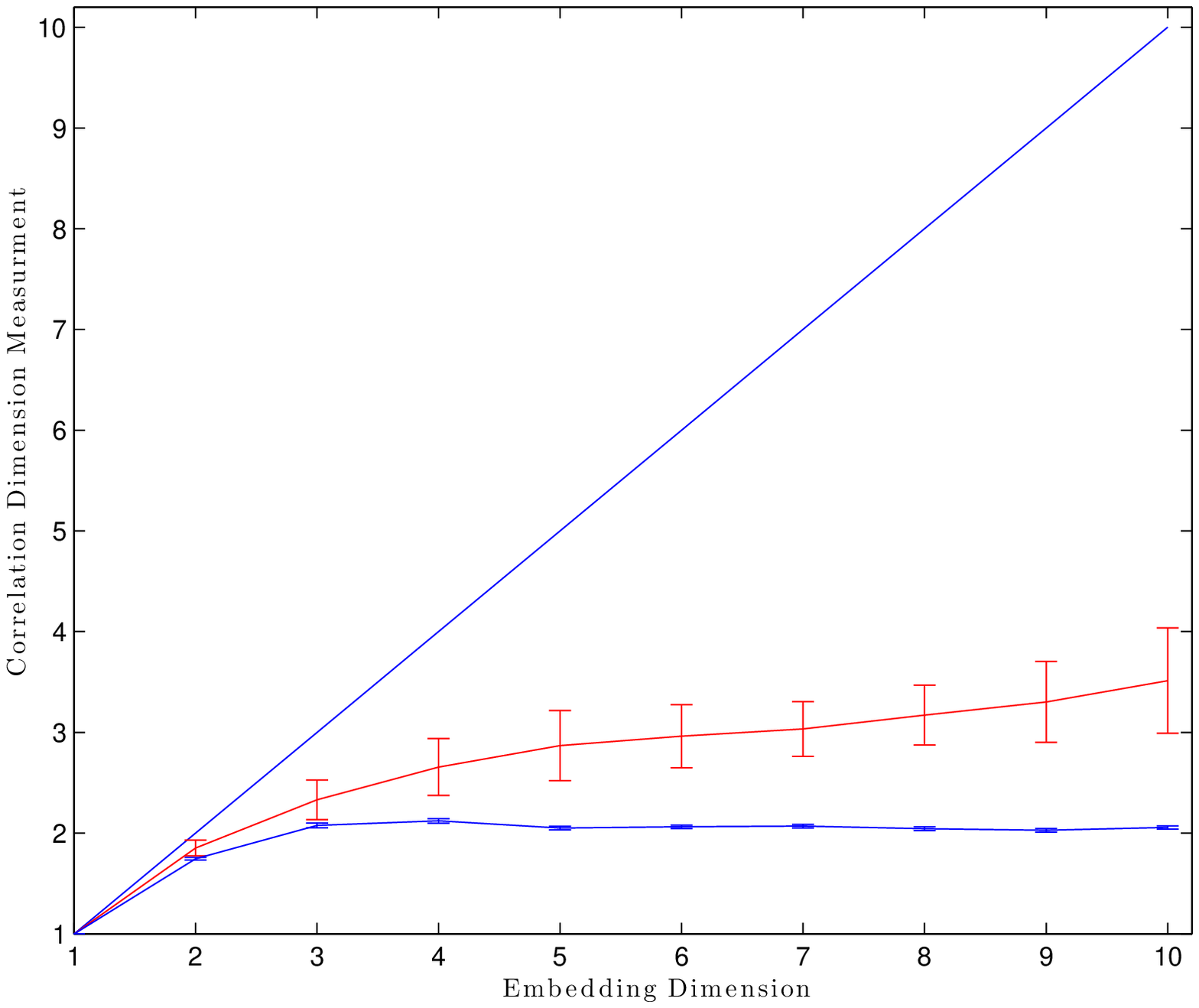} 
			     		\label{fig:Dim}}
			}
		\caption{(a) $\ln{C}$ versus $\ln{R}$ for PSR B1828$-$11. The dashed horizontal blue lines mark the rough scaling region, 
		$C=\frac{10}{n_{s}}$ to $0.10$. The solid horizontal red lines mark the flattest scaling region that is larger than a third of the rough 
		scaling region. (b) The flat blue line is the correlation dimension measurements, within the scaling in \emph{(a)}, versus embedding dimension. 
		The line at $45^{\circ}$ is what a purely random time series would be in that given $m$. The sloped red line is the mean and standard 
		deviation of 10 surrogate data sets (see text).}
		\label{fig:CorrDimMea}
	\end{figure}

	When this technique is applied to the four pulsars from Fig. \ref{fig:PulAtt}, two pulsars plateau very nicely. PSR B1828$-$11, seen in Fig. 
	\ref{fig:CorrDimMea}, averages to a correlation dimension of $2.06\pm0.03$ \footnote{The error calculation is presented in appendix \ref{sec:Error}}, 
	and PSR B1540$-$06 averages to $2.50\pm0.09$, for embedding dimensions greater than 3. The other two pulsars fail to converge to a constant. We 
	attribute this non-convergence to the sparse coverage, where each pass around the attractor is too far apart to get a proper measurement. 
	
\subsubsection{Surrogate data}
	We want to guarantee that any plateaus are due to geometric correlations and cannot be produced by random processes. Therefore, 
	we would like to test several different time series with the similar mean, variance, and autocorrelation function as the original data set. 
	These data sets are known as \emph{surrogate data}.
	
	The idea of surrogate data was first introduce in \cite{Theiler199277}, but we use an improved version that was presented in 
	\cite{1999chao.dyn..9041S} for our surrogate data sets. They start by shuffling the order of the original time series, and then take the Fourier 
	transform of this shuffled series. Keeping the phase 	angle of the shuffled set, they replace the amplitudes with the Fourier amplitudes of 
	the original times series. Then they reverse the Fourier transform, and create the desired surrogate data set. \cite{1999chao.dyn..9041S} 
	iteratively do this until changes in the Fourier spectrum are reduced. Fortunately, this is all done in the program \emph{surrogates} 
	in the \textsc{tisean} package \citep{TISEAN}, which we used for our analysis.
	
	We form 10 surrogate data sets for each pulsar and run them through the same non-linear analysis as the post processed time series. We then find 
	the average and standard deviation for the surrogates' correlation dimension for each embedding dimension to trace out the region were we should 
	expect other surrogates to lie. We are confident that the original time series is not due to a random process if its correlation dimension 
	measurements lie away from this region.
	
\begin{figure} 
		 \centering
		 \mbox{
				\subfigure[]
			 	{\includegraphics[width=.52\linewidth]{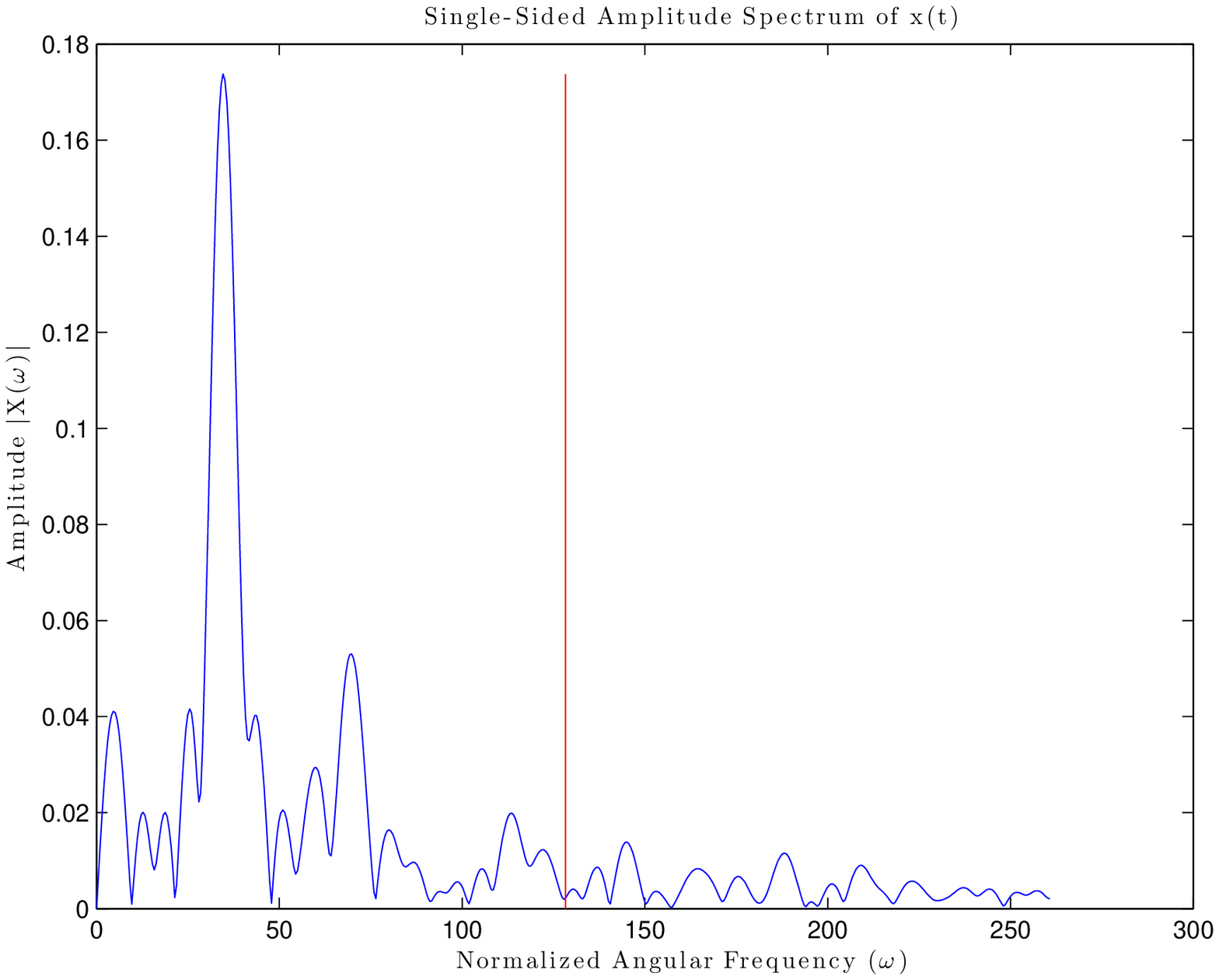}} 
				\subfigure[]
				{\includegraphics[width=.5\linewidth]{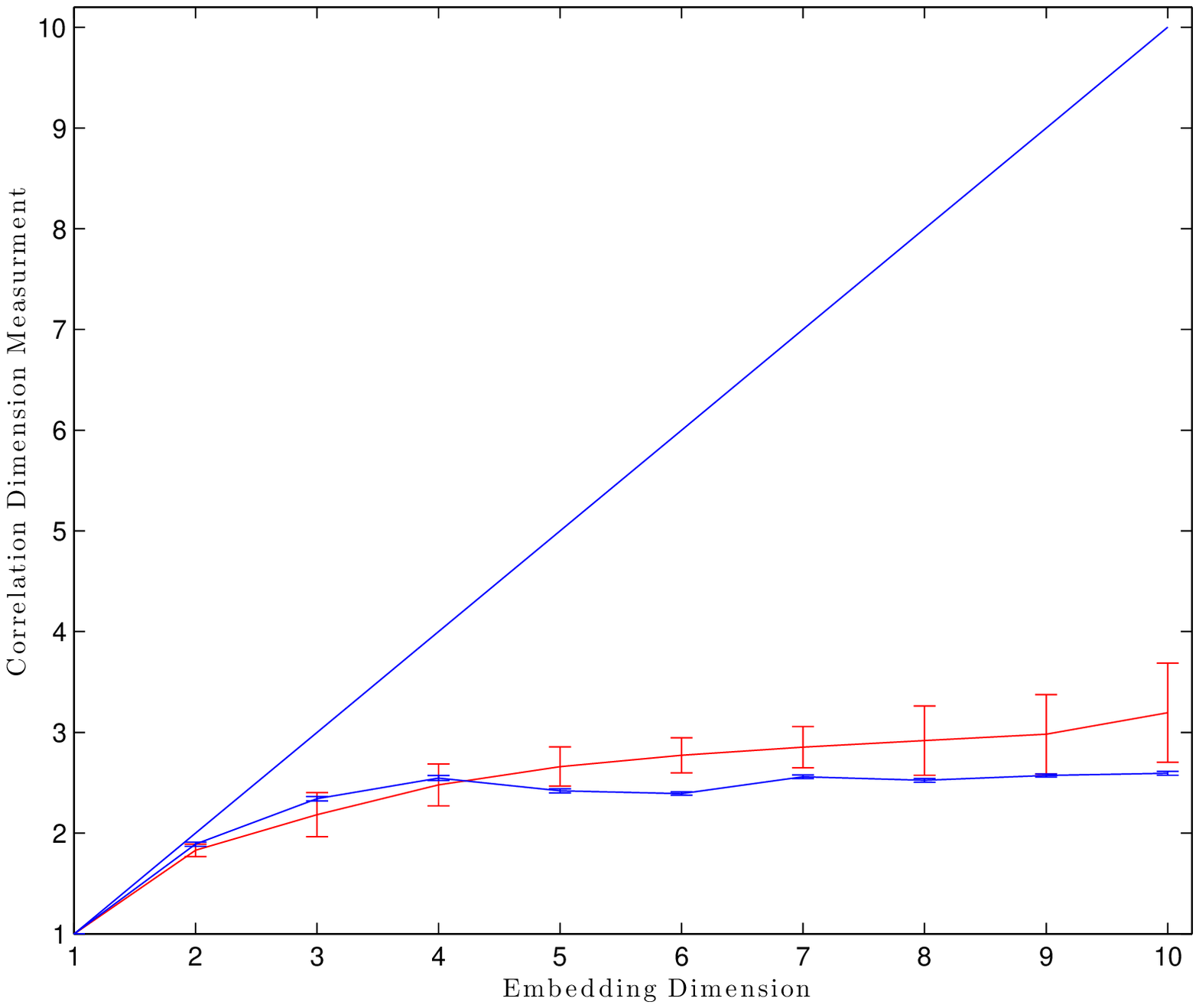}}
			}
		 \caption{ (a) The amplitudes of the Fourier series versus normalized frequency for PSR B1540$-$06. The vertical red line marks where we set the cut of frequency. (b) The blue error bars show the correlation dimension measurement, and the red error bars show the surrogate data region for PSR 
		 B1540$-$06.}
		\label{fig:PulDim}
	\end{figure}

	As seen in Fig. \ref{fig:CorrDimMea}, PSR B1828$-$11 is well outside the surrogate region, suggesting that it is a true dimension measurement. The 
	measurements for PSR B1540$-$06, seen in Fig. \ref{fig:PulDim}, marginally misses this boundary, but because this pulsar is highly periodic, the 
	Fourier spectrum is dominated by a single frequency. This restricts the complexity of the surrogate to where little change in dimension is expected.
	We will see more on this in Section \ref{sec:BT}.   
		
\subsubsection {Correlation Benchmark Testing} \label{sec:BT}
	Though each part of the algorithm works independently, we want to ensure that it all works together correctly. Therefore we
	 need to run through the process with a time series from known attractors under similar conditions to our original data, in order to see if we
	 receive similar dimensions. For this analysis we chose the Lorenz and R\"ossler attractors, whose dimensions have been
	  documented as being $2.07\pm0.09$ and $1.99\pm0.07$ respectively \citep{Sprott:2003fk}. 
	
	We first generate the $x$ component time series for a corresponding attractor, making sure that our initial conditions are well
	 on the attractor. Then, we extract a section from this time series out to the same number of turning points as used in the pulsar
	 time series. This extracted section was then normalized on both axes, as we did in Section \ref{sec:MtG}. We then resample 
	 the normalized time series by applying a cubic spline to the normalized times for the pulsar time series. This ensures that the 
	 new time series has the same time spacing and number of turning points as the pulsar time series. This new time 
	 series is then run through the algorithm to see if the proper dimension is achieved.
	
	\begin{figure} 
		\begin{tabular}{c | c}
			\includegraphics[width=.5\linewidth,trim=1100 0 0 0 ,clip=true]{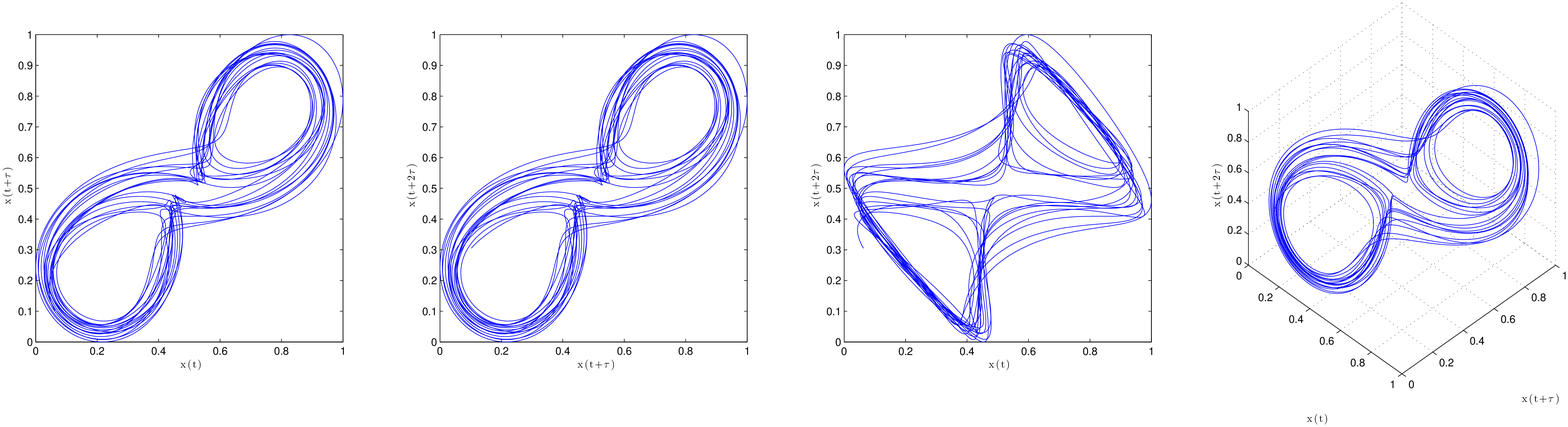} & 
			\includegraphics[width=.5\linewidth,trim=1100 0 0 0 ,clip=true]{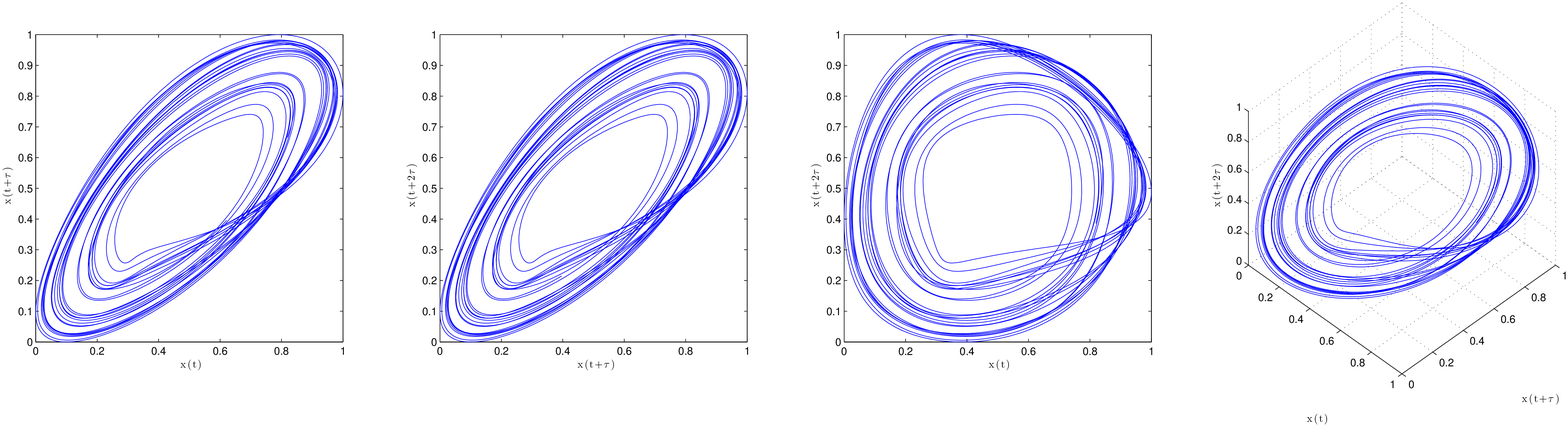} \\
			\includegraphics[width=.5\linewidth]{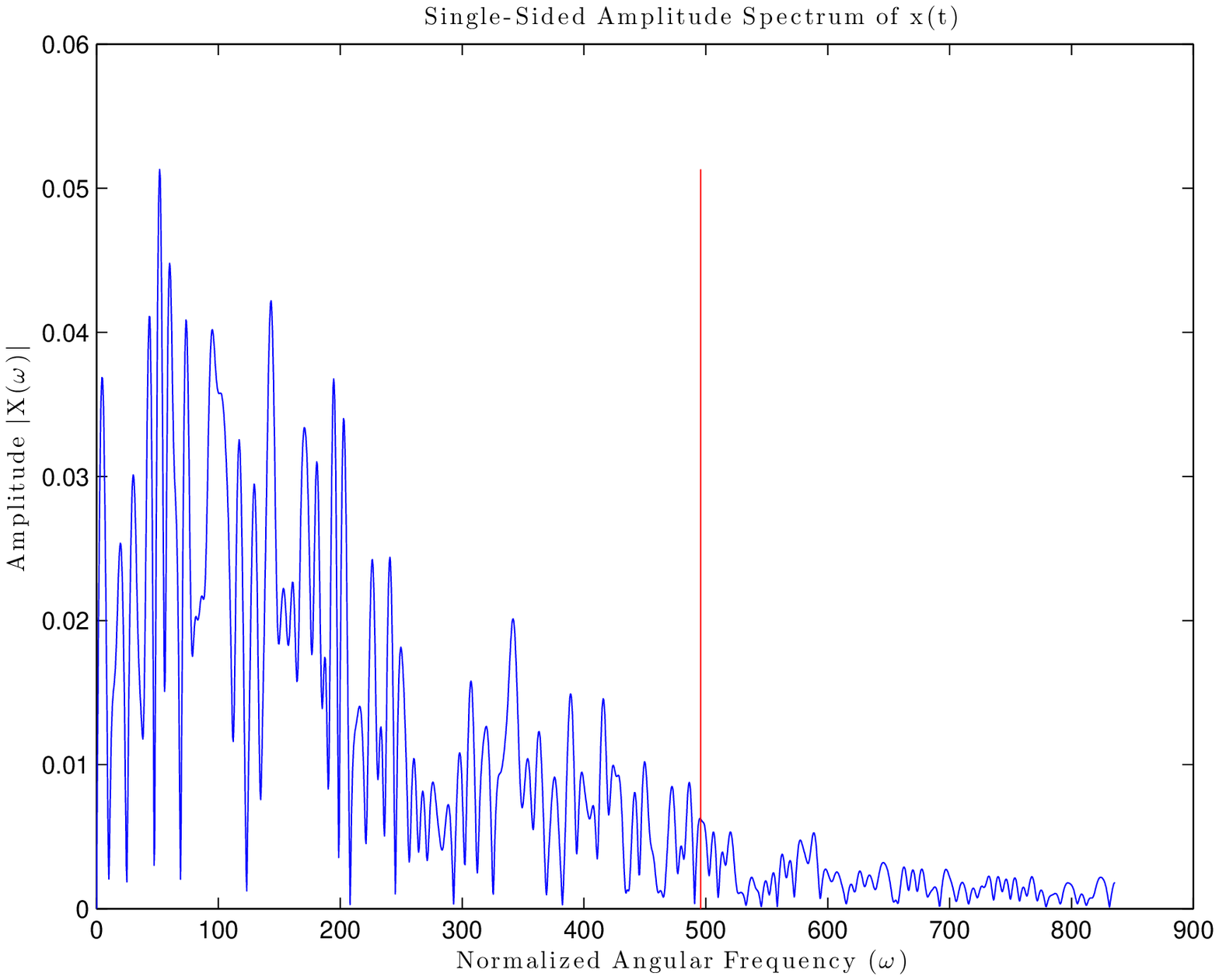} & 
				\includegraphics[width=.5\linewidth]{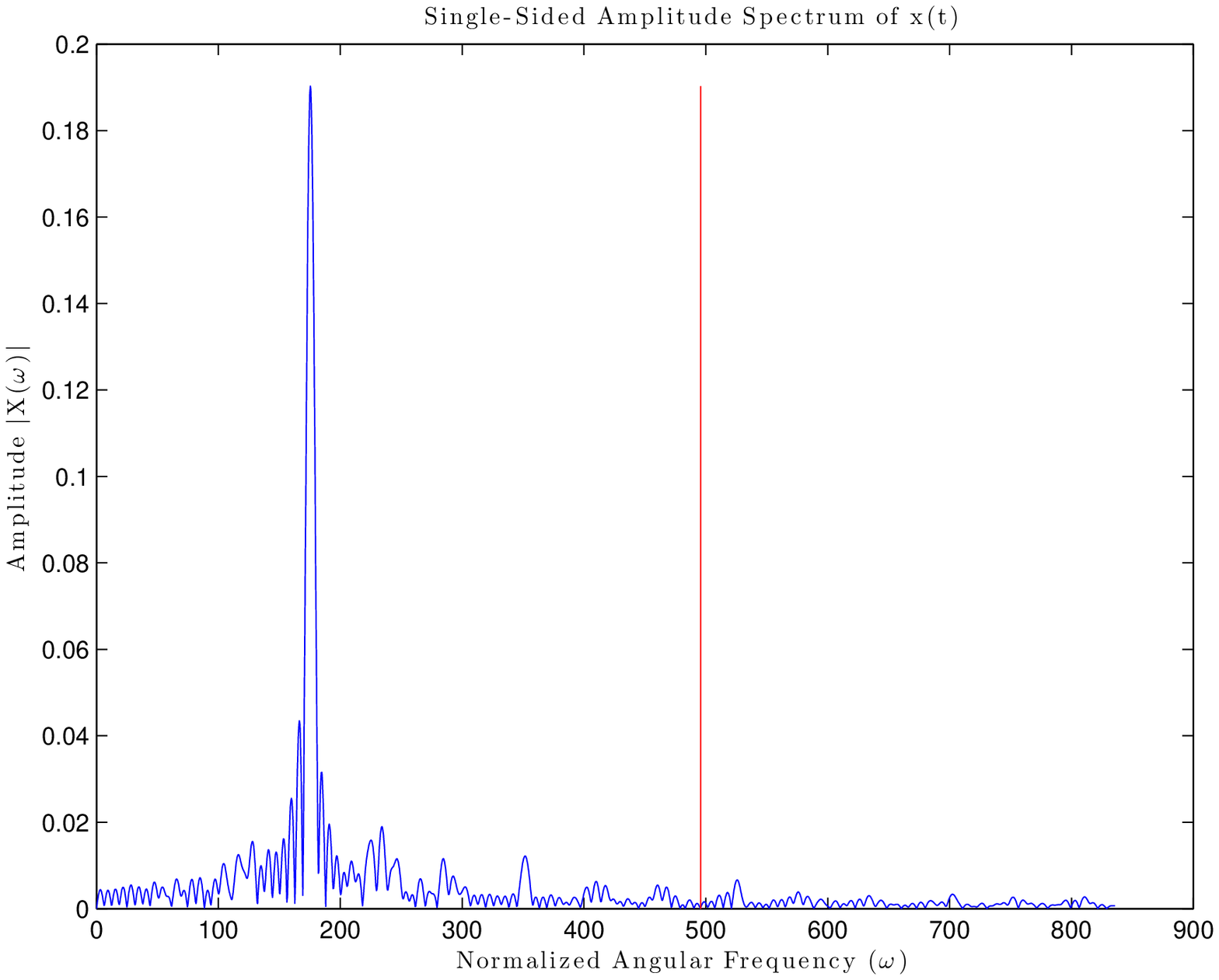} \\
			\includegraphics[width=.5\linewidth]{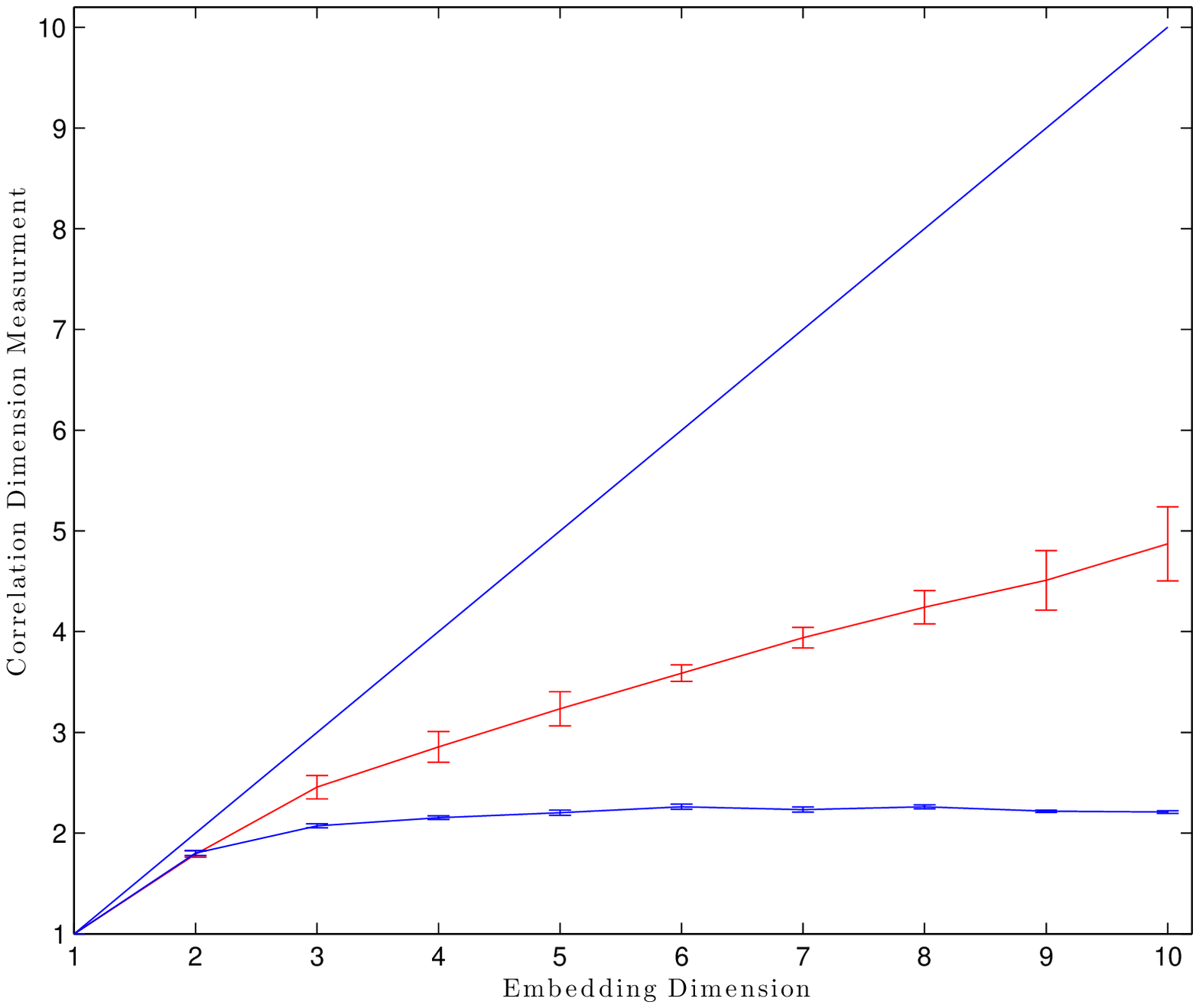} & 
				\includegraphics[width=.5\linewidth]{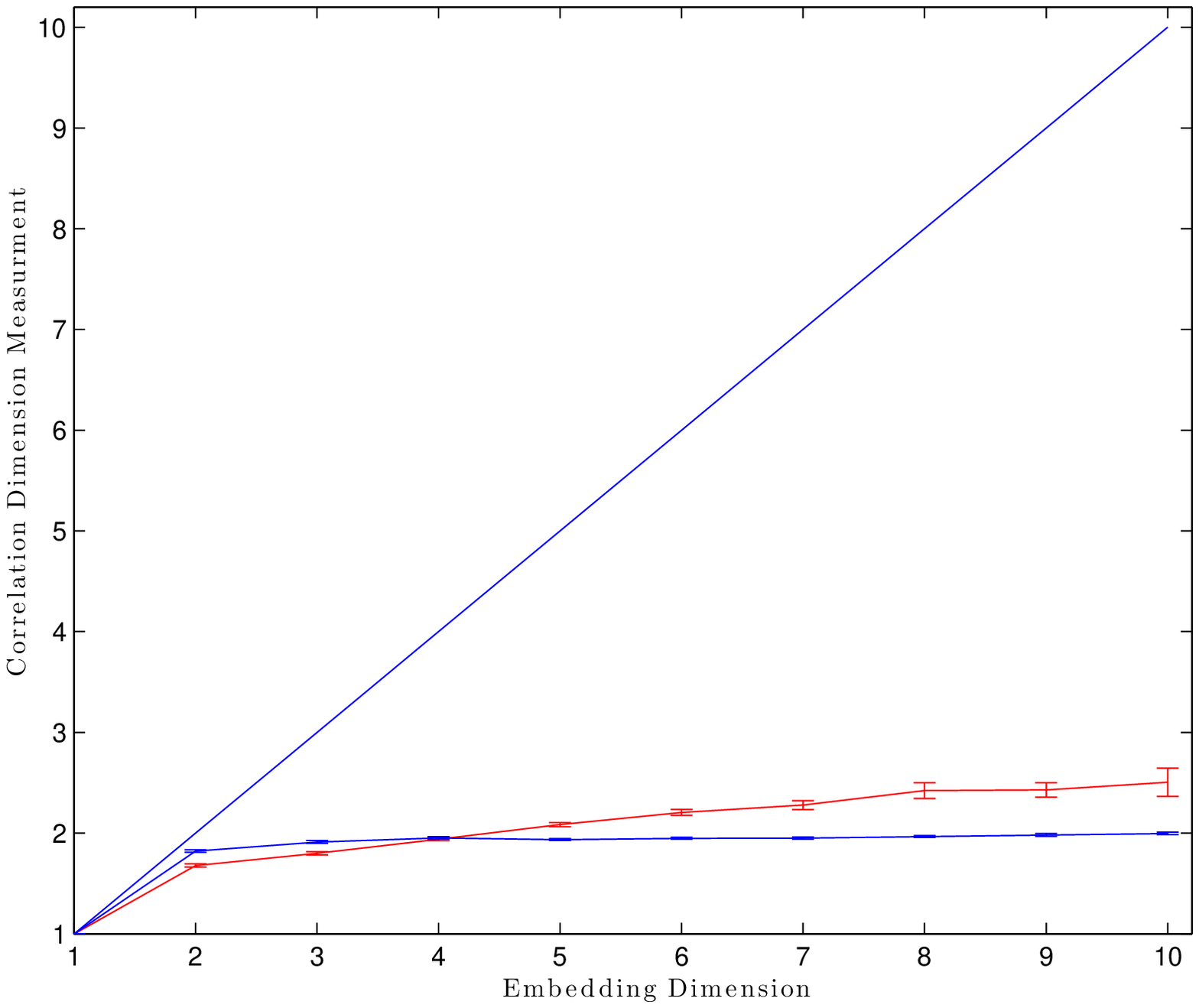} \\
		\end{tabular}
		\caption{Benchmark results for PSR B1828$-$11. \emph{Left column} is the Lorenz attractor results. \emph{Right
		 column} is the R\"ossler attractor results.\emph{Top row} are the reconstructed attractors. \emph{Middle row} are the
		 Fourier transforms where the vertical red line marks the cut off frequency from Section \ref{sec:Nr}. \emph{Bottom row} 
		 are the correlation dimension measurements (blue error bars) with the surrogate region (red error bars). }
		\label{fig:BenMar}
	\end{figure}
	
	An example from the correlation dimension measurements are seen in Fig. \ref{fig:BenMar}, where we can see that due to the
	 simplicity of the R\"ossler attractor in the frequency domain the surrogate data region only deviates by a small amount. We 
	 refer to this as a {\it borderline} detection. We also classify detections as being {\it inside} or {\it outside} based on their 
	 proximity to the surrogate region. Regardless, the correlation dimensions for both attractors plateau for all of the pulsar time series. 
	
	\begin{table}
		
		\begin{tabular}{ l c c c c }
		
		 & \multicolumn{2}{c }{{\bf{Lorenz}}} & \multicolumn{2}{c}{{\bf{R\"ossler}}}\\
		\hline
		 & Correlation & Surrogate & Correlation & Surrogate\\
		Pulsar & Dimension & Region & Dimension & Region\\
		\hline 
		B1828$-$11  & $2.20 \pm 0.06$ & O & $1.96 \pm 0.02$ & B\\
		B0740$-$28  & $2.11 \pm 0.03$ & O & $2.16 \pm 0.08$ & B\\
		B1826$-$17  & $2.00 \pm 0.10$ & O & $1.96 \pm 0.04$ & B\\
		B1642$-$03  & $2.05 \pm 0.07$ & O & $2.00 \pm 0.10$ & B\\
		B1540$-$06  & $1.90 \pm 0.20$ & O & $2.30 \pm 0.10$ & B\\
		J2148$+$63  & $2.33 \pm 0.05$ & O & $2.13 \pm 0.02$ & B\\
		B0919$+$06 & $1.90 \pm 0.10$ &  O & $2.00 \pm 0.10$& I  \\
		B1714$-$34  & $1.85 \pm 0.01$ & I & $2.87 \pm 0.07$  & I \\
		B1818$-$04  & $1.70 \pm 0.05$ & O & $2.26 \pm 0.08$ & B\\
		B2044$+$2740  & $1.83 \pm 0.03$ & O & $2.10 \pm 0.10$ & O\\
		B1903$+$07  & $2.11 \pm 0.09$ & O & $2.00 \pm 0.20$ & I\\
		B0950$+$08  & $1.68 \pm 0.09$ & O & $2.00 \pm 0.20$ & B\\
		\hline
		{\bf{Average}} & {$2.00 \pm 0.20$} && {$2.00 \pm 0.10$} &\\
		
		\end{tabular}

		\caption{ The weighted average for the correlation dimension for $m>3$ for the corresponding pulsar and attractor. The
		 surrogate region column marks weather the plateau line was either inside \emph{(I)}, outside \emph{(O)}, or borderline to
		 \emph{(B)} the surrogate data region. The average row is the weighted average of the correlation dimension 
		 measurements that were either marked as \emph{O} or \emph{B}.}
	\label{tab:CorrRe}
	\end{table}

	The results from all of the series are listed in Table \ref{tab:CorrRe}. We can see that the algorithm does rather well 
	considering we start with less than 300 data points. Although the algorithm is not refined enough to pinpoint the dimension, it is 
	sufficient to determine the number of governing variables. If we round the dimension measurement to the nearest 
	integer and then add one we will receive the proper number of variables, for any series that is either outside or borderline to 
	the surrogate region.

\subsection{Measuring the butterfly effect} \label{sec:BE}

	Though our correlation dimension measurement can narrow the number of governing variables, it is not definitive enough on its own
	to say that our attractors are chaotic. Therefore, we need to search for other signs of chaos in our topologies to ensure that they 
	are indeed strange attractors. 

\subsubsection{Lyapunov exponent}
	As mentioned in Section \ref{sec:CB}, the \emph{butterfly effect} is only present in chaotic systems. We should only see an 
	exponential divergence in nearby locations over time if the system is non-linear. The exponent of this increase is a characteristic 
	of the system and quantifies the strength of chaos \citep{BookTISEAN}. This is known as the \emph{Lyapunov exponent}.
	
	There are as many Lyapunov exponents as there are axes. Here we are only interested in measuring the maximum exponent
	 because it gives us the most information about our system. We start by looking at the distance between two locations, 
	 $\delta_0=\| \bf{x}_{t_1}- \bf{x}_{t_2} \|$, then recording this separation over time, 
	 $\delta_{\Delta t}=\| \bf{x}_{t_1+\Delta t}- \bf{x}_{t_2+\Delta t} \|$. The Lyapunov exponent, $\lambda$, would be 
	 \begin{equation}
	 	\delta_{\Delta t} \approx \delta_0 e^ {\lambda \Delta t}.
		\label{eq:LyEx}
	\end{equation}
	 Since the separation between two points cannot be greater than the size of the attractor itself, Equation \ref{eq:LyEx} will only hold true 
	 for $\Delta t$ values where $\delta_{\Delta t}$ is smaller than the attractor. 
	 
	 If $\lambda$ is positive, the system will be highly sensitive to initial conditions and therefore chaotic. If $\lambda$ is negative, the 
	 system would eventually converge to a single fixed point. If $\lambda$ is zero then this would represent a
	 limit cycle where the path keeps repeating itself and is said to be \emph{marginally stable} \citep{BookTISEAN}.
	 
	 In actuality, the separations do not grow everywhere on the attractor and locally they can even shrink, and with contributions from 
	 experimental noise it is more robust to use the average to obtain the Lyanpunov exponent. 
	 
	 The algorithm for this averaging was introduced independently by \cite{Rosenstein1993117} and \cite{Kantz199477}. One 
	  first picks a centre point located at $\bf{x}_{t_0}$ and takes note of the data points within a radius of $\epsilon$. This is known as the
	 neighbourhood. For a fixed $\Delta t$, the mean $\delta_{\Delta t}$ is calculated across the whole neighbourhood. The logarithm of this 
	 mean distance is then averaged over all points in the attractor. Therefore, one needs to compute
	 \begin{equation}
	 	S(\Delta t)=\frac{1}{N} \sum_{t_0=t_1}^{t_N} \ln{\left( \frac{1}{|{U({\bf{x}_{t_0}})}|} \sum_{{\bf{x}_t} \in {U(\bf{x}_{t_0})}} |{\bf{x}_{t_0+\Delta t}-\bf{x}_{t+\Delta t}|}\right)},
	 \end{equation}
	where $U(\bf{x}_{t_0})$ is the neighbourhood centred on $\bf{x}_{t_0}$. For our analysis we chose an $\epsilon$ that is twice as 
	large as our average distance to the nearest neighbour, while accommodating for the Theiler window. If the exponential relationship is 
	present, we should see an overall linear behaviour when $S(\Delta t)$ is plotted with respect to $\Delta t$. In order to conserve statistical
	accuracy we do not look beyond time-scales that are half the total time. 
		
	Similar to the correlation dimension, the Lyanpunov exponent is invariant to the number of embedding dimensions as long as $m > d$. Therefore we 
	calculate $S(\Delta t)$ from $m = 2$ to $10$ to ensure that the slopes in the linear regions remain constant. When noise is present in 
	deterministic systems, it causes a process similar to diffusion where $\delta_{\Delta t}$ expands proportionate to $\sqrt{\Delta t}$ on small 
	scales \citep{BookTISEAN}. This causes $S(\Delta t)$ to have a $\frac{1}{2}ln{(\Delta t)}$ behaviour, which produces a steep increase over 
	short time intervals.   	 
	
	 \begin{figure*}
	 \centering
		\mbox{
			\subfigure[]
	  			{\includegraphics[width=.5\linewidth]{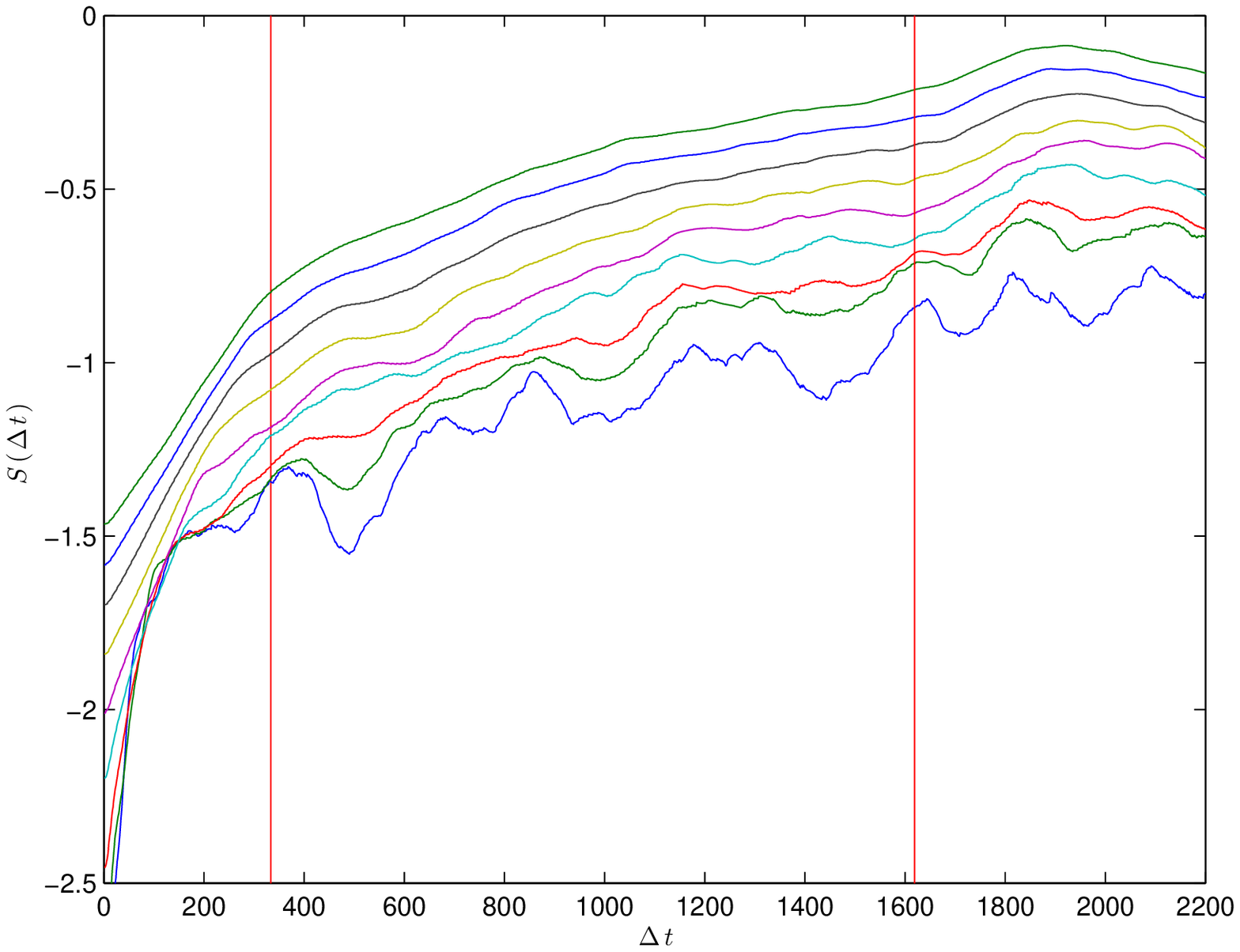}} 
	     			
			\quad
			 \subfigure[]		 	
			 	{\includegraphics[width=.5\linewidth]{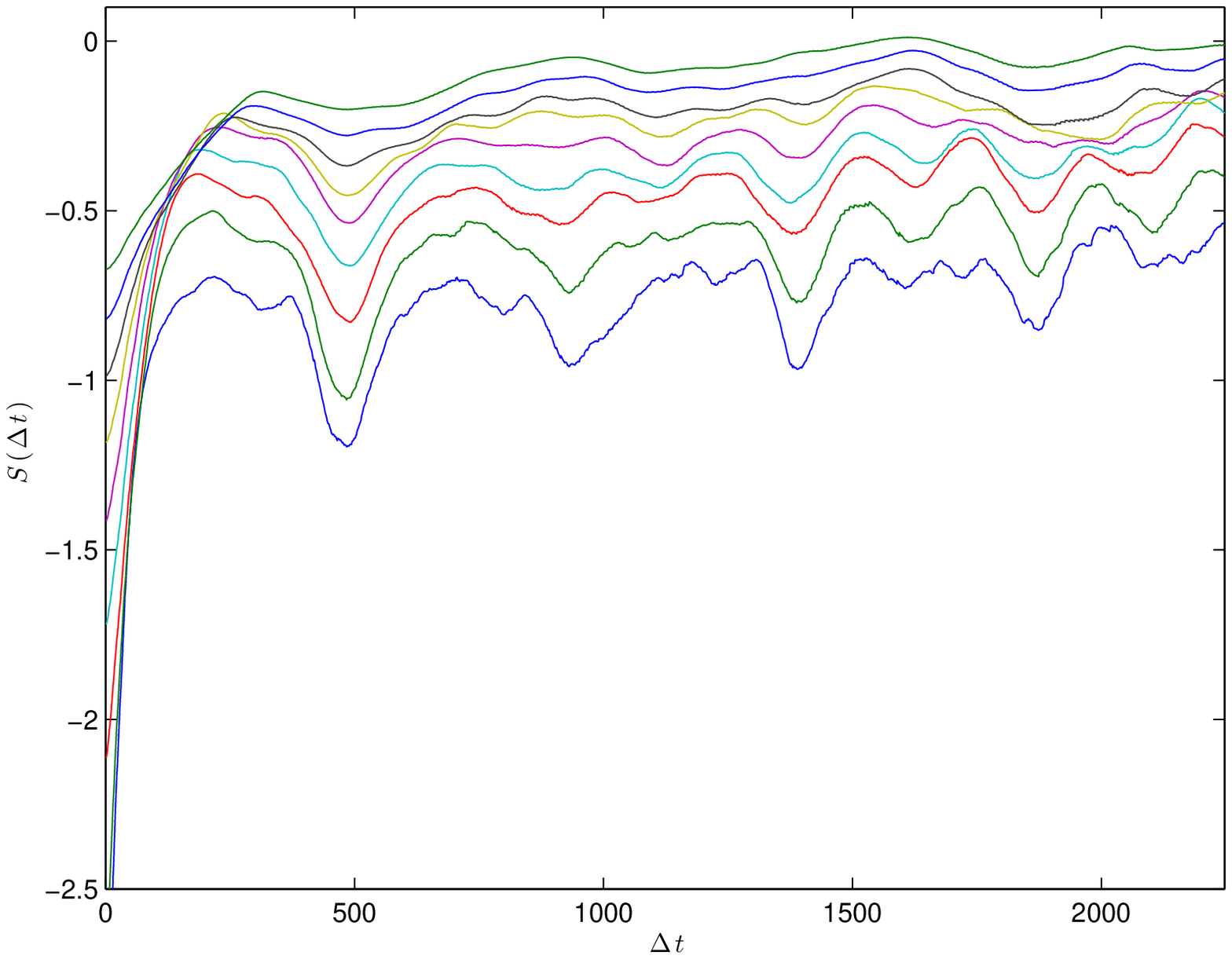}} 				
			}
			 \caption{(a) $S(\Delta t)$ versus time elapsed for PSR B1828$-$11. Each curve is for a different embedding dimension from 2-10 starting 
		 at the lowest curve. The vertical red lines mark the region where least square regression was applied. This corresponds to an average maximum
		 Lyapunov exponent of $\lambda_{\rm \rm max}=(4.0\pm0.3)\times10^{-4}$ inverse days. (b) $S(\Delta t)$ for a surrogate data set for B1828$-$11.} 
		\label{fig:SDt}
	\end{figure*}
	
	When this is applied to PSR B1828$-$11, seen in Fig. \ref{fig:SDt}, three distinct regions appear. On small scales, we can see a convergence 
	region due to a combination of a noise floor and non-normality \citep{TISEAN}. Because of these effects, it takes the neighbourhood a while to align 
	in the direction of the largest Lyapunov exponent. Once the neighbourhood has converged, we see a similar linear behaviour across all the embedding
	dimensions. This continues until the curves starts to saturate to a value on the scale of the attractor for that particular embedded dimension.
	
	The positive slope of the linear region suggests that B1828$-$11 is a chaotic system with a maximum Lyapunov exponent of
	 $(4.0\pm0.3)\times10^{-4}$ inverse days. When the same procedure is done to a surrogate data set of B1828$-$11, no linear region is observed.
	 Once the surrogate neighbourhood passes the convergent region it directly saturates around a constant value. This constant saturation would suggest
	 a Lyapunov exponent of zero, which is what is expected for a linear dynamical equation. The surrogate's strikingly different behaviour would imply that 
	 the linear region in B1828$-$11 is a real detection and is not a consequence of a random process. 
	 
	 When this is applied to the other pulsars there are no other definitive detections. Either the convergent region is dominant causing a direct
	 saturation, or the total time of the measurement is too small to positively state a separation between the convergent and linear regions. We 
	 will expand on these ideas more in the following section.  
	
\subsubsection{Lyapunov Benchmark Testing}

	Again we want to ensure that the algorithm is working in its entirety to produce the known values of the Lyapunov exponent. Therefore we follow 
	the same procedure as Section \ref{sec:BT}, matching the same number of turning points and sampling spacing of the pulsar. Because PSR 
	B1828$-$11 is our only detection, we concentrate our attention on its benchmark results. 
	
	The Lyapunov exponent changes with different parameters in the governing equations. Because of this we use the most common chaotic parameters 
	for the Lorenz and R\"ossler attractors\footnote{Lorenz: $\sigma=10$, $r=28$, $b=8/3$; R\"ossler : $a=b=0.2$, $c=5.7$}. Under these 
	conditions, the maximum Lyapunov exponent for the Lorenz attractor is $\lambda_{\rm \rm max} \simeq 0.9056$ and for the R\"ossler is $
	\lambda_{\rm max} \simeq 0.0714$ 
	 \citep{Sprott:2003fk}.
	 
	 \begin{figure} 
		\begin{tabular}{c | c}
			\includegraphics[width=.5\linewidth]{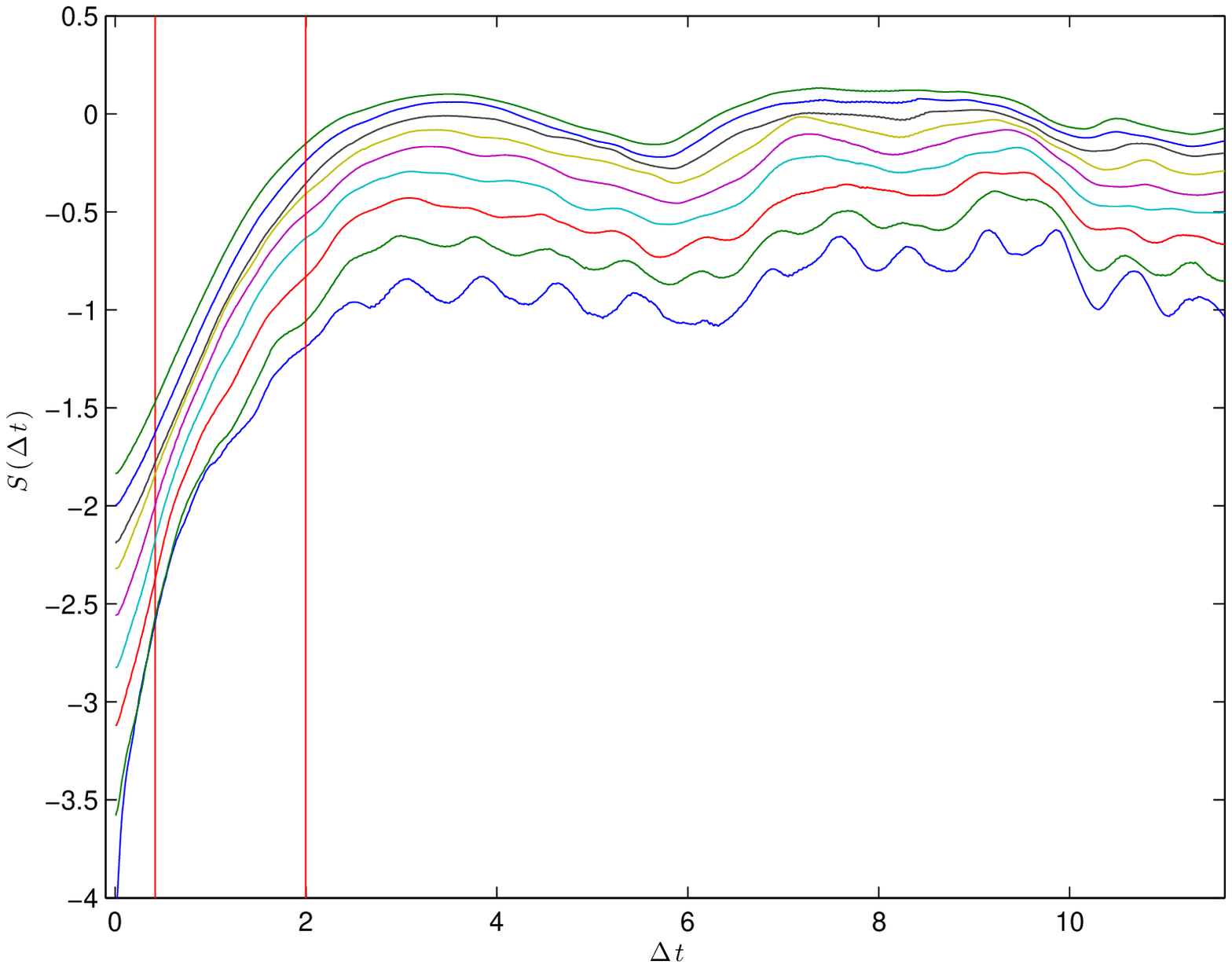} & 
				\includegraphics[width=.5\linewidth]{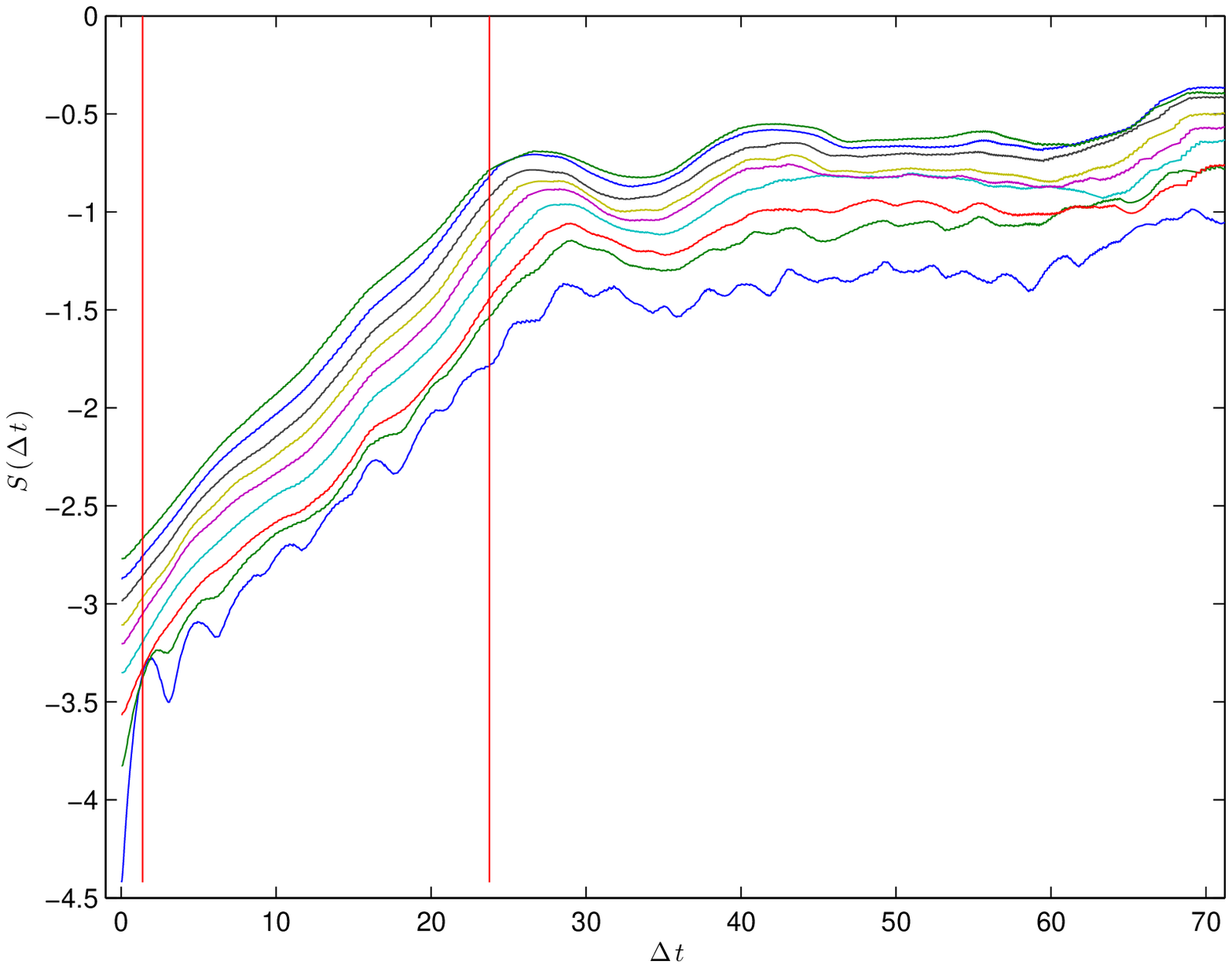} \\
			\includegraphics[width=.5\linewidth]{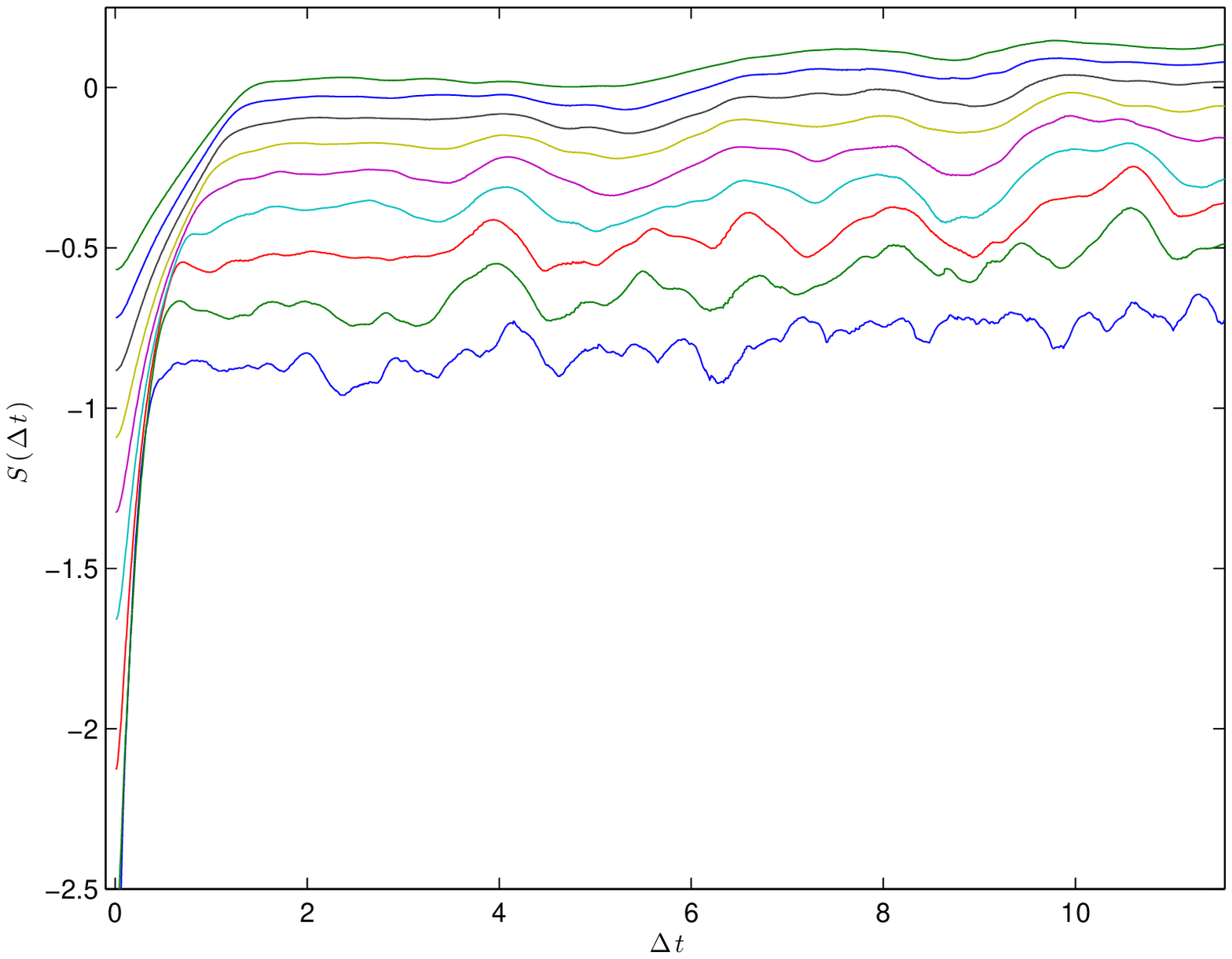} & 
				\includegraphics[width=.5\linewidth]{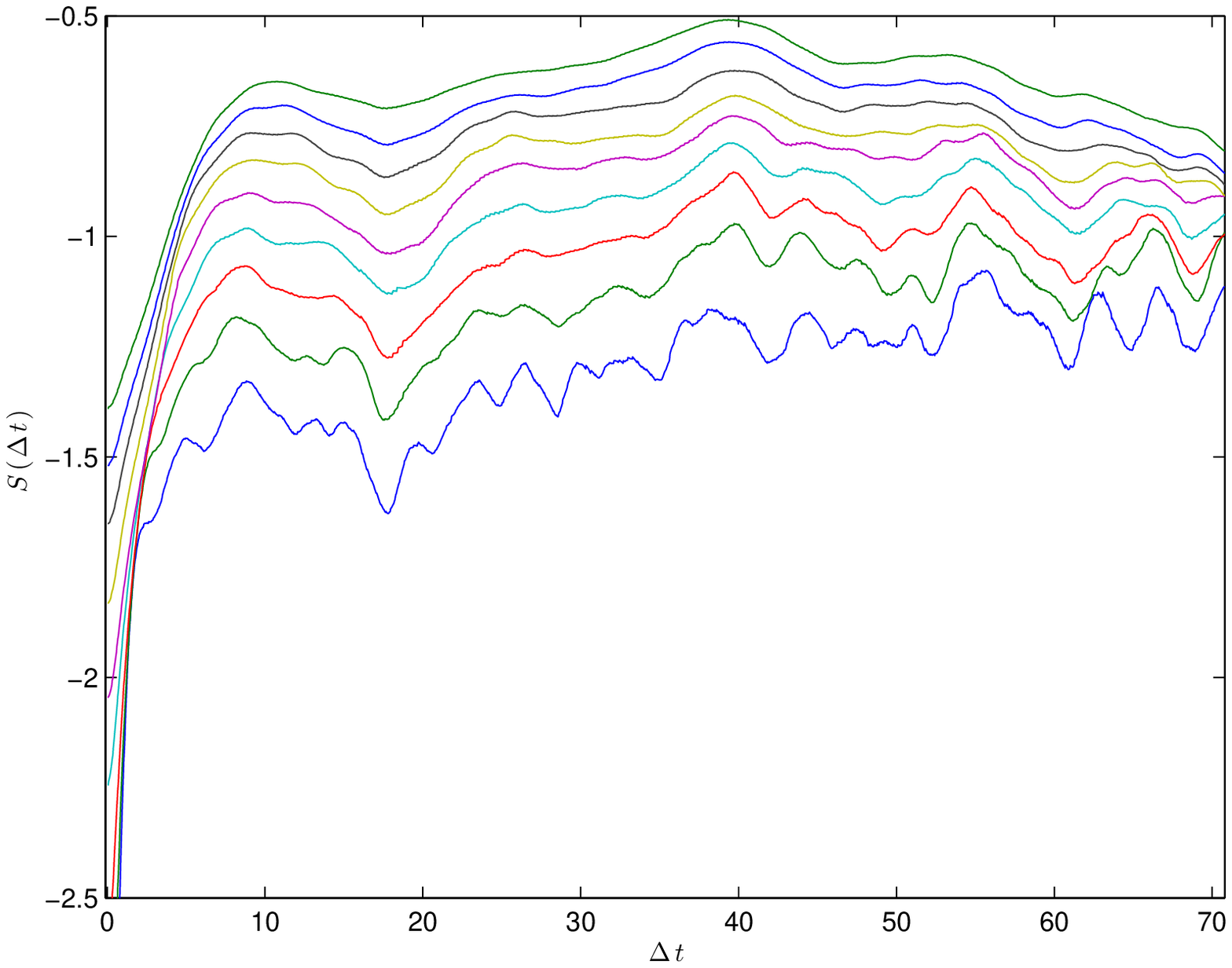} \\
					\end{tabular}
		\caption{Benchmark results for PSR B1828$-$11. \emph{Left column} is the Lorenz attractor results. \emph{Right
		 column} is the R\"ossler attractor results.\emph{Top row} are the $S(\Delta t)$ for reconstructed attractors. \emph{Bottom row} 
		 are the $S(\Delta t)$ surrogate data for the appropriate attractors. }
		\label{fig:BenMarLy}
	\end{figure}

	The results for the benchmark testing of B1828$-$11, seen in Fig. \ref{fig:BenMarLy}, seems to produce conflicting results. The algorithm estimates 
	a maximum exponent for the R\"ossler attractor to be a reasonable $\lambda_{\rm max}=0.080 \pm 0.003$, but for the Lorenz attractor it estimates
	an unreasonable value of $\lambda_{\rm max}=0.28 \pm 0.05$. 
	
	The cause of this discrepancy can be seen in the surrogate results. There we can see that the convergence time-scales are affecting our estimates.
	For the R\"ossler attractor, the time for the neighbourhood to converge is about 5 time units, while the exponential behaviour will last on time 
	scales of $\frac{1}{\lambda_{\rm max}} \approx 14$ units. Therefore, the R\"ossler attractor is outlasting the convergence time and will be able 
	to demonstrate its exponential behaviour. The Lorenz attractor with these parameter values is more sensitive to initial conditions and will only exhibit 
	its exponential behaviour on a time-scale of $\frac{1}{\lambda_{\rm max}} \approx 1$ unit, which is on the same scale of its convergence time. Therefore 
	the $S(\Delta t)$ for the Lorenz attractor is not portraying its true behaviour. 
	 
	This convergence time is partly inherent to the system and partly due to noise. Because of this there are two ways to correct this behaviour. The 
	first is to simply reduce the noise in the time series. The other way is to increase the sampling period of the measurements in order to start with 
	smaller neighbourhoods which will give the exponential behaviour several orders of magnitude to appear. Unfortunately, we are unable to comfortably 
	remove any more noise in the algorithm. Also, to reduce the neighbourhoods by orders of magnitude, we have to increase the number of data points by
	orders of magnitudes, which is currently beyond our system's capabilities. 
	
	Because of this convergence time, our algorithm is sensitive to lower levels of chaos with smaller Lyapunov exponents and is not able to
	distinguish between higher levels of chaos and the convergence region.   
	 
\section {Conclusions}

	By using a careful combination of turning point analysis, cubic splining, and Fourier transforms, we have constructed an algorithm that re-samples an 
	unevenly spaced time series without losing structural information. We have demonstrated this through an array of benchmark testing with known 
	chaotic time series under similar conditions to a given pulsar time series. This testing has shown that there are no significant changes in the 
	correlation dimension or the maximum Lyapunov exponents, when it was detectable. 
	
	These techniques were applied to the pulsar spin-down rates from \cite{DataHome}, where PSR B1828$-$11 exhibits clear chaotic 
	behaviour. We have shown that the measurements of its correlation dimension and maximum Lyapunov exponents are largely invariant across
	embedding dimensions. This, combined with its strikingly different reactions compared to its surrogate data sets, has shown that the chaotic 	
	characteristics in this pulsar are not caused by random processes.
	
	 The positive measurement of $\lambda_{\rm max}=(4.0\pm0.3)\times10^{-4}$ inverse days confirms that B1828$-$11 is chaotic in nature. For a system of 
	 equations to be chaotic there needs to be a minimum of three dynamical equations with three governing variables. The correlation 
	 dimension measurement of B1828$-$11, $D=2.06\pm0.03$, implies that there are a total of three governing variables, meeting the minimum 
	 requirements for chaos. One governing variable is clearly the spin-down rate of the pulsar. At this time we can only speculate on the other 
	 two. We know that the magnetic fields of a pulsar can change its dynamics and change the pulse profile. It has also been shown that changes to the 
	 superfluid interior can affect the long term dynamics of a pulsar \citep{Ho:2012mx}. Therefore, these seem to be great 
	 candidates. Beyond this, the subject is still a mystery and needs to be explored with non-linear simulations. Regardless of the model chosen, it is possible 
	 to perform some of the methods presented here, on the simulated data, to see if similar chaotic behaviours are present. 
	 		 
	 Knowing that B1828$-$11 is governed by three variables, the reconstructed attractor in Fig. \ref{fig:PulAtt} would be an accurate depiction of its 
	 strange attractor. Because this is visually similar to the other attractors in Fig. \ref{fig:PulAtt}, we would find it peculiar if their dynamics were not 
	 somehow related. Unfortunately, the techniques in this paper were unable to confirm this relationship. If these pulsars continue to be observed with an 
	 increase cadence, this would improve the correlation dimension and Lyapunov exponent measurements, perhaps to the point where these 
	 similarities could be quantified. With these measurements and a working model, estimates for the parameters can be given, which will give us further 
	 insight into the interior and/or exterior of these pulsars and how this relates to their dynamics.

\bibliographystyle{mn2e}
\bibliography{Nonlinear_ref}

\begin{thebibliography}{}

\bibitem[\protect\citeauthoryear{Anderson \& Itoh}{Anderson \&
  Itoh}{1975}]{ANDERSON:1975fk}
Anderson P.~W.,  Itoh N.,  1975, Nature, 256, 25

\bibitem[\protect\citeauthoryear{Argoul, Arneodo, Richetti, Roux \&
  Swinney}{Argoul et~al.}{1987}]{doi:10.1021/ar00144a002}
Argoul F.,  Arneodo A.,  Richetti P.,  Roux J.~C.,    Swinney H.~L.,  1987,
  Accounts of Chemical Research, 20, 436

\bibitem[\protect\citeauthoryear{Backer}{Backer}{1970}]{BACKER:1970fk}
Backer D.~C.,  1970, Nature, 228, 42

\bibitem[\protect\citeauthoryear{Brockwell \& Davis}{Brockwell \&
  Davis}{1996}]{BookYellowStats}
Brockwell P.~J.,  Davis R.~A.,  1996, Time series: theory and methods, 2nd ed
  edn.
Springer, New York, pp 312,313

\bibitem[\protect\citeauthoryear{Camilo, Ransom, Chatterjee, Johnston \&
  Demorest}{Camilo et~al.}{2012}]{0004-637X-746-1-63}
Camilo F.,  Ransom S.~M.,  Chatterjee S.,  Johnston S.,    Demorest P.,  2012,
  The Astrophysical Journal, 746, 63

\bibitem[\protect\citeauthoryear{{Delaney} \& {Weatherall}}{{Delaney} \&
  {Weatherall}}{1998}]{1998AAS...192.6802D}
{Delaney} T.,  {Weatherall} J.~C.,  1998, in American Astronomical Society
  Meeting Abstracts \#192 Vol.~30 of Bulletin of the American Astronomical
  Society, {Using Chaos to Test Pulsar Emission Models}.
p.~921

\bibitem[\protect\citeauthoryear{DeLaney \& Weatherall}{DeLaney \&
  Weatherall}{1999}]{0004-637X-519-1-291}
DeLaney T.,  Weatherall J.~C.,  1999, The Astrophysical Journal, 519, 291

\bibitem[\protect\citeauthoryear{Gold}{Gold}{1969}]{GOLD:1969uq}
Gold T.,  1969, Nature, 221, 25

\bibitem[\protect\citeauthoryear{Grassberger \& Procaccia}{Grassberger \&
  Procaccia}{1983}]{Grassberger1983189}
Grassberger P.,  Procaccia I.,  1983, Physica D: Nonlinear Phenomena, 9, 189

\bibitem[\protect\citeauthoryear{{Harding}, {Shinbrot} \& {Cordes}}{{Harding}
  et~al.}{1990}]{1990ApJ...353..588H}
{Harding} A.~K.,  {Shinbrot} T.,    {Cordes} J.~M.,  1990, Astrophysical
  Journal, 353, 588

\bibitem[\protect\citeauthoryear{{Hegger}, {Kantz} \& {Schreiber}}{{Hegger}
  et~al.}{1998}]{TISEAN}
{Hegger} R.,  {Kantz} H.,    {Schreiber} T.,  1998, in eprint
  arXiv:chao-dyn/9810005 {Practical implementation of nonlinear time series
  methods: The TISEAN package}.
p. 10005

\bibitem[\protect\citeauthoryear{Ho \& Andersson}{Ho \&
  Andersson}{2012}]{Ho:2012mx}
Ho W.~C.,  Andersson N.,  2012

\bibitem[\protect\citeauthoryear{Hobbs, Lyne \& Kramer}{Hobbs
  et~al.}{2006}]{1009-9271-6-S2-31}
Hobbs G.,  Lyne A.,    Kramer M.,  2006, Chinese Journal of Astronomy and
  Astrophysics, 6, 169

\bibitem[\protect\citeauthoryear{{Jones}}{{Jones}}{2012}]{2012MNRAS.420.2325J}
{Jones} D.~I.,  2012, Monthly Notices of the Royal Astronomical Society, 420,
  2325

\bibitem[\protect\citeauthoryear{Kantz}{Kantz}{1994}]{Kantz199477}
Kantz H.,  1994, Physics Letters A, 185, 77

\bibitem[\protect\citeauthoryear{Kantz \& Schreiber}{Kantz \&
  Schreiber}{2004}]{BookTISEAN}
Kantz H.,  Schreiber T.,  2004, Nonlinear time series analysis, 2nd ed edn.
Cambridge University Press, Cambridge, UK

\bibitem[\protect\citeauthoryear{Kendall \& Stuart}{Kendall \&
  Stuart}{1966}]{BookRedStats}
Kendall M.~G.,  Stuart A.,  1966, The advanced theory of statistics.
C. Griffin, London, pp 351,352

\bibitem[\protect\citeauthoryear{Kramer, Lyne, O'Brien, Jordan \&
  Lorimer}{Kramer et~al.}{2006}]{Kramer28042006}
Kramer M.,  Lyne A.~G.,  O'Brien J.~T.,  Jordan C.~A.,    Lorimer D.~R.,  2006,
  Science, 312, 549

\bibitem[\protect\citeauthoryear{Lorenz}{Lorenz}{1963}]{Lorenz:1963fk}
Lorenz E.~N.,  1963, Journal of the Atmospheric Sciences, 20, 130

\bibitem[\protect\citeauthoryear{Lorenz}{Lorenz}{1993}]{Lorenz:1993vn}
Lorenz E.~N.,  1993, The essence of chaos.
University of Washington Press, Seattle

\bibitem[\protect\citeauthoryear{Lyne, Hobbs, Kramer, Stairs \& Stappers}{Lyne
  et~al.}{2010}]{DataHome}
Lyne A.,  Hobbs G.,  Kramer M.,  Stairs I.,    Stappers B.,  2010, Science,
  329, 408

\bibitem[\protect\citeauthoryear{Manchester \& Taylor}{Manchester \&
  Taylor}{1977}]{Manchester:1977fk}
Manchester R.~N.,  Taylor J.~H.,  1977, Pulsars.
W. H. Freeman, San Francisco

\bibitem[\protect\citeauthoryear{Mathieu \& Scott}{Mathieu \&
  Scott}{2000}]{BookTurbFlow}
Mathieu J.,  Scott J.,  2000, An introduction to turbulent flow.
Cambridge University Press, Cambridge

\bibitem[\protect\citeauthoryear{Olsen \& Degn}{Olsen \&
  Degn}{1985}]{CambridgeJournals:4334828}
Olsen L.~F.,  Degn H.,  1985, Quarterly Reviews of Biophysics, 18, 165

\bibitem[\protect\citeauthoryear{{Petit} \& {Tavella}}{{Petit} \&
  {Tavella}}{1996}]{1996A&amp;A...308..290P}
{Petit} G.,  {Tavella} P.,  1996, Astronomy and Astrophysics, 308, 290

\bibitem[\protect\citeauthoryear{Press}{Press}{2007}]{Press:2007fk}
Press W.~H.,  2007, Numerical recipes: the art of scientific computing, 3rd ed
  edn.
Cambridge University Press, Cambridge, UK

\bibitem[\protect\citeauthoryear{Provenzale, Smith, Vio \& Murante}{Provenzale
  et~al.}{1992}]{Provenzale199231}
Provenzale A.,  Smith L.,  Vio R.,    Murante G.,  1992, Physica D: Nonlinear
  Phenomena, 58, 31

\bibitem[\protect\citeauthoryear{Rosen, McLaughlin \& Thompson}{Rosen
  et~al.}{2011}]{2041-8205-728-1-L19}
Rosen R.,  McLaughlin M.~A.,    Thompson S.~E.,  2011, The Astrophysical
  Journal Letters, 728, L19

\bibitem[\protect\citeauthoryear{Rosenstein, Collins \& Luca}{Rosenstein
  et~al.}{1993}]{Rosenstein1993117}
Rosenstein M.~T.,  Collins J.~J.,    Luca C. J.~D.,  1993, Physica D: Nonlinear
  Phenomena, 65, 117

\bibitem[\protect\citeauthoryear{R\"ossler}{R\"ossler}{1976}]{Rossler}
R\"ossler O.,  1976, Physics Letters A, 57, 397

\bibitem[\protect\citeauthoryear{Sauer, Yorke \& Casdagli}{Sauer
  et~al.}{1991}]{Sauer:uq}
Sauer T.,  Yorke J.~A.,    Casdagli M.,  1991, Journal of Statistical Physics,
  65, 579

\bibitem[\protect\citeauthoryear{{Scargle}}{{Scargle}}{1992}]{1992scma.conf..411S}
{Scargle} J.~D.,  1992, in {E.~D.~Feigelson \&amp; G.~J.~Babu} ed., Statistical
  Challenges in Modern Astronomy {Chaotic processes in astronomical data.}.
pp 411--436

\bibitem[\protect\citeauthoryear{{Schreiber} \& {Schmitz}}{{Schreiber} \&
  {Schmitz}}{1999}]{1999chao.dyn..9041S}
{Schreiber} T.,  {Schmitz} A.,  1999, in eprint arXiv:chao-dyn/9909041
  {Improved surrogate data for nonlinearity tests}.
p.~9041

\bibitem[\protect\citeauthoryear{Smale}{Smale}{1967}]{0202.55202}
Smale S.,  1967, Bull. Am. Math. Soc., 73, 747

\bibitem[\protect\citeauthoryear{Sprott}{Sprott}{2003}]{Sprott:2003fk}
Sprott J.~C.,  2003, Chaos and time-series analysis.
Oxford University Press, Oxford

\bibitem[\protect\citeauthoryear{Strogatz}{Strogatz}{1994}]{BookStrogatz}
Strogatz S.~H.,  1994, Nonlinear dynamics and Chaos: with applications to
  physics, biology, chemistry, and engineering.
Addison-Wesley Pub., Reading, Mass.

\bibitem[\protect\citeauthoryear{Takens}{Takens}{1981}]{Takens:1981fk}
Takens F., , 1981, Detecting strange attractors in turbulence

\bibitem[\protect\citeauthoryear{Theiler}{Theiler}{1986}]{Theiler:1986fk}
Theiler J.,  1986, Phys. Rev. A; Physical Review A, 34, 2427

\bibitem[\protect\citeauthoryear{Theiler, Eubank, Longtin, Galdrikian \&
  Farmer}{Theiler et~al.}{1992}]{Theiler199277}
Theiler J.,  Eubank S.,  Longtin A.,  Galdrikian B.,    Farmer J.~D.,  1992,
  Physica D: Nonlinear Phenomena, 58, 77

\end{thebibliography}

\clearpage
\appendix
\section{Turning Point Variance Derivation}
\label{sec:App}
	\emph{This derivation is presented in \cite{BookRedStats} but is rewritten here for the reader's convenience. }
	\\
	
	The number of turing points $p$ can be seen as a summation
	\begin {equation}
		p=\sum_{i=1}^{n-2}X_i,
	\end{equation}
	where $X_i$ is equal to one if $u_{i+1}$ is a turning point, and zero if not. We have already shown
	that the expectation value of the sum 
	\begin {equation}
	E(p)=\sum{}E(X_i)=\frac{2}{3}(n-2).
	\label{eq:EX}
	\end{equation}
	In order to calculate the variance we need to find the expectation of the square of the number of turning points
	\begin{equation}
	E(p^2)=E\left\{\left(\sum_{1}^{n-2}X_i\right)^2\right\}.
	\end{equation}
	This can be expanded to 
	\begin{multline}
		E(p^2)=E \left\{ \sum_{n-2}X_i^2 +2\sum_{n-3}X_iX_{i+1} +2\sum_{n-4}X_iX_{i+2} \right.\\
		\left.+\sum_{(n-4)(n-5)} X_iX_{i+k} \right\}, k\ne0,1,2,
	\end{multline} 
	where the suffixes to the $\sum$ signs indicate the number of terms over which summation takes place. We then have
	\begin{multline}
	E(p^2)=(n-2)E(X_i^2)+2(n-3)E(X_iX_{i+1})+2(n-4)E(X_iX_{i+2})\\+(n-4)(n-5)E(X_iX_{i+k}).
	\label{eq:long}
	\end {multline}
	Since $X_i^2=X_i$, we have
	\begin{equation} E(X_i^2)=\frac{2}{3}. \end{equation}
	For $k>2$, $X_i$ and $X_{i+k}$ are independent, for they have no values in common. Thus
	
	\begin{equation} E(X_iX_{i+k})=E(X_i)E(X_{i+k})=\frac{4}{9}. \end{equation}
	
	We still need to evaluate $E(X_iX_{i+1})$ and $E(X_iX_{i+2})$. For the first term to contribute, there needs to be two 
	consecutive turning points. In order to do this, we need a minimum of four data points, and define their values as
	 $u_{1}<u_{2}<u_{3}<u_{4}$. This will generate $4!$ or $24$ permutations, but one would find that only 10 permutations will
	 contain two turning points. This leads to an expectation value of 
	
	\begin{equation}
	E(X_iX_{i+1})=\frac{10}{24}=\frac{5}{12}.
	\end{equation}
	 
	 Similarly, for $E(X_iX_{i+2})$ five data points with five different values are needed. This produces $5!$ or $120$
	 permutations, but only 54 will produce a turning point in the second and fourth locations. This leads us to 
	\begin{equation}
	E(X_iX_{i+2})=\frac{54}{120}=\frac{9}{20}.
	\end{equation}
	Now, substituting these values into \ref{eq:long} we find
	\begin{align}
	E(p^2)&=\frac{2}{3}(n-2)+\frac{5}{6}(n-3)+\frac{9}{10}(n-4)+\frac{4}{9}(n-4)(n-5)\nonumber \\
	E(p^2)&=\frac{40n^2-144n+131}{90}.
	\label{eq:Ep2}
	\end {align}
	Finally, using the definition of the variance 
	\begin{equation}
	\sigma_{p}^2=E(p^2)-E(p)^2,
	\end{equation}
	 we insert our results from Equation \ref{eq:EX} and \ref{eq:Ep2} to conclude that 
	\begin{align}
	\sigma_{T}^2&=\frac{40n^2-144n+131}{90}-\frac{4(n-2)^2}{9}\nonumber\\
	&\nonumber\\
	\sigma_{T}^2&=\frac{16n-29}{90}. 
	\end{align}

\section{Correlation Dimension Error calculations} \label{sec:Error}

	The correlation sum, $C_i$, is the average of the pointwise measurements around the attractor for a certain test radius, $R_i$. Because of this, we can 
	use the standard deviation of a mean as the uncertainty of the correlation sum for that particular radius, 
	\begin{equation}
		\sigma_{C_i}=\frac{\sigma_{\rm pointwise}}{\sqrt{N-w}}
		\label{}
	\end{equation}
	where $N$ is the number of location vectors and $w$ is the Theiler window.
	
	Next we use standard propagation of error techniques, to form the first order estimate of the uncertainty for the natural logarithm of the correlation sum, 
	\begin{equation}
		\sigma_{\ln{C_i}}=\frac{\sigma_{C_i}}{C_i}.
		\label{}
	\end{equation}
	
	Within the desired scaling region, we follow the outline in \cite{Press:2007fk} for linear regression of least-squares to estimate the correlation 
	dimension and its uncertainty. \cite{Press:2007fk} breaks this calculation into several summations. We have rewritten these sums for this application
	as follows:  
	
	\begin{equation}
		S \equiv \sum_{i=N_{0}}^{N_{C}} \frac{1}{\sigma_{\ln{C_i}}^2}, \\
	\end{equation}
	
	\begin{equation}
		S_{\ln{R}} \equiv \sum_{i=N_{0}}^{N_{C}} \frac{{\ln{R}}_i}{\sigma_{\ln{C_i}}^2}, \\
	\end{equation}
	
	\begin{equation}
		S_{\ln{C}} \equiv \sum_{i=N_{0}}^{N_{C}} \frac{{\ln{C}}_i}{\sigma_{\ln{C_i}}^2},\\
	\end{equation}

	\begin{equation}
		S_{\ln{R}^2} \equiv \sum_{i=N_{0}}^{N_{C}} \frac{({\ln{R}}_i)^2}{\sigma_{\ln{C_i}}^2} \\
	\end{equation}
	and
	\begin{equation}
		S_{\ln{R}\ln{C}} \equiv \sum_{i=N_{0}}^{N_{C}} \frac{{\ln{R}}_i {\ln{C}}_i}{\sigma_{\ln{C_i}}^2}, \\
	\end{equation}
	where $N_0$ and $N_C$ is the minimum and maximum index of the correlation sum measurements for a particular embedding dimension within the 
	desired scaling region. With these summations, we are then able to calculate the slope. A common denominator value, 

	\begin{equation}
		\Delta \equiv S S_{\ln{R}^2} - (S_{\ln{R}})^2,  \\
	\end{equation}
	is used to calculate the correlation dimension,
	\begin{equation}
		d_m= \frac{S S_{\ln{R}\ln{C}} - S_{\ln{R}} S_{\ln{C}}}{\Delta},  \\
	\end{equation}
	and variance,
	\begin{equation}
		\sigma_{d_m}^2={\frac{S}{\Delta}},
	\end{equation}
	for each embedding dimension.
	
	To average the correlation dimension across the embedding dimensions, we use a weighted mean, where each measurement 
	is weighted based on the inverse variance. For simplicity, we use normalized weighting coefficients,  
	\begin{equation}
		w_m=\frac{\sigma_{d_m}^{-2}}{\sum_{m=3}^{10} \sigma_{d_m}^{-2}}.
	\end{equation}
	With this weighting, the mean correlation dimension,
	\begin{equation}
		<d>= \sum_{m=3}^{10} w_m d_m,
	\end {equation}
	and the variance, 
	\begin{equation}
		\sigma_{<d>}^2= {\sum_{m=3}^{10} w_m ( d_m-<d>)^2},
	\end{equation}
	are calculated. 
\end{document}